%% file: main.tex
\documentclass[11pt]{article}

\usepackage{geometry}
\usepackage{amsmath, amssymb, amsthm}
\usepackage{graphicx}
\usepackage{booktabs}
\usepackage{natbib}
\usepackage{setspace}
\usepackage{hyperref}
\usepackage{float}
\usepackage{array}
\usepackage{longtable}
\usepackage{pdflscape}
\usepackage{subcaption}
\usepackage{fancyvrb}
\usepackage{enumitem}
\usepackage{caption}
\usepackage{placeins}
\usepackage[strings]{underscore}

\hypersetup{
  colorlinks=true,
  linkcolor=blue,
  citecolor=blue,
  urlcolor=blue
}

\graphicspath{{figs/}}

\newtheorem{proposition}{Proposition}
\newtheorem{corollary}{Corollary}[proposition]
\newtheorem{remark}{Remark}

\newcommand{\draftfigure}[4][width=\linewidth]{
\begin{figure}[htbp]
\centering
\IfFileExists{#4}{
  \includegraphics[#1]{#4}
}
{
  \fbox{\parbox{0.85\textwidth}{\centering Missing figure:\\ \texttt{#4}}}
}
\caption{#2}
\label{#3}
\end{figure}
}

\newcommand{\csvtable}[3]{
\begin{table}[H]
\centering
\tiny
\caption{#1}
\label{#2}
\begin{minipage}{0.98\textwidth}
\IfFileExists{#3}{
{\tiny \VerbatimInput{#3}}
}
{
\fbox{\parbox{0.9\textwidth}{Missing CSV table: \texttt{#3}}}
}
\end{minipage}
\end{table}
}

\newcommand{\latextablefile}[3]{
\begin{table}[htbp]
\centering
\captionsetup{skip=2pt}
\tiny
\caption{#1}
\label{#2}
\IfFileExists{#3}{
\begingroup
\renewenvironment{table}[1][]{\ignorespaces}{\ignorespacesafterend}
\resizebox{\textwidth}{!}{\input{#3}}
\endgroup
}
{
\fbox{\parbox{0.9\textwidth}{Missing LaTeX table: \texttt{#3}}}
}
\end{table}
}

\newcommand{\fulltableinput}[1]{
\IfFileExists{#1}{
\input{#1}
}{
\begin{table}[H]
\centering
\fbox{\parbox{0.9\textwidth}{Missing full-table file: \texttt{#1}}}
\end{table}
}
}

\title{Mislearning of Factor Risk Premia under Structural Breaks\\
{\large A Misspecified Bayesian Learning Framework}}

\author{Yimeng Qiu}
\date{March 10, 2026}

\begin{document}

\maketitle

\begin{abstract}
While asset-pricing models increasingly recognize that factor risk premia are subject to structural change, existing literature typically assumes that investors correctly account for such instability. This paper studies how investors instead learn under a misspecified model that underestimates structural breaks. We propose a minimal Bayesian framework in which this misspecification generates persistent prediction errors and pricing distortions, and we introduce an empirically tractable measure of mislearning intensity $(\Delta_t)$ based on predictive likelihood ratios.

The empirical results yield three main findings. First, in benchmark factor systems, elevated mislearning does not forecast a deterministic short-run collapse in performance; instead, it is associated with stronger long-horizon returns and Sharpe ratios, consistent with an equilibrium premium for acute model uncertainty. Second, in a broader anomaly universe, this pricing relation does not generalize uniformly: mislearning is more strongly associated with future drawdowns, downside semivolatility, and other measures of instability, with substantial heterogeneity across anomaly families. Third, the cross-sectional relation between instability and mislearning is inherently conditional: while a monotonic link between break-proneness and average mislearning does not hold in the full cross-section, it re-emerges in low-friction (low-IVOL) environments where break-state severity is more comparable across assets.

\end{abstract}

\vspace{0.5em}

\newpage
\tableofcontents
\newpage

\section{Introduction}

Empirical asset pricing studies repeatedly document that factor risk premia are time-varying and subject to structural instability. Allowing for structural breaks often leads to rejection of the assumption of stable premia. For example, \citet{SmithTimmermann2022} document discrete breaks in cross-sectional risk premia and show that several classic factor premia decline in the later sample. Similarly, \citet{Chib2024} find that allowing for multiple breaks substantially reduces the set of priced factors.

This paper asks a behavioral question: \emph{how do investors learn about factor risk premia in environments with structural change?}

We propose a minimal framework of \emph{misspecified Bayesian learning}. Investors follow Bayesian updating but operate under an incorrect model class. Specifically, they filter the price of risk assuming a stable low-drift process while the true process contains occasional structural shifts. When structural changes occur, the Kalman gain under the stable model is too small, leading to slow belief adjustment. This generates persistent prediction errors and mispricing.

\subsection{Measuring Mislearning}

We introduce an empirically tractable measure of mislearning intensity based on predictive density comparisons between two models:
\begin{itemize}
\item a \textbf{stable model}, interpreted as the investor's belief system; and
\item a \textbf{break model}, interpreted as the researcher's benchmark for structural instability.
\end{itemize}

The log predictive likelihood ratio between the two models approximates relative entropy (KL divergence) and serves as a measure of mislearning.

\subsection{Testable Predictions}

The model generates three main predictions:
\begin{enumerate}[label=(\arabic*)]
\item Mislearning spikes around structural breaks.
\item High mislearning need not imply an immediate short-horizon collapse in realized performance. Instead, it should be most informative about how pricing distortions are resolved over medium to longer horizons. In lower-friction settings, elevated mislearning can be followed by higher long-horizon compensation, consistent with an uncertainty premium. In settings with more severe limits to arbitrage and price-impact frictions, the same state variable may instead load more strongly on subsequent instability and downside risk than on unconditional Sharpe compensation.
\item Factors with more unstable risk premia exhibit greater \emph{mislearning exposure}, though this exposure need not be one-dimensional. In particular, break frequency and break-state severity may vary independently across factor taxonomies. As a result, the mapping from break-proneness to average mislearning is inherently conditional and may not appear as a uniform monotonic relation in the presence of cross-sectional heterogeneity in break-state severity.
\end{enumerate}

\subsection{Relation to Literature}

Our framework differs from rational learning models such as \citet{Veronesi1999}, which assume correctly specified models. Instead, investors operate within a misspecified model class. The framework is related to behavioral models such as \citet{BSV1998} but focuses on structural instability in factor premia.

More broadly, the paper speaks to the literature on disappearing or unstable risk premia, model uncertainty, and belief-based asset pricing. Relative to recent work on time-varying factor premia, our emphasis is not merely that premia vary over time, but that a stable learning rule can become systematically wrong when the underlying data-generating process experiences discrete shifts.

\subsection{Market Structure and the Role of Passive Ownership}
\label{sec:institutional}

A secondary institutional question is whether market structure changes how a given belief error is transmitted into later outcomes. Using aggregate ICI data, we proxy the slow-moving share of rule-based capital by lagged passive ownership. The evidence does not suggest that passive ownership creates mislearning or robustly absorbs it on impact. Instead, in the benchmark FF6 and q5 systems, higher passive ownership is associated with weaker subsequent compensation following elevated mislearning, especially in cumulative returns. We therefore interpret passive ownership as a secondary market-structure modifier rather than as a primitive source of mislearning.

\subsection{Contribution}

The paper makes three contributions.

First, it provides a minimal misspecified-learning asset-pricing framework that directly maps structural instability in factor risk premia into a measurable state variable. The framework shows how slow belief updating under model misspecification generates persistent prediction errors and pricing distortions following structural breaks.

Second, it delivers a structural reinterpretation of the cross-sectional relation between instability and mislearning. Proposition~4 shows that unconditional mislearning arises from two distinct components: the frequency of structural breaks and the severity of mislearning conditional on break states. A stronger implication (Corollary~4.1) predicts a monotonic relation between break-proneness and average mislearning when break-state severity is comparable across assets. Empirically, this condition fails in the full cross-section. We show that this failure is driven by systematic heterogeneity in break-state severity. In particular, idiosyncratic volatility (IVOL), a standard proxy for limits to arbitrage, strongly predicts break-state severity ($\mu_{1,k}$) but has no predictive power for break frequency ($\pi_k$). This separation allows us to use IVOL as an ex-ante screening variable: in low-IVOL (low-friction) environments, where severity is more homogeneous, the monotonic relation between break-proneness and mislearning re-emerges, whereas it breaks down in high-IVOL environments. These results establish that the cross-sectional implications of mislearning are inherently conditional on the limits-to-arbitrage environment.

Third, it provides an empirical implementation based on predictive density comparisons and documents how mislearning maps into future outcomes. In benchmark factor systems, elevated mislearning is associated with stronger long-horizon returns and Sharpe ratios, consistent with an uncertainty premium. In a broader anomaly universe, mislearning is instead more strongly linked to future instability, including drawdowns and downside semivolatility, with substantial heterogeneity across anomaly families.

\section{True Process: Factor Risk Premia with Structural Breaks}

Consider $K$ factor returns. To properly align the state transition with the observation timing in a standard state-space formulation, we specify:
\begin{equation}
f_{t+1} = \lambda_{t+1} + u_{t+1}, \qquad u_{t+1} \sim \mathcal{N}(0,\Sigma_u),
\label{eq:true_obs}
\end{equation}
where $\lambda_{t+1}$ represents the conditional expected factor return driving the realization at $t+1$ (such that $\mathbb{E}_t[f_{t+1}] = \mathbb{E}_t[\lambda_{t+1}]$).

The true state evolution is
\begin{equation}
\lambda_{t+1} = A\lambda_t + \eta_{t+1} + J_{t+1},
\label{eq:true_state}
\end{equation}
where $\eta_{t+1}\sim \mathcal{N}(0,\Sigma_\eta)$ and $J_{t+1}$ represents structural breaks:
\begin{equation}
J_{t+1} =
\begin{cases}
0 & \text{with probability } 1-p,\\
\zeta_{t+1} & \text{with probability } p,
\end{cases}
\qquad
\zeta_{t+1}\sim \mathcal{N}(\mu_J,\Sigma_J).
\label{eq:true_jump}
\end{equation}

This specification captures a simple but empirically relevant environment: most of the time expected premia evolve gradually, but occasionally they shift discretely due to changes in macro conditions, investor clientele, market structure, or factor crowding.

\section{Investor Beliefs: Misspecified Bayesian Learning}

Investors observe factor returns but believe that the latent state evolves smoothly without jumps:
\begin{equation}
f_{t+1} = \lambda_{t+1} + u_{t+1}, \qquad u_{t+1}\sim \mathcal{N}(0,\Sigma_u),
\label{eq:belief_obs}
\end{equation}
\begin{equation}
\lambda_{t+1} = A\lambda_t + \tilde{\eta}_{t+1}, \qquad \tilde{\eta}_{t+1}\sim \mathcal{N}(0,\tilde{\Sigma}_\eta),
\label{eq:belief_state}
\end{equation}
with the crucial misspecification that:
\begin{equation}
\tilde{\Sigma}_\eta \ll \Sigma_\eta,
\label{eq:belief_smallvar}
\end{equation}
and no explicit jump component.

Thus investors are Bayesian, but within a misspecified model class: they underestimate state volatility and ignore the possibility of breaks. Under this belief system, posterior beliefs follow the standard Kalman filter.

Let
\begin{equation}
\lambda_t \mid \mathcal{F}_t \sim \mathcal{N}(\hat{\lambda}_t, P_t),
\end{equation}
where $\mathcal{F}_t=\sigma(f_1,\dots,f_t)$. The belief recursion is
\begin{align}
\hat{\lambda}_{t+1|t} &= A\hat{\lambda}_t,\\
P_{t+1|t} &= AP_tA^\top + \tilde{\Sigma}_\eta,\\
K_{t+1} &= P_{t+1|t}\left(P_{t+1|t}+\Sigma_u\right)^{-1},\\
\hat{\lambda}_{t+1} &= \hat{\lambda}_{t+1|t} + K_{t+1}\left(f_{t+1} - \hat{\lambda}_{t+1|t}\right),\\
P_{t+1} &= (I-K_{t+1})P_{t+1|t}.
\end{align}

Because $\tilde{\Sigma}_\eta$ is too small, the steady-state Kalman gain $K$ is too small, and belief adjustment is too slow following a structural shift.

\paragraph{Maintained misspecification in the subjective state variance.}
Throughout, we treat $\tilde{\Sigma}_\eta$ as a fixed hyperparameter of the investor's perceived model class rather than an object that is re-estimated in real time. This maintained rigidity is consistent with evidence that investors neglect low-probability contingencies and rely on parsimonious forecasting representations instead of continuously re-specifying rare-break processes \citep{GSV2012,FLM2010}. It is also consistent with the fact that learning about jump frequencies or state-variance parameters is much slower and less precise than filtering the latent state itself, especially when breaks are infrequent and the sample is finite \citep{CDJL2016}. Accordingly, the restriction $\tilde{\Sigma}_\eta \ll \Sigma_\eta$ is best interpreted as a tractable representation of persistent model-class rigidity, not as a literal claim that investors can never revise beliefs about state volatility.

\section{Mislearning Intensity}

Define the one-step-ahead predictive densities under the two models for the realized return at time $t$, conditional strictly on prior information $\mathcal{F}_{t-1}$:
\[
p_S(f_t \mid \mathcal{F}_{t-1}) \qquad\text{and}\qquad p_B(f_t \mid \mathcal{F}_{t-1}),
\]
where $S$ denotes the stable investor-belief model and $B$ denotes the break-aware benchmark.

The mislearning measure, evaluated at the end of period $t$, is
\begin{equation}
\Delta_t = \log \frac{p_B(f_t \mid \mathcal{F}_{t-1})}{p_S(f_t \mid \mathcal{F}_{t-1})}.
\label{eq:delta}
\end{equation}

By defining $\Delta_t$ strictly using information up to time $t$, we ensure that the state variable is fully observable at the time of portfolio formation, effectively eliminating any look-ahead bias when predicting future performance $Perf_{t \to t+h}$.

Large positive $\Delta_t$ indicates that the stable model assigns much lower probability to the realized return than the break model does. In that sense, $\Delta_t$ measures the severity of local model misspecification.

A rolling version is also useful to capture persistent regimes:
\begin{equation}
\overline{\Delta}_t(m)=\frac{1}{m}\sum_{j=0}^{m-1}\Delta_{t-j}.
\label{eq:rolling_delta}
\end{equation}

\section{Asset Pricing Implications}

To formalize the pricing distortions induced by mislearning, we embed the filtering problem into a stylized equilibrium framework. Consider a market with a risk-free asset and $K$ risky factors. A representative investor has constant absolute risk aversion (CARA) with a coefficient $\gamma$ and maximizes expected utility over next-period wealth. 

Assume that $f_{t+1}$ denotes the vector of risky factor excess returns and that the investor solves a one-period CARA-normal portfolio problem. In this section, we use $f_{t+1}$ throughout to denote factor excess returns, so the equilibrium implications are stated directly in return space. Under the subjective stable model,
\[
\mathbb{E}_t^S[f_{t+1}] = \mathbb{E}_t^S[\lambda_{t+1}] = A\hat{\lambda}_t.
\]
Strictly speaking, the investor's conditional covariance matrix of returns is
\[
\mathrm{Var}_t^S(f_{t+1}) = P_{t+1|t} + \Sigma_u,
\]
where $P_{t+1|t}$ is the filtered state uncertainty. For tractability, however, we abstract from state-uncertainty variation in portfolio demand and approximate the subjective return covariance by $\Sigma_u$ alone. Equivalently, we treat filtered state uncertainty as second order relative to return noise, i.e. $P_{t+1|t} \ll \Sigma_u$. Under this approximation, the investor's optimal portfolio demand vector at time $t$ is
\begin{equation}
x_t = \frac{1}{\gamma} \Sigma_u^{-1} \mathbb{E}_t^S[f_{t+1}]
= \frac{1}{\gamma} \Sigma_u^{-1} A\hat{\lambda}_t,
\label{eq:optimal_demand}
\end{equation}
where $\gamma$ is the coefficient of absolute risk aversion.

To endogenize the time-varying nature of the true risk premium $\lambda_t$, we introduce a time-varying, exogenous supply of factor assets, $S_t$, which can be interpreted as institutional allocation shifts, mechanical rebalancing, or noise trader demand. The market clearing condition requires $x_t = S_t$, which implies that the subjective risk premium required to clear the market is:
\begin{equation}
A\hat{\lambda}_t = \gamma \Sigma_u S_t
\label{eq:market_clearing}
\end{equation}
Crucially, suppose the unobserved supply follows $S_t = \bar{S} + \nu_t$, where the supply shock $\nu_t$ occasionally experiences discrete structural shifts due to sudden shifts in aggregate demand or market-clearing supply. Because market clearing implies
\[
m_t^S := \mathbb{E}_t^S[f_{t+1}] = \gamma \Sigma_u S_t,
\]
these supply shifts move the market-clearing component of subjective expected excess returns. The investor, however, continues to filter the premium under the misspecified stable model, so the subjective mean $m_t^S = A\hat{\lambda}_t$ need not coincide with the true conditional mean
\[
m_t^T := \mathbb{E}_t[f_{t+1}].
\]
Their difference,
\[
w_t := m_t^T - m_t^S,
\]
is the belief wedge that drives predictable subsequent return reversals.

\begin{proposition}[Slow updating after breaks]
Suppose the true process follows \eqref{eq:true_state}--\eqref{eq:true_jump}, while investors filter using \eqref{eq:belief_obs}--\eqref{eq:belief_smallvar}. If a nonzero break occurs at time $t^\star$, then the posterior mean error $\hat{\lambda}_t-\lambda_t$ remains systematically biased for multiple periods after $t^\star$. The persistence of this bias increases as $\tilde{\Sigma}_\eta$ becomes smaller.
\end{proposition}

\textit{See Appendix \ref{app:proofs} for the formal proof.}

\begin{proposition}[Mislearning spikes near structural breaks]
When realized returns are more consistent with the predictive density of the break model than that of the stable model, mislearning intensity $(\Delta_t)$ rises. This increase is monotonically larger when the mean shift of the structural break is larger. Furthermore, \textbf{conditional on a sufficiently large break-consistent realization}, a more rigid stable model (i.e., a smaller subjective state variance) strictly magnifies the likelihood gap and produces larger mislearning spikes.
\end{proposition}

\begin{proposition}[Uncertainty Premium and Future Performance]
\label{prop:uncertainty_premium}
When mislearning intensity ($\Delta_{t}$) spikes following a structural break, the asset enters a regime of elevated model uncertainty and belief divergence. In equilibrium, investors demand a higher risk premium to compensate for this ambiguity. Consequently, while $\Delta_{t}$ exhibits only weak predictive power for future realized volatility prior to standard risk controls, it is associated with elevated future Sharpe ratios in long-horizon regressions. This dynamic reflects an equilibrium compensation for acute model risk, rather than a deterministic collapse in short-term factor performance.
\end{proposition}

\noindent \textit{Interpretation and scope: low-friction benchmark and outcome bifurcation.}
To clarify why the pricing implication in Proposition~\ref{prop:uncertainty_premium} need not generalize uniformly across assets, consider the equilibrium expected excess-return decomposition established in Appendix~\ref{app:proofs} (Theorem~1):
\begin{equation}
\mathbb{E}_t[f_{t+1}] = w_t + \gamma \Sigma_u S_t,
\label{eq:bifurcation}
\end{equation}
where
\[
m_t^S := \mathbb{E}_t^S[f_{t+1}], \qquad
m_t^T := \mathbb{E}_t[f_{t+1}], \qquad
w_t := m_t^T - m_t^S.
\]
The term $w_t$ captures the belief wedge generated by misspecified filtering, while $\gamma \Sigma_u S_t$ captures the market-clearing component associated with exogenous supply shocks. Proposition~\ref{prop:uncertainty_premium} is most naturally interpreted as a \textbf{low-friction benchmark}: when the price-impact component remains small relative to the gradual correction embodied in $w_t$, elevated mislearning is resolved primarily through belief adjustment and appears as a compensated uncertainty premium.

When limits to arbitrage are more severe, however, the same mislearning shock need not be expressed through future Sharpe compensation. If the price-impact component becomes sufficiently important relative to the belief-correction component, elevated mislearning is more likely to be realized through price dislocation, realized volatility, drawdowns, downside semivolatility, or broader instability. This distinction provides a model-consistent interpretation of why the benchmark factor systems are more consistent with an uncertainty-premium mechanism, whereas the broader anomaly universe more often reflects the instability channel. Empirically, the IVOL-screened cross-sectional tests below exploit this logic by using IVOL as a reduced-form proxy for environments in which the low-friction benchmark is more or less likely to hold.

\begin{proposition}[Cross-factor mislearning exposure: decomposition]
\label{prop:cross_factor_decomposition}
Let $B_{k,t}$ denote the break-state indicator for factor $k$, and define
\[
\pi_k = \Pr(B_{k,t}=1),\qquad
\mu_{1,k} = \mathbb{E}[\Delta_{k,t}\mid B_{k,t}=1],\qquad
\mu_{0,k} = \mathbb{E}[\Delta_{k,t}\mid B_{k,t}=0].
\]
Then the unconditional average mislearning exposure of factor $k$ satisfies
\[
\mathbb{E}[\Delta_{k,t}] = \pi_k \mu_{1,k} + (1-\pi_k)\mu_{0,k}.
\]
Moreover, for any fixed spike threshold $c$,
\[
\Pr(\Delta_{k,t}>c)
=
\pi_k q_{1,k}(c) + (1-\pi_k)q_{0,k}(c),
\]
where
\[
q_{1,k}(c)=\Pr(\Delta_{k,t}>c\mid B_{k,t}=1),
\qquad
q_{0,k}(c)=\Pr(\Delta_{k,t}>c\mid B_{k,t}=0).
\]
Thus cross-factor variation in unconditional mislearning can arise through either break frequency or break-state severity, and the same is true for the frequency of large mislearning spikes. This decomposition holds without imposing any restriction on the relation between break frequency and break-state severity across factors.
\end{proposition}

\begin{corollary}[Conditional monotone exposure]
\label{cor:prop4_monotone}
Suppose the non-break component of mislearning is comparable across factors, so that $\mu_{0,k}=\bar{\mu}_0$, and define the break-state severity gap
\[
\delta_k := \mu_{1,k}-\mu_{0,k}.
\]
If $\delta_k \ge 0$ for all $k$ and $\delta_k$ is constant across factors or weakly increasing in $\pi_k$, then $\mathbb{E}[\Delta_{k,t}]$ is weakly increasing in $\pi_k$. Likewise, for any fixed threshold $c$, if
\[
q_{0,k}(c)=\bar q_0(c)
\qquad\text{and}\qquad
\eta_k(c):=q_{1,k}(c)-q_{0,k}(c)\ge 0
\]
is constant across factors or weakly increasing in $\pi_k$, then $\Pr(\Delta_{k,t}>c)$ is weakly increasing in $\pi_k$. This result provides a sufficient condition rather than a necessary one.
\end{corollary}
\noindent \textit{Empirical interpretation.}
Proposition~\ref{prop:cross_factor_decomposition} implies that cross-factor average mislearning can differ because factors break more often, because mislearning is more severe conditional on break states, or both. Corollary~\ref{cor:prop4_monotone} adds a stronger sufficient condition under which break-proneness maps monotonically into average mislearning and spike frequency. In the data, this stronger implication need not hold uniformly across the full cross-section because the break-state severity gap can itself be heterogeneous. A practical screening variable for this heterogeneity is the limits-to-arbitrage environment. In particular, Appendix~\ref{app:prop4-reducedform} shows that IVOL predicts break-state conditional severity $\mu_{1,k}$ but does not predict break-proneness $\pi_k$. This motivates using IVOL-sorted subsamples to evaluate where the corollary is more likely to hold empirically.

\begin{remark}[Cross-sectional implication]
Proposition~\ref{prop:cross_factor_decomposition} shows that once cross-factor heterogeneity in break-state severity is admitted, the low-friction benchmark in Proposition~\ref{prop:uncertainty_premium} need not generalize uniformly across assets. The empirical analysis exploits this distinction by using IVOL as a reduced-form proxy for the limits-to-arbitrage environment, thereby identifying settings in which the conditional monotonic implication in Corollary~\ref{cor:prop4_monotone} is more likely to hold.
\end{remark}

\begin{remark}[Passive Ownership as a Secondary Market-Structure Modifier]
A higher share of rule-based capital need not create mislearning. It can, however, reduce the share of belief-responsive capital that corrects a given belief wedge. Empirically, we therefore study whether lagged passive ownership weakens the subsequent compensation associated with elevated mislearning in benchmark factor systems.
\end{remark}

\section{Empirical Design}

\subsection{Data}

The empirical implementation relies primarily on public monthly factor return data: MKT-RF, SMB, HML, RMW, CMA, and UMD. The core factor dataset is supplemented by volatility controls and a passive-ownership proxy constructed from ICI data.

\subsection{Estimating Predictive Densities}

Two models are estimated for each factor:
\begin{itemize}
\item a \textbf{stable state-space model} estimated via Kalman filtering; and
\item a \textbf{break-allowing model} estimated as a Markov-switching benchmark.
\end{itemize}

The stable model produces one-step-ahead predictive density $p_S(f_t\mid\mathcal{F}_{t-1})$, while the break model produces $p_B(f_t\mid\mathcal{F}_{t-1})$. These directly feed into the construction of $\Delta_t$ in \eqref{eq:delta}.

\subsection{Baseline Predictive Tests}

The baseline predictive specification is
\begin{equation}
Perf_{t\rightarrow t+h} = a + b \Delta_t + \epsilon_{t+h},
\label{eq:baseline_predictive}
\end{equation}
where $Perf_{t\rightarrow t+h}$ denotes future performance over horizon $h$, such as future cumulative return, future Sharpe ratio, future realized volatility, or a tail-event indicator.

Controlled specifications add standard risk controls, such as rolling volatility and market uncertainty. All forward outcome variables are constructed to exclude the contemporaneous return at time $t$. In dense monthly benchmark panels, this corresponds closely to the calendar interval from $t+1$ onward. In broader anomaly panels with occasional missing observations, the forward horizon is formed from the next available return observations after $t$. This preserves the no-look-ahead timing of the predictive design even when the realized forward window is not a perfectly contiguous calendar block in calendar time.

\subsection{Cross-Sectional Diagnostics for Proposition~\ref{prop:cross_factor_decomposition}}

To evaluate Proposition~\ref{prop:cross_factor_decomposition} and Corollary~\ref{cor:prop4_monotone} in the anomaly universe, we construct anomaly-level cross-sectional objects from the break-probability and mislearning panels. For each anomaly $k$, we define break-proneness
\[
\pi_k = \frac{1}{T_k}\sum_t \Pr(B_{k,t}=1),
\]
which we interpret as the empirical counterpart of the theoretical break probability. We further define break-state conditional mislearning severity as
\[
\mu_{1,k}
=
\frac{\sum_t \Pr(B_{k,t}=1)\Delta_{k,t}}
{\sum_t \Pr(B_{k,t}=1)},
\]
and the corresponding non-break component as
\[
\mu_{0,k}
=
\frac{\sum_t (1-\Pr(B_{k,t}=1))\Delta_{k,t}}
{\sum_t (1-\Pr(B_{k,t}=1))}.
\]
We also compute the unconditional average mislearning of each anomaly as the full-sample mean of $\Delta_{k,t}$.

To study when the conditional monotonic implication in Corollary~\ref{cor:prop4_monotone} is more likely to hold, we construct an anomaly-level limits-to-arbitrage proxy based on monthly return data. Specifically, IVOL is measured as the standard deviation of residuals from a time-series regression of each anomaly's return on the Fama--French three factors. Reduced-form cross-sectional diagnostics then evaluate whether IVOL is associated with break-state conditional severity $\mu_{1,k}$, with break-proneness $\pi_k$, or with both. Full reduced-form results are reported in Appendix~\ref{app:prop4-reducedform}, while the IVOL-tertile validation of Corollary~\ref{cor:prop4_monotone} is reported in Section~\ref{sec:prop4-ivol} and Appendix~\ref{app:prop4-ivol}.

\subsection{Passive Ownership, Timing, and Institutional Tests}

We study whether market structure, as proxied by lagged passive ownership, changes how mislearning is reflected in subsequent outcomes. The passive variables used here should be interpreted as macro-level proxies for the prevalence of rule-based capital rather than as direct measures of passive trading or portfolio holdings.

\subsubsection{Systemic Investing Proxies}

We use manually collected data from the Investment Company Institute's (ICI) Long-Term Fund Trends report. Our matched benchmark-factor sample spans 121 months, from January 2016 to January 2026.

\paragraph{Primary Measure}
Our primary institutional proxy is the total index-fund asset share, denoted $PassiveShare^{Total}_t$. We interpret this variable as a slow-moving ownership-composition measure for the share of rule-based capital in the aggregate fund sector. During our sample period, this measure rises from 28.45\% to 52.69\%.

\paragraph{Robustness Measure}
For robustness only, we construct a detrended proxy, $PassiveShare^{Detrended}_t$, using a strictly one-sided Hodrick--Prescott filter with monthly smoothing parameter $\lambda = 129{,}600$. At each date, the cyclical component is computed using only information available up to that date.

\paragraph{Timing and Alignment}
Because publication-lag metadata for the ICI releases are not fully available in the current implementation, all primary institutional regressions use lagged passive proxies, i.e., information dated $t-1$ when forming predictions at $t$. Same-month passive specifications are not used as paper evidence.

\paragraph{Data Processing}
The ICI series are standardized to end-of-month timestamps and merged onto the benchmark factor panels by month. We preserve the raw economic magnitudes of these proxies and do not winsorize or standardize them further.

\subsubsection{Institutional Tests}

We use two institutional specifications.

\paragraph{(i) Break-Onset Interaction}
As a falsification benchmark, we test whether lagged passive ownership dampens mislearning at break onset:
\begin{equation}
\Delta_{k,t}
=
a_k + b\,Break_{k,t} + c\,SystemicIntensity_{t-1}
+ d\left(Break_{k,t}\times SystemicIntensity_{t-1}\right)
+ \epsilon_{k,t}.
\end{equation}
If passive ownership acts as an absorber, the interaction term $d$ should be negative. If it does not, onset buffering is not supported.

\paragraph{(ii) Outcome-Mapping Interaction}
Our main institutional test asks whether lagged passive ownership changes how elevated mislearning maps into subsequent outcomes:
\begin{equation}
Perf_{k,t\rightarrow t+h}
=
a_k + b\,\Delta_{k,t} + c\,SystemicIntensity_{t-1}
+ d\left(\Delta_{k,t}\times SystemicIntensity_{t-1}\right)
+ X_t'\theta + \epsilon_{k,t+h},
\end{equation}
where the main-text outcome is the 12-month future cumulative return. Appendix evidence reports supporting results for the 12-month Sharpe ratio and one downside-risk outcome.

\paragraph{Identification Note}
Because passive ownership varies only over time and is common across factors within a month, our baseline extension is reported without month fixed effects. To address trend concerns, we additionally report month-fixed-effect interaction-only robustness in the appendix.

\subsection{Inference in the Institutional Regressions}

Because forward cumulative returns overlap over 12-month horizons, the passive-extension interaction regressions may inherit serial dependence from overlapping observations. Accordingly, the passive extension relies on overlap-robust or dependence-robust inference where appropriate. The institutional interpretation is based primarily on sign stability across FF6 and q5 and across appendix robustness checks rather than on any single auxiliary specification.

\section{Empirical Results}

\subsection{Descriptive Evidence}

Figure \ref{fig:factor-series} plots the monthly factor return series used in the analysis. The six factors display substantial time variation, occasional large dislocations, and pronounced heterogeneity across styles.

\draftfigure[width=0.95\textwidth]
{Monthly factor return series for the six benchmark factors.}
{fig:factor-series}
{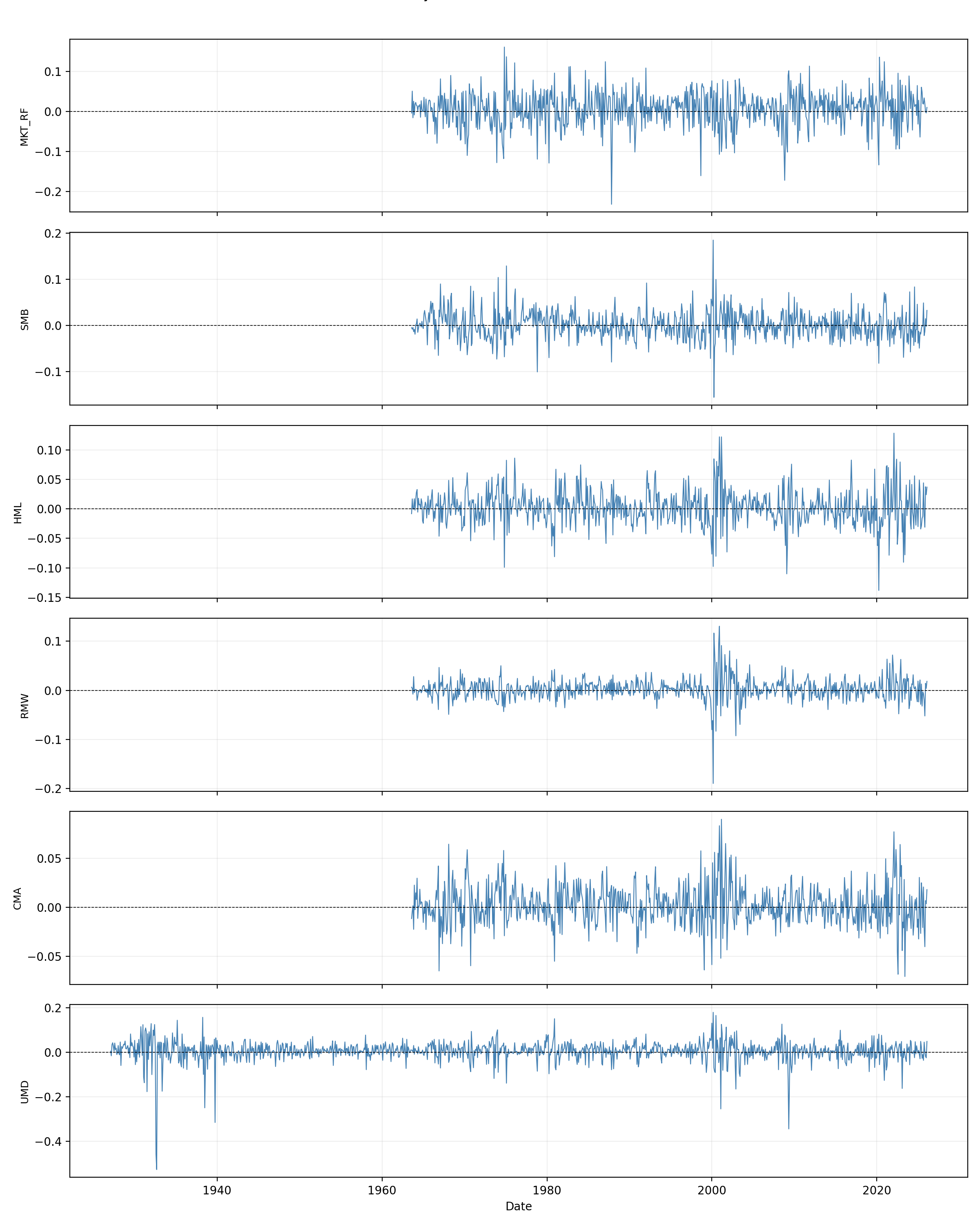}
\FloatBarrier

To keep the main text concise, the factor-by-factor filtered-state plots, break-probability plots, and full $\Delta_t$ series are reported in Appendix \ref{app:factor-figures}.

\subsection{Cross-Factor Heterogeneity}
\label{sec:cross_factor_heterogeneity}

We begin with the FF6 factor set and evaluate both Proposition~\ref{prop:cross_factor_decomposition} and Corollary~\ref{cor:prop4_monotone}. Proposition~\ref{prop:cross_factor_decomposition} itself is a decomposition result: cross-factor differences in unconditional mislearning can arise through break frequency, break-state severity, or both. Corollary~\ref{cor:prop4_monotone} adds a stronger sufficient condition under which more break-prone factors also exhibit higher average mislearning and more frequent large-mislearning spikes. Empirically, however, this stronger monotonic implication need not hold cleanly if break-state severity differs materially across factors.

To evaluate this implication more symmetrically, Table~\ref{tab:ff6_prop4_revised} reports a revised FF6 heterogeneity summary. For each FF6 factor, the table reports: (i) the full-sample mean break probability, (ii) the share of months classified as break states using a posterior break-probability threshold of 0.5, (iii) the average mislearning intensity conditional on break months, (iv) spike frequency based on a pooled 90th-percentile threshold of $\Delta_t$, and (v) the 12-month future-Sharpe coefficient from the relevant predictive specification.

The revised FF6 evidence supports the presence of substantial cross-factor heterogeneity, but it also shows that the relation between break-proneness and average mislearning severity is not perfectly monotone. Some factors enter break states more frequently, while others exhibit more severe conditional mislearning once in the break state. This implies that break frequency and break-state severity are empirically distinct dimensions. Accordingly, the FF6 evidence is somewhat more supportive than the q5 case, but it still suggests that the stronger monotonic implication in Corollary~\ref{cor:prop4_monotone} should be interpreted with caution.

\FloatBarrier
\input{tabs/ff6_prop4_revised_main.tex}

\subsection{Baseline Predictive Regressions}

Table \ref{tab:baseline-predictive} reports a condensed baseline specification that retains only pooled estimates (core results). Table \ref{tab:controlled-predictive} reports the corresponding condensed controlled specification.

\input{tabs/baseline_predictive_v1.tex}
\input{tabs/controlled_predictive_v1.tex}

\FloatBarrier
\paragraph{Economic Interpretation of Predictive Regressions:}
Contrary to the naive hypothesis that mislearning mechanically leads to a persistent deterioration in short-term factor returns, the bias-free results in Table \ref{tab:controlled-predictive} reveal a compelling long-term equilibrium dynamic. In the short run ($h=3, 6$), the predictive coefficients on returns and Sharpe ratios are statistically indistinguishable from zero, reflecting the noisy and turbulent nature of belief adjustments immediately following a regime shift. However, over a longer investment horizon ($h=12$), we document a highly significant \emph{positive} relationship between $\Delta_t$ and both future cumulative returns ($p = 0.003$) and future Sharpe ratios ($p = 0.006$). This empirical finding strongly supports an \textbf{uncertainty premium mechanism}: when model misspecification is severe, the perceived ambiguity of the asset's true data-generating process forces the market to price in a substantial risk premium. Therefore, periods of high mislearning act as proxy indicators for elevated model uncertainty, which are subsequently compensated by higher risk-adjusted returns over a one-year horizon.

Complete factor-level results are moved to Appendix \ref{app:predictive-full-tables} and split into separate tables by horizon ($h=3,6,12$) for readability.

\subsection{q-Factor Robustness: Evidence from Hou--Xue--Zhang Factors}

To assess whether the empirical insights depend on the specific factor taxonomy, we re-estimate the entire mislearning pipeline on the Hou--Xue--Zhang q5 factor set (market, size, investment, profitability, and expected growth). We compute q5-based mislearning series, estimate predictive regressions, and compare unrestricted baseline, common-sample baseline, and controlled specifications. The q5 factor returns are obtained from the official Global-Q data library, which currently provides q-factor and q5 factor returns in the 1967--2024 sample, together with a technical document describing the construction of the factors. The passive-investing measures are compiled from the official ICI Monthly Active and Index Data releases and the corresponding Active and Index Investing statistical reports.\citep{GlobalQFactors2025,GlobalQTech2025,HMXZ2021,ICI2026Monthly,ICI2026Release}

Several patterns emerge. First, the state-variable property is preserved: mislearning exhibits little predictive power for 3- or 6-month outcomes, consistent with the view that acute model uncertainty is not immediately priced. Second, at the 12-month horizon, the pooled coefficients retain a positive tilt, especially for future Sharpe ratios, suggesting that elevated mislearning is associated with a longer-horizon uncertainty premium. Third, this long-horizon tilt survives when the baseline is restricted to the common-sample window, indicating that it is not merely an artifact of differing sample lengths. At the same time, the positive long-horizon relation is not uniformly shared across all q-factors, and is strongest for the market q-factor.

\input{tabs/q5_robust_summary.tex}

Overall, the q5 analysis lends partial but meaningful support to the dynamic pricing interpretation of mislearning. It confirms the state-variable and long-horizon uncertainty-premium dimensions of the mechanism while also highlighting substantial heterogeneity across alternative factor taxonomies. The full q5 predictive regression tables are reported in Appendix~\ref{app:q5-robustness-tables}.

\FloatBarrier

\subsection{q-Factor Cross-Factor Heterogeneity}
\label{sec:q5-prop4}

We next examine whether the decomposition logic in Proposition~\ref{prop:cross_factor_decomposition} and the stronger monotonic implication in Corollary~\ref{cor:prop4_monotone} extend to the Hou--Xue--Zhang q5 factor set. Under Proposition~\ref{prop:cross_factor_decomposition}, greater mislearning exposure can arise through higher break-state frequency, more severe break-state conditional mislearning, or more frequent large-mislearning spikes. Corollary~\ref{cor:prop4_monotone} adds a sufficient condition under which these dimensions move monotonically with break-proneness, but this need not hold empirically if break-state severity varies across factors.

Table~\ref{tab:q5_prop4_revised} reports a revised q5 heterogeneity summary. For each q5 factor, the table shows: (i) the full-sample mean break probability, (ii) the share of months classified as break states using a posterior break-probability threshold of 0.5, (iii) the average mislearning intensity conditional on break months, (iv) spike frequency based on a pooled 90th-percentile threshold of $\Delta_t$, and (v) the 12-month future-Sharpe coefficient from the common-sample baseline predictive regression. This design allows us to separate unconditional factor instability from the severity of mislearning during break episodes.

The evidence provides partial support for the broader cross-factor logic in Proposition~\ref{prop:cross_factor_decomposition}, but only limited support for the stronger monotonic implication in Corollary~\ref{cor:prop4_monotone}. On the one hand, the q5 factors clearly display economically meaningful heterogeneity in all reported dimensions. Break-state exposure differs substantially across factors, and the frequency of large mislearning spikes is far from uniform. On the other hand, the mapping from break-proneness to mislearning severity is not monotone. In particular, the market q-factor has the highest unconditional break probability and break-state share, but the lowest break-state conditional average mislearning. By contrast, the size factor has the lowest break-state exposure but the highest average mislearning conditional on break months, reflecting that infrequent breaks can nonetheless be associated with severe model misspecification when they do occur. Similarly, the ROE factor exhibits the highest pooled spike frequency without having the highest break probability.

Accordingly, the q5 results do not deliver a full replication of the monotonic implication in Corollary~\ref{cor:prop4_monotone}. They confirm that cross-factor heterogeneity remains present in an alternative factor taxonomy, but they do not establish a clean one-to-one mapping from break-proneness to average mislearning severity. We therefore interpret the q5 evidence as a qualitative robustness check rather than as a high-powered cross-sectional test. A broader anomaly universe is better suited for evaluating the conditional monotonic logic more sharply.

\FloatBarrier
\input{tabs/q5_prop4_revised_main.tex}
\FloatBarrier

\subsection{Anomaly-Universe Evidence}
\label{sec:anomaly}

\paragraph{Pooled Predictive Evidence in the Anomaly Universe}

We next extend the analysis to a broad anomaly universe consisting of 212 long--short portfolios. A natural question is whether the benchmark long-horizon Sharpe result generalizes beyond the benchmark factor systems. Appendix~\ref{app:anomaly-predictive} reports pooled 12-month predictive regressions under alternative inference specifications.

The pooled anomaly-universe evidence does not reproduce the benchmark Sharpe-ratio result. Across inference choices, the coefficient on $\Delta_t$ in the 12-month future-Sharpe regression remains economically small and statistically insignificant. This null is therefore not driven by the choice of standard-error estimator or clustering scheme. Put differently, the anomaly-universe Sharpe result is weak for substantive rather than inferential reasons.

\paragraph{Family-Level Heterogeneity}

The pooled anomaly result masks substantial heterogeneity across anomaly families. To investigate this, we classify anomalies into economically interpretable groups, including value, momentum, investment, profitability/quality, accrual/accounting, risk/volatility, growth/issuance, reversal/microstructure, and residual categories. The classification uses transparent name-based rules together with corrected exact-match overrides for ambiguous cases, which helps avoid the substring-based misclassification problems that can arise in broader anomaly taxonomies.

\begin{figure}[!ht]
\centering
\includegraphics[width=0.75\textwidth]{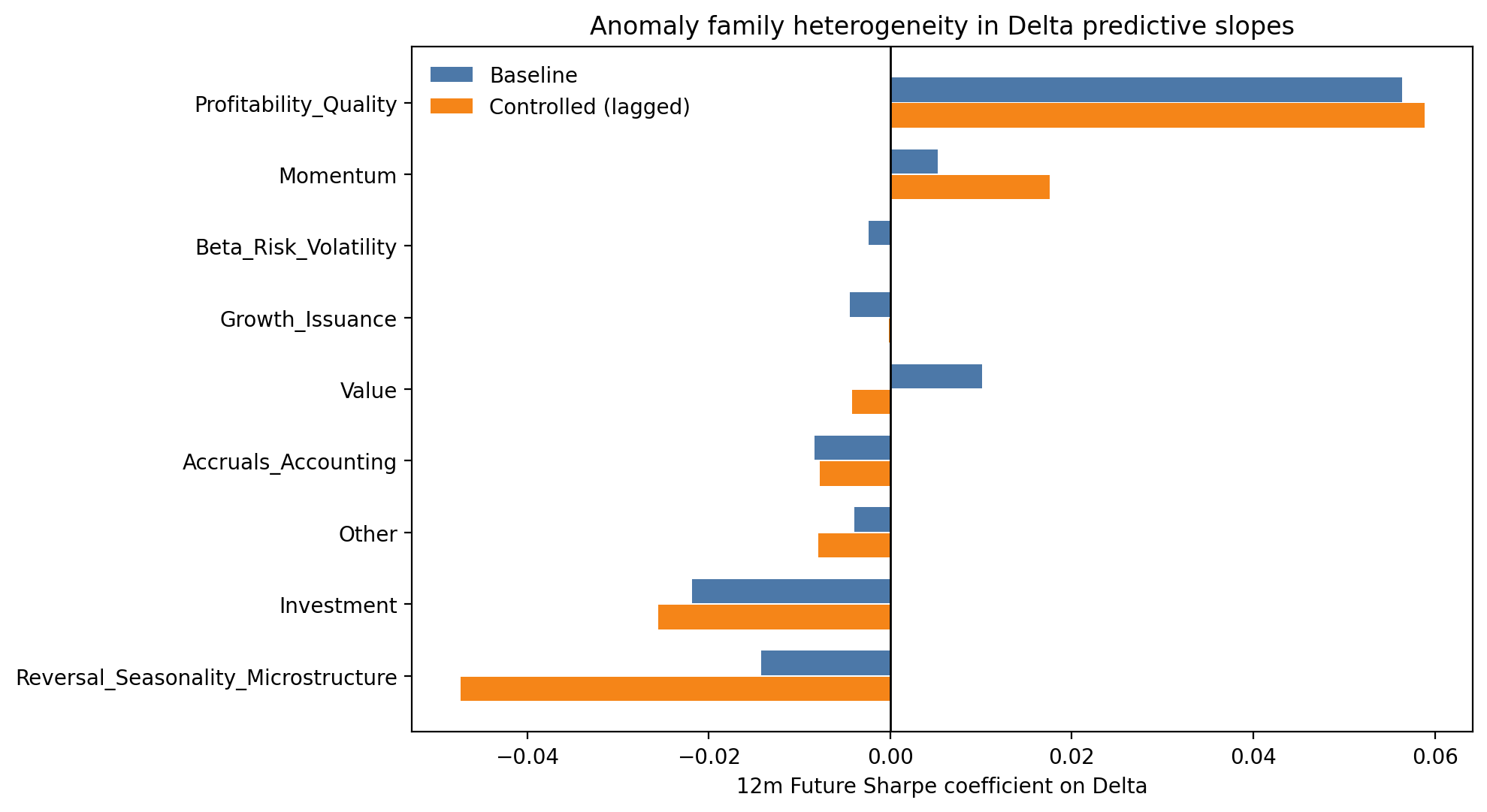}
\caption{Family-level coefficients of $\Delta_t$ in 12-month future-Sharpe regressions across anomaly groups.}
\label{fig:anomaly-family-sharpe}
\end{figure}

Table~\ref{tab:anomaly-family-regressions} reveals strong cross-family heterogeneity. In particular, the profitability/quality family displays a large and highly significant positive relation between $\Delta_t$ and the 12-month future Sharpe ratio, whereas investment and reversal/microstructure families display negative slopes. Figure~\ref{fig:anomaly-family-sharpe} visualizes this dispersion. The implication is that mislearning is not uniformly priced across the anomaly universe; rather, its long-horizon pricing effect depends on the economic structure of the anomaly family.

\input{tabs/anomaly_stage5diag_family_regressions.tex}

\FloatBarrier

\paragraph{Alternative Outcomes: Tail Risk, Drawdowns, and Break-State Payoffs}

The weak pooled Sharpe result raises the possibility that the Sharpe ratio is simply not the most informative outcome in a large anomaly universe. We therefore consider alternative forward outcomes that more directly capture instability and downside risk. Table~\ref{tab:anomaly-alt-outcomes} reports predictive regressions for future drawdowns, downside semivolatility, and volatility ratios. Additional family-level outcome summaries are reported in Appendix~\ref{app:anomaly-predictive}.

In sharp contrast to the pooled Sharpe result, the anomaly-universe evidence becomes considerably stronger once the outcome is allowed to reflect future instability rather than unconditional risk-adjusted mean returns. Higher $\Delta_t$ predicts significantly larger future drawdowns, higher downside semivolatility, and larger volatility ratios. Appendix~\ref{app:anomaly-predictive} further shows that the strongest family-level break-state payoff response arises in the profitability/quality family. These results suggest that in a broad anomaly universe, mislearning is better interpreted as a state variable for future instability and conditional break-state payoffs than as a universal predictor of long-horizon Sharpe ratios.

\input{tabs/anomaly_stage5diag_alt_outcomes.tex}

\FloatBarrier

\paragraph{Model Fit, Delta Quality, and Extreme-Value Diagnostics}

A remaining concern is that the anomaly-universe results could be driven by model failure, pathological Delta distributions, or a small number of extreme observations. Appendix~\ref{app:anomaly-family} reports detailed fit-quality diagnostics and the cross-anomaly Delta distribution, while Appendix~\ref{app:anomaly-predictive} reports extreme-value robustness checks.

The break model fits broadly successfully across anomalies, with no widespread estimation failures and only a small number of degenerate break-state cases. Although a subset of anomalies displays heavy-tailed or skewed Delta distributions, cleaning procedures such as winsorization, trimming, and leave-top-percent-out exercises do not restore a positive pooled Sharpe relation. By contrast, the strongest alternative-outcome results, especially those involving drawdowns and downside semivolatility, remain robust after cleaning. Hence the anomaly-universe evidence is not driven by generic model failure or by a handful of extreme observations.

\paragraph{Interpretation within the anomaly universe}

Taken together, the anomaly-universe evidence suggests that the pricing implications of mislearning are not uniform across assets. In particular, the benchmark factor systems provide the cleanest setting in which elevated mislearning predicts a long-horizon uncertainty premium. In contrast, within a broader anomaly universe, the empirical manifestation of mislearning shifts toward three dimensions: pronounced family heterogeneity, stronger sensitivity of downside- and instability-based outcomes, and greater state dependence around break episodes.

These findings indicate that, in a large cross-section of anomalies, mislearning operates less as a universal predictor of unconditional Sharpe ratios and more as a state variable governing conditional risk and instability. \textbf{This maps directly into the theoretical outcome bifurcation outlined in Section 5: when limits to arbitrage are high, the price-impact component of mislearning overwhelms the gradual uncertainty premium.}

\FloatBarrier

\subsection{IVOL-Screened Cross-Sectional Validation of Corollary~\ref{cor:prop4_monotone}}
\label{sec:prop4-ivol}

The anomaly universe also provides a sharper cross-sectional environment for evaluating the conditional monotonic implication in Corollary~\ref{cor:prop4_monotone}. A key issue is that the severity gap entering the corollary need not be comparable across anomalies. As a practical screening device, we use IVOL as an ex-ante proxy for the cross-anomaly frictions \textbf{(the empirical counterpart to $\Sigma_u$ in Equation \ref{eq:bifurcation})} that make break-state severity heterogeneous. Reduced-form anomaly-level diagnostics in Appendix~\ref{app:prop4-reducedform} show that IVOL predicts break-state conditional severity $\mu_{1,k}$, but does not predict break-proneness $\pi_k$. This makes IVOL a natural screening variable for sorting the cross-section into environments in which the corollary is more or less likely to hold.

We therefore partition anomalies into IVOL tertiles and re-estimate the cross-sectional relation between break-proneness and unconditional mislearning within each group. The resulting pattern provides \emph{partial support} for Corollary~\ref{cor:prop4_monotone}. In the Low-IVOL group, both the slope of average mislearning on break-proneness and the slope of mislearning spike frequency on break-proneness are positive and economically clean. The spike-frequency relation is strongest in the Low-IVOL subsample, while the average-mislearning relation is strongest in the Medium-IVOL subsample. Accordingly, the data do not support an unconditional monotonic law for the full anomaly cross-section, but they do indicate that the original Prop~4 logic becomes more clearly visible in lower-friction environments. Full results are reported in Appendix~\ref{app:prop4-ivol}.

\subsection{Passive Ownership as a Slow-Moving Market-Structure Modifier}
\label{sec:passive-mapping}

\paragraph{Break onset is not the channel.}
We first ask whether passive ownership dampens the immediate mislearning shock at structural breaks. The evidence does not support that interpretation: the lagged break-onset interaction is not negative and is statistically weak. The institutional interpretation therefore does not rely on an absorber mechanism; Appendix~\ref{app:common-sample-passive} reports the corresponding benchmark diagnostic.

\paragraph{Main result: weaker cumulative-return compensation.}
Our main institutional test studies whether lagged passive ownership changes how elevated mislearning maps into subsequent compensation. Table~\ref{tab:passive-stockonly} reports the stock-only benchmark mapping for future 12-month cumulative returns. In both FF6 and q5, the interaction between mislearning and lagged passive ownership is negative: the estimated coefficient is -0.8535 in FF6 and -0.5262 in q5. Higher passive ownership is therefore associated with weaker subsequent compensation following elevated mislearning in the benchmark factor systems.

\input{tabs/passive_stockonly_main.tex}

\paragraph{Interpretation and scope.}
Appendix~\ref{app:passive-robustness} shows that this negative cumulative-return interaction remains negative with month fixed effects and after excluding the March 2020--June 2020 episode. Figure~\ref{fig:passive-leaveoneout} further shows that the interaction does not flip sign in leave-one-year-out exercises. Appendix~\ref{app:passive-support} reports narrower supporting evidence for the 12-month Sharpe ratio and a downside-risk outcome. At the same time, the detrended passive proxy is too imprecise to support a strong high-frequency cyclical interpretation. We therefore interpret passive ownership as a slow-moving market-structure composition variable. In the current data, this interpretation is better supported than either an on-impact absorber mechanism or a short-run flow-based propagation channel.

\subsection{Additional Background Tables}

For completeness, we also report model-fit and model-comparison diagnostics. These are supplementary to the identification results, but they document the relative fit of the stable and break-aware specifications. They are reported in Appendix \ref{app:model-fit}.

\section{Conclusion}

This paper studies how investors learn about factor risk premia when the true environment is subject to structural breaks but investors update beliefs using a misspecified stable model. We develop a minimal Bayesian framework in which this misspecification generates persistent forecast errors and pricing distortions, and we propose a tractable empirical proxy for mislearning based on predictive likelihood comparisons between stable and break-aware models.

Three main conclusions emerge.

First, in benchmark factor systems, mislearning behaves as a state variable associated with a long-horizon uncertainty premium. Periods of elevated mislearning do not forecast a deterministic short-run collapse in factor performance. Instead, they are followed by stronger future cumulative returns and Sharpe ratios, consistent with equilibrium compensation for model uncertainty.

Second, this pricing relation does not generalize uniformly to a broader anomaly universe. There, mislearning is more strongly associated with future instability---including drawdowns, downside semivolatility, and related tail-risk outcomes---than with unconditional long-horizon Sharpe ratios. At the same time, the anomaly evidence reveals substantial cross-sectional heterogeneity: the economic manifestation of mislearning depends on anomaly family, payoff dimension, and break-state conditions. More broadly, the cross-sectional evidence indicates that Proposition~\ref{prop:cross_factor_decomposition} is best interpreted as a decomposition result with a conditional monotonic implication. In particular, the original Prop~4 logic becomes more clearly visible once the anomaly universe is screened into lower-friction environments using IVOL.

Third, a limited institutional extension suggests that market structure can condition how mislearning is resolved. Using lagged aggregate passive ownership from ICI, we find in the FF6 and q5 benchmark systems that higher passive ownership is associated with weaker subsequent cumulative-return compensation following elevated mislearning. This evidence does not support an on-impact absorber mechanism. It is best interpreted as a slow-moving ownership-composition effect within the benchmark factor systems.

Taken together, the results suggest that mislearning is a conditional pricing force whose consequences depend on belief distortions and the surrounding market environment. Mislearning does not map into a single universal outcome: in some settings it is compensated, while in others it is realized through instability. Future work may extend this framework to richer belief dynamics, more granular measures of delegated capital, and explicit microfoundations for how market structure mediates the transmission of model misspecification into asset prices.
\clearpage
\appendix
\setcounter{table}{0}
\renewcommand{\thetable}{A\arabic{table}}
\setcounter{figure}{0}
\renewcommand{\thefigure}{A\arabic{figure}}

\section{Additional Factor-Level Figures}
\label{app:factor-figures}

This appendix collects factor-level figures used to document cross-factor variation in stable-state estimates, break probabilities, and mislearning dynamics.

\subsection{Stable-Model Filtered States}

\begin{figure}[htbp]
\centering
\begin{subfigure}{0.48\textwidth}
\includegraphics[width=\linewidth]{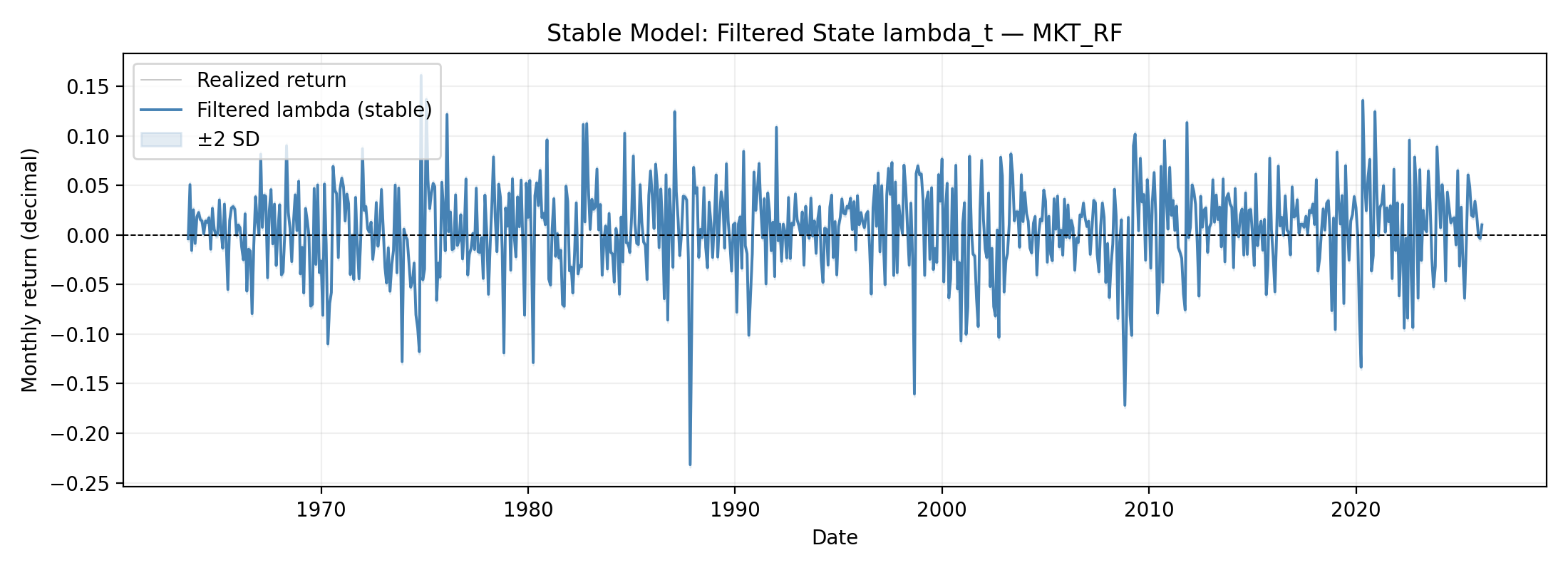}
\caption{MKT-RF}
\end{subfigure}
\hfill
\begin{subfigure}{0.48\textwidth}
\includegraphics[width=\linewidth]{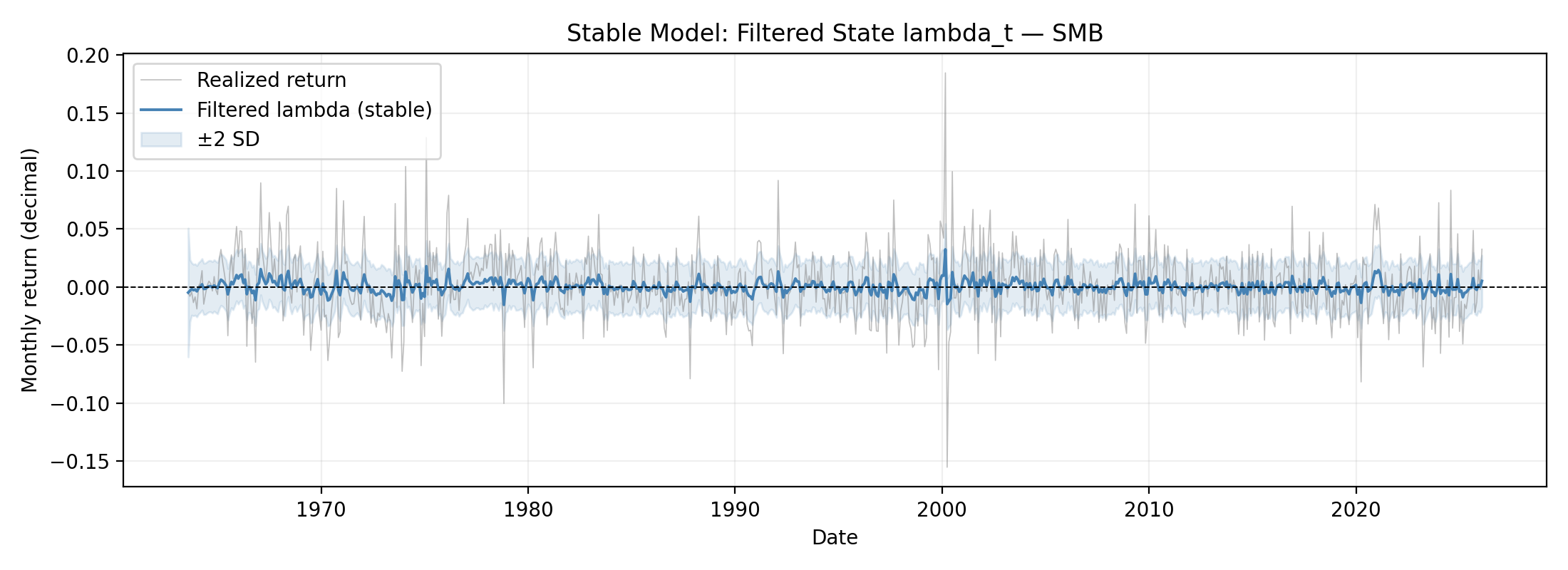}
\caption{SMB}
\end{subfigure}

\vspace{0.8em}

\begin{subfigure}{0.48\textwidth}
\includegraphics[width=\linewidth]{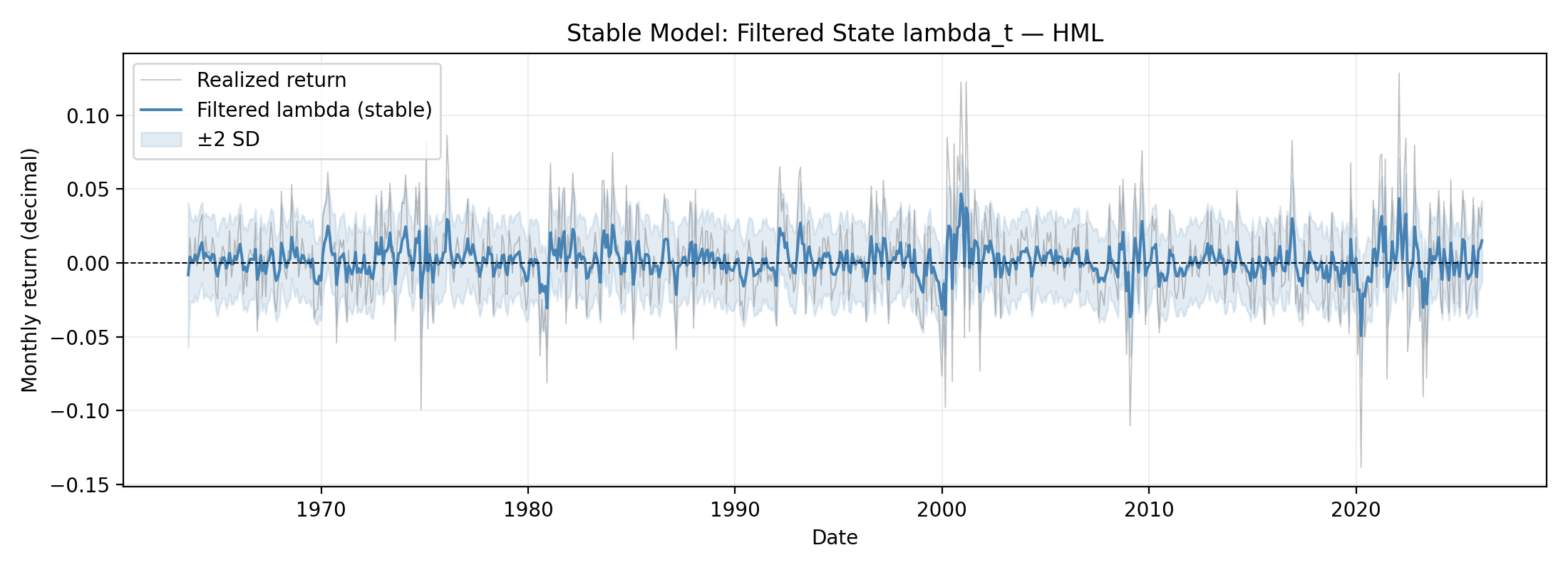}
\caption{HML}
\end{subfigure}
\hfill
\begin{subfigure}{0.48\textwidth}
\includegraphics[width=\linewidth]{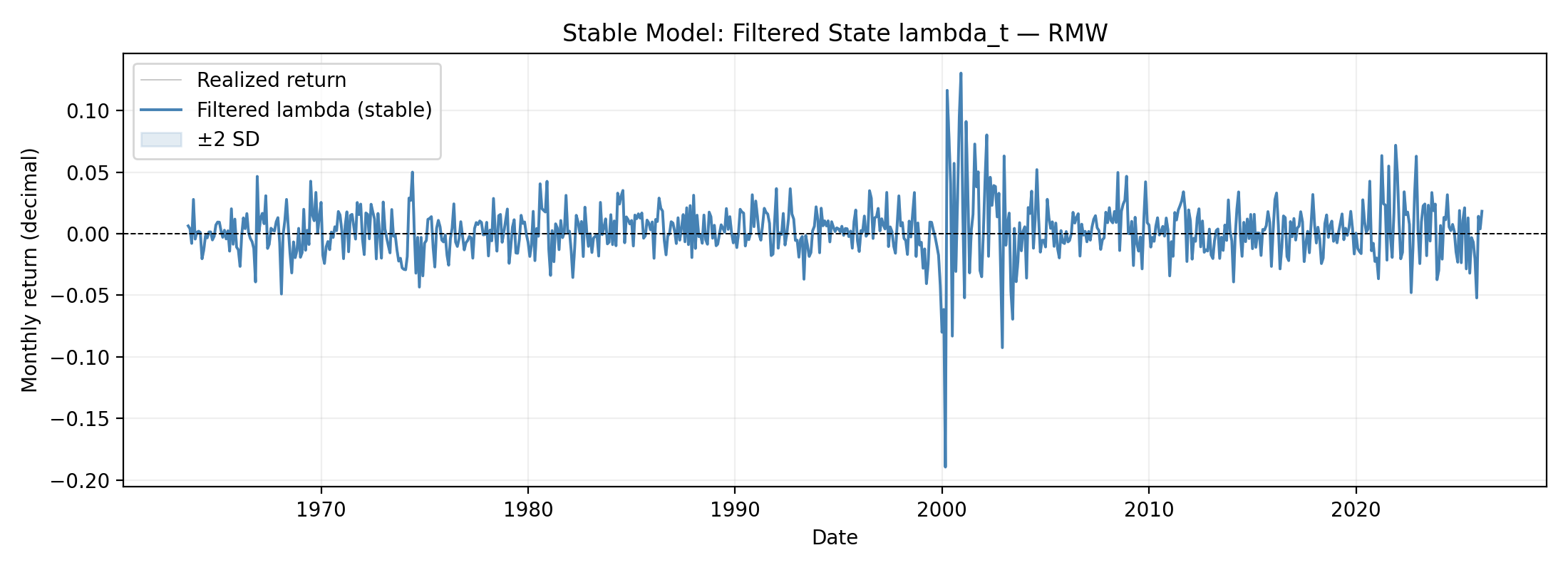}
\caption{RMW}
\end{subfigure}

\vspace{0.8em}

\begin{subfigure}{0.48\textwidth}
\includegraphics[width=\linewidth]{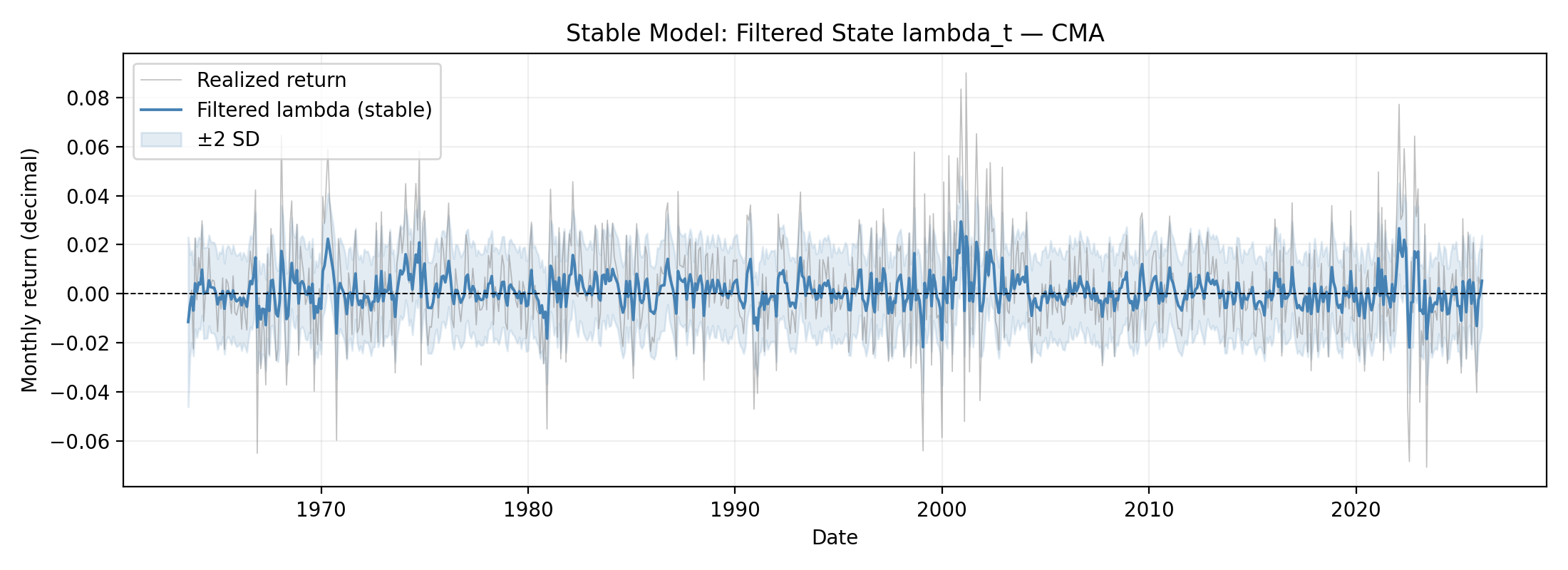}
\caption{CMA}
\end{subfigure}
\hfill
\begin{subfigure}{0.48\textwidth}
\includegraphics[width=\linewidth]{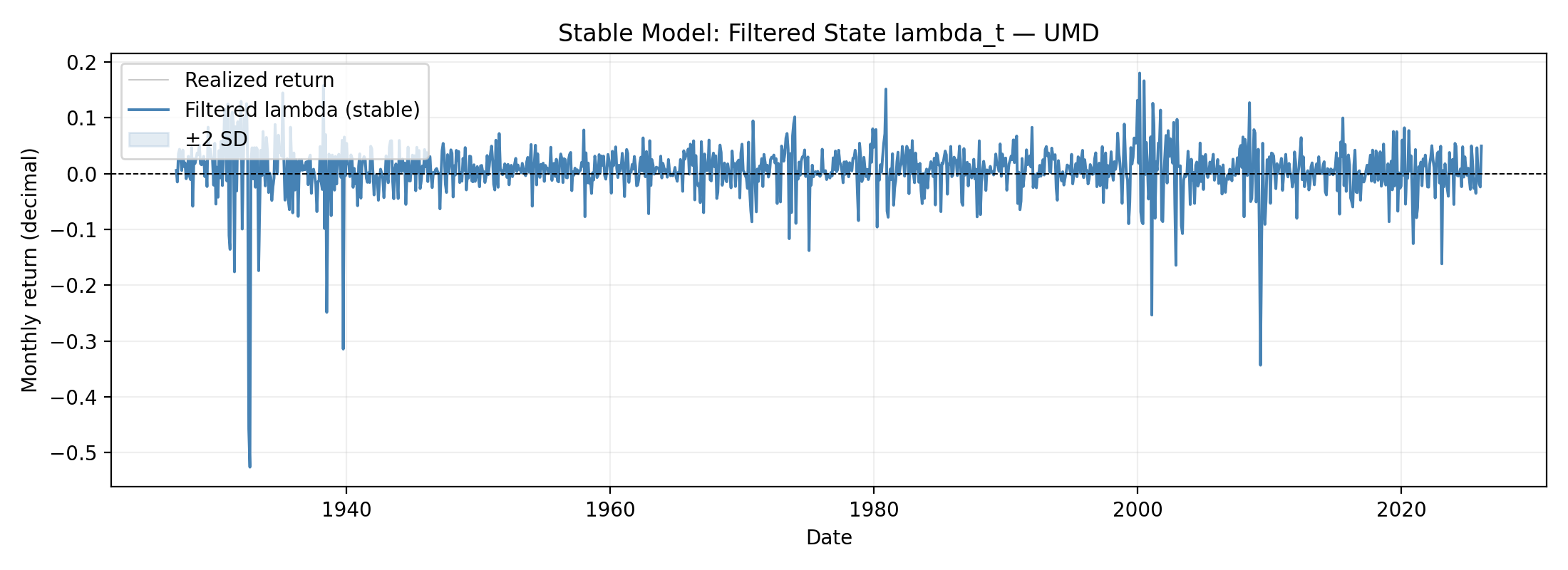}
\caption{UMD}
\end{subfigure}
\caption{Stable-model filtered state estimates with uncertainty bands.}
\label{fig:app-lambda-stable}
\end{figure}

\FloatBarrier

\clearpage

\subsection{Break Probabilities}

\begin{figure}[htbp]
\centering
\begin{subfigure}{0.48\textwidth}
\includegraphics[width=\linewidth]{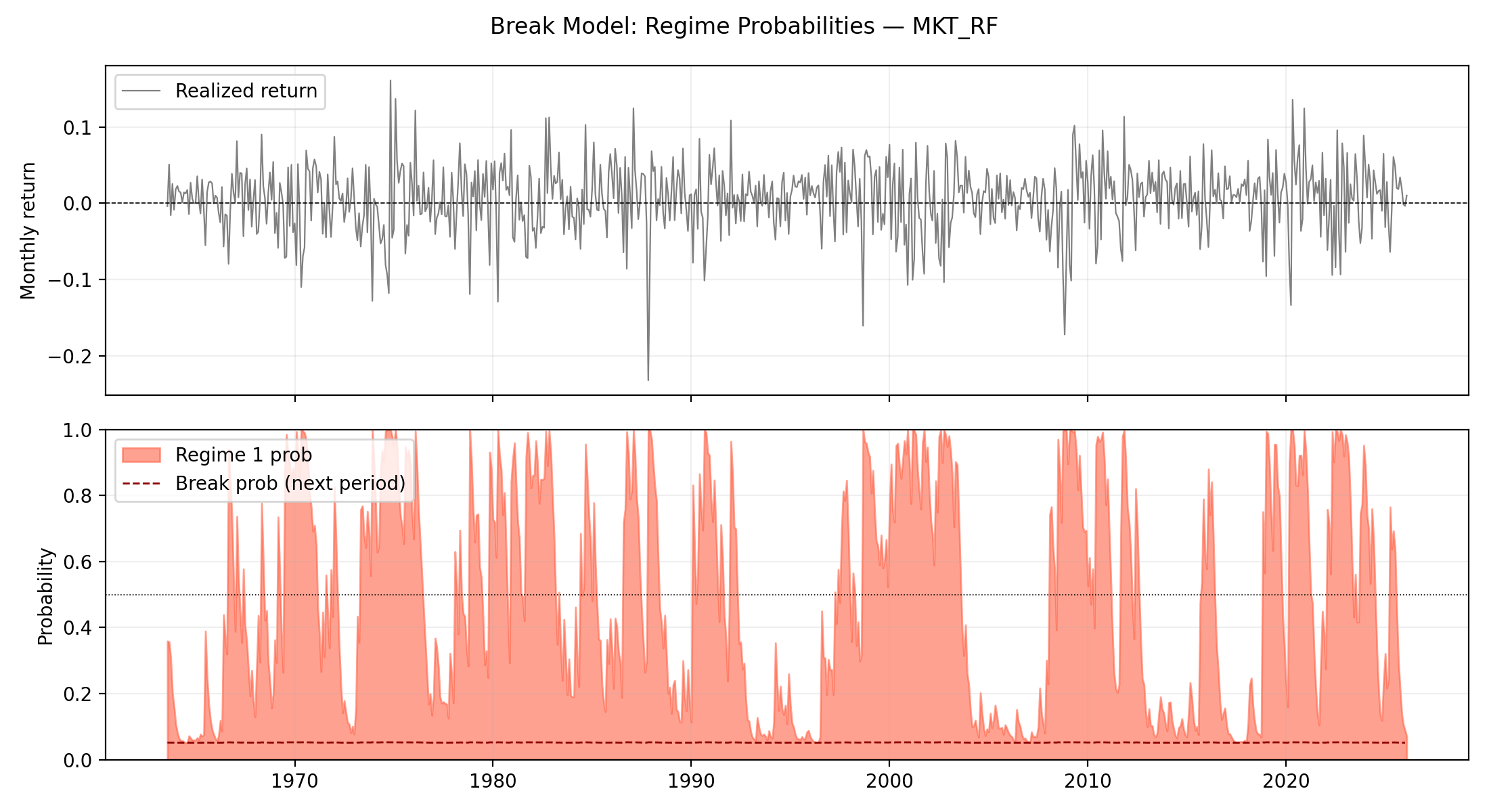}
\caption{MKT-RF}
\end{subfigure}
\hfill
\begin{subfigure}{0.48\textwidth}
\includegraphics[width=\linewidth]{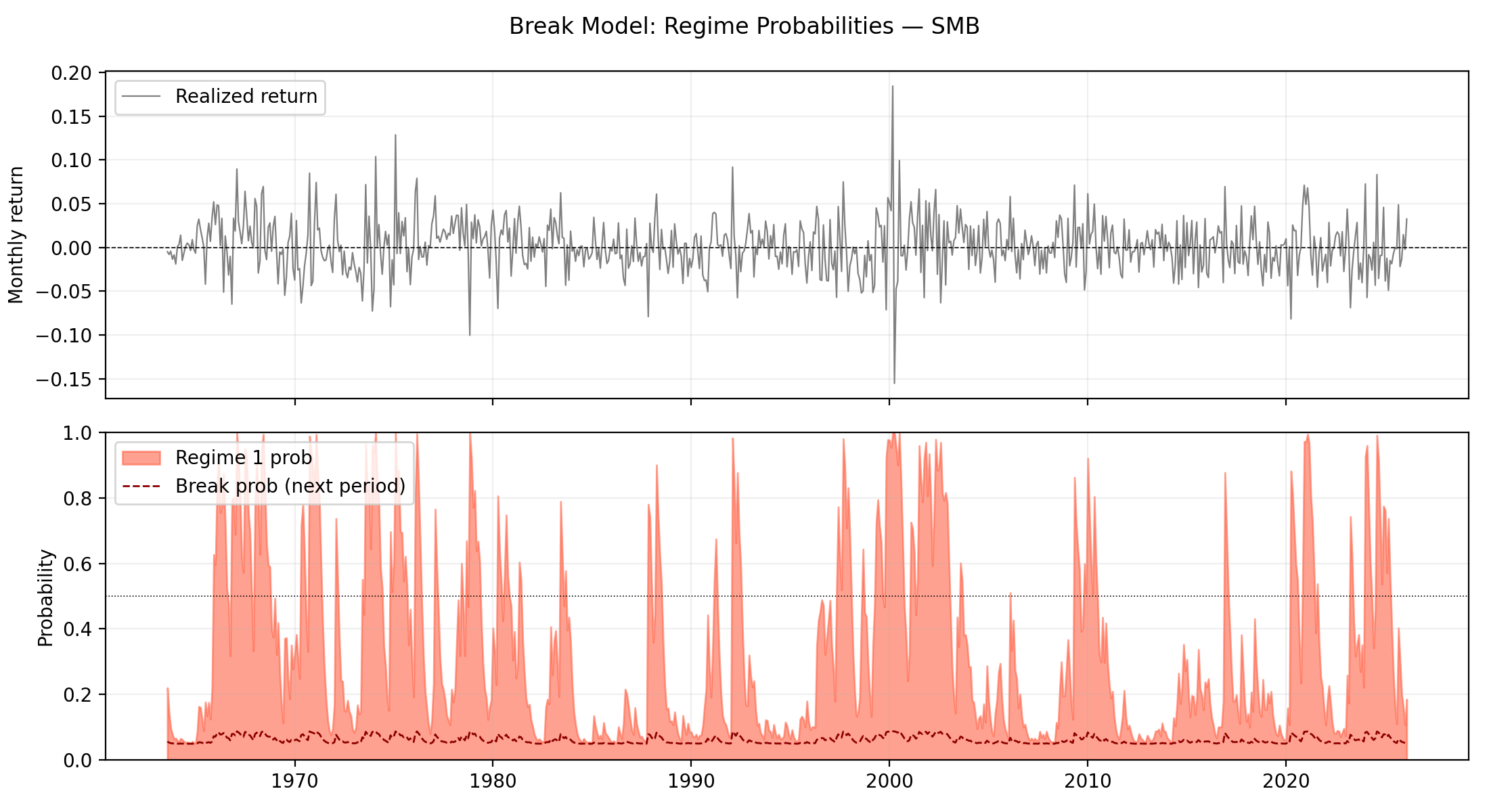}
\caption{SMB}
\end{subfigure}

\vspace{0.8em}

\begin{subfigure}{0.48\textwidth}
\includegraphics[width=\linewidth]{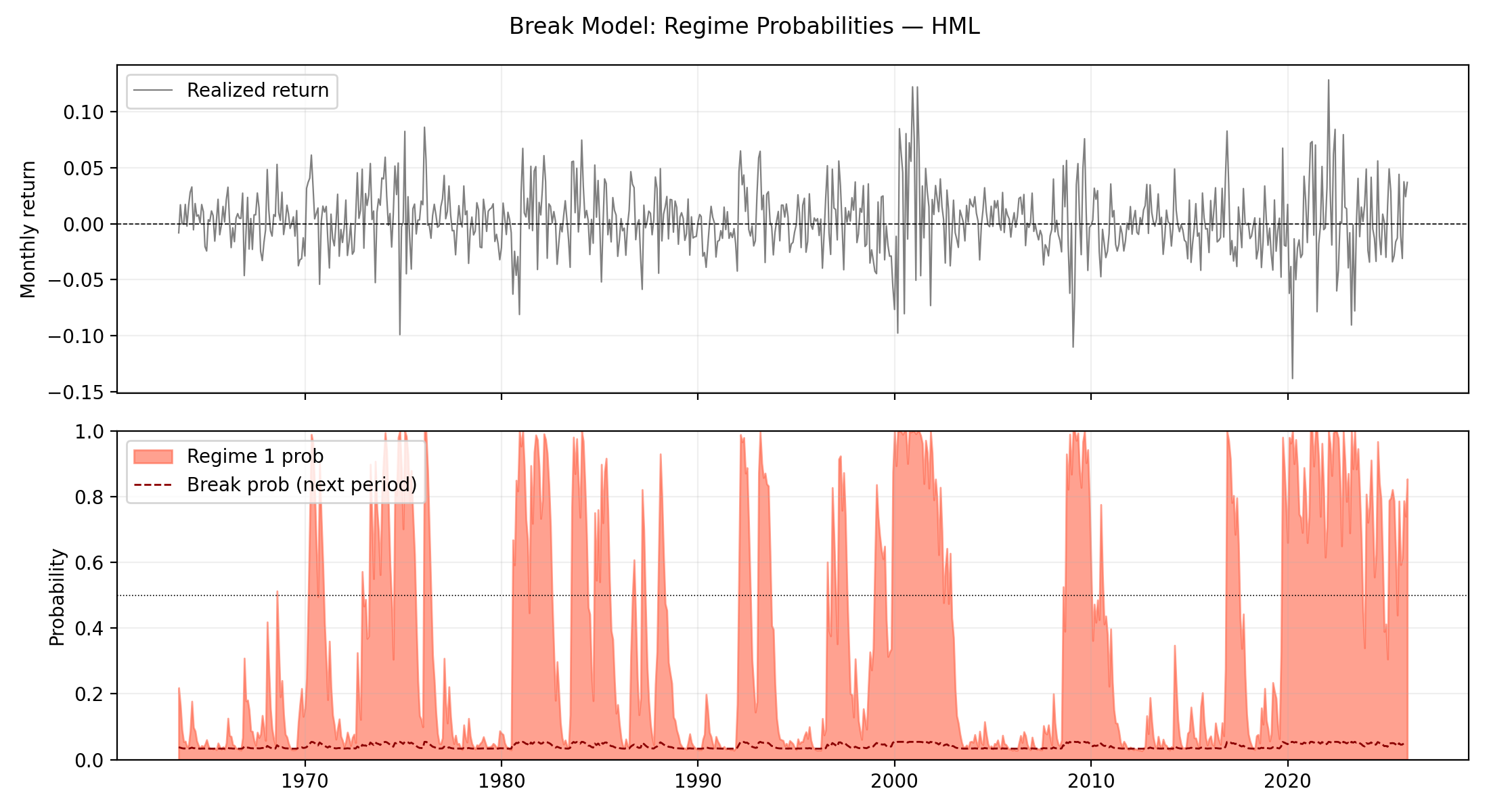}
\caption{HML}
\end{subfigure}
\hfill
\begin{subfigure}{0.48\textwidth}
\includegraphics[width=\linewidth]{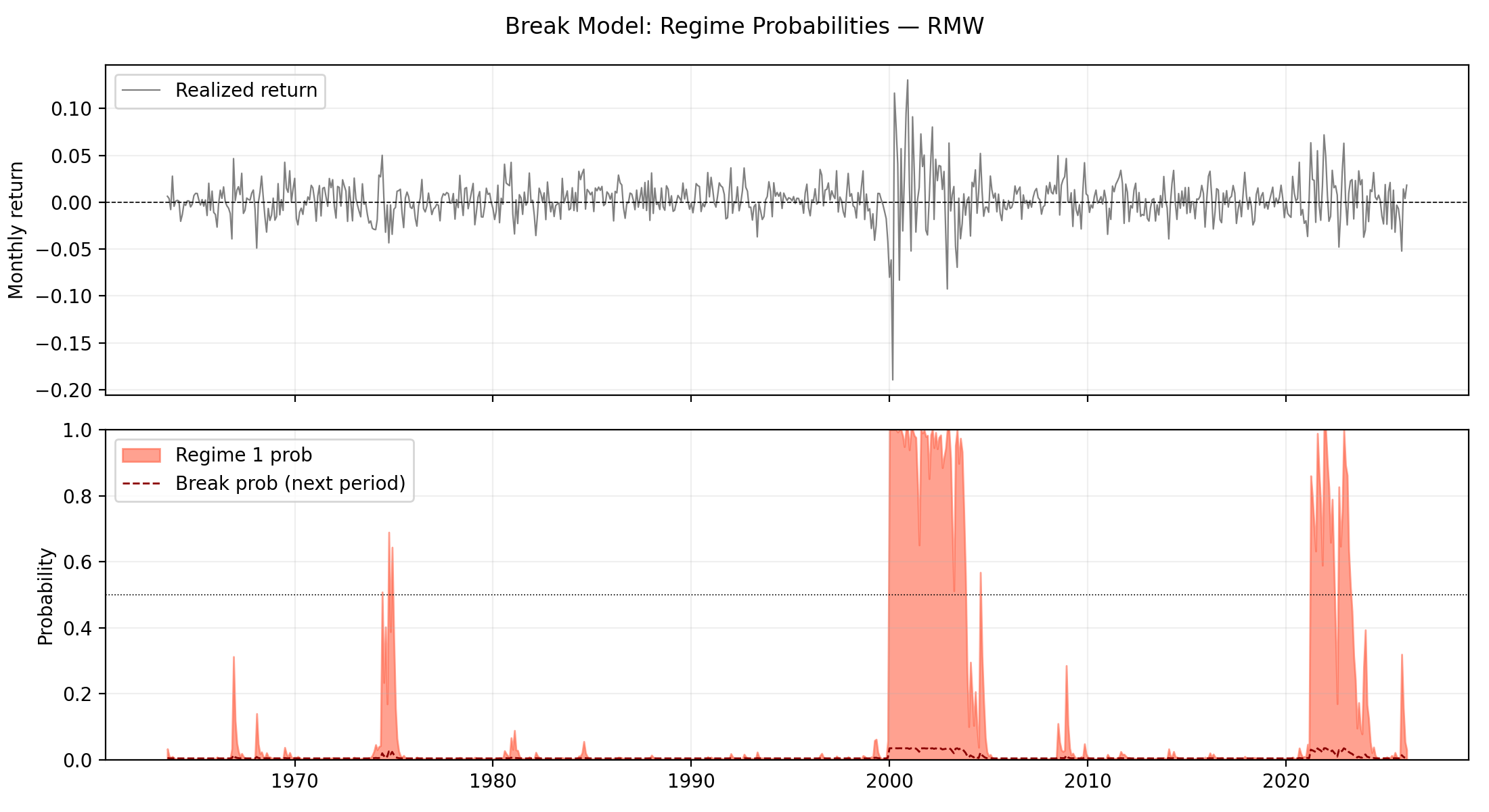}
\caption{RMW}
\end{subfigure}

\vspace{0.8em}

\begin{subfigure}{0.48\textwidth}
\includegraphics[width=\linewidth]{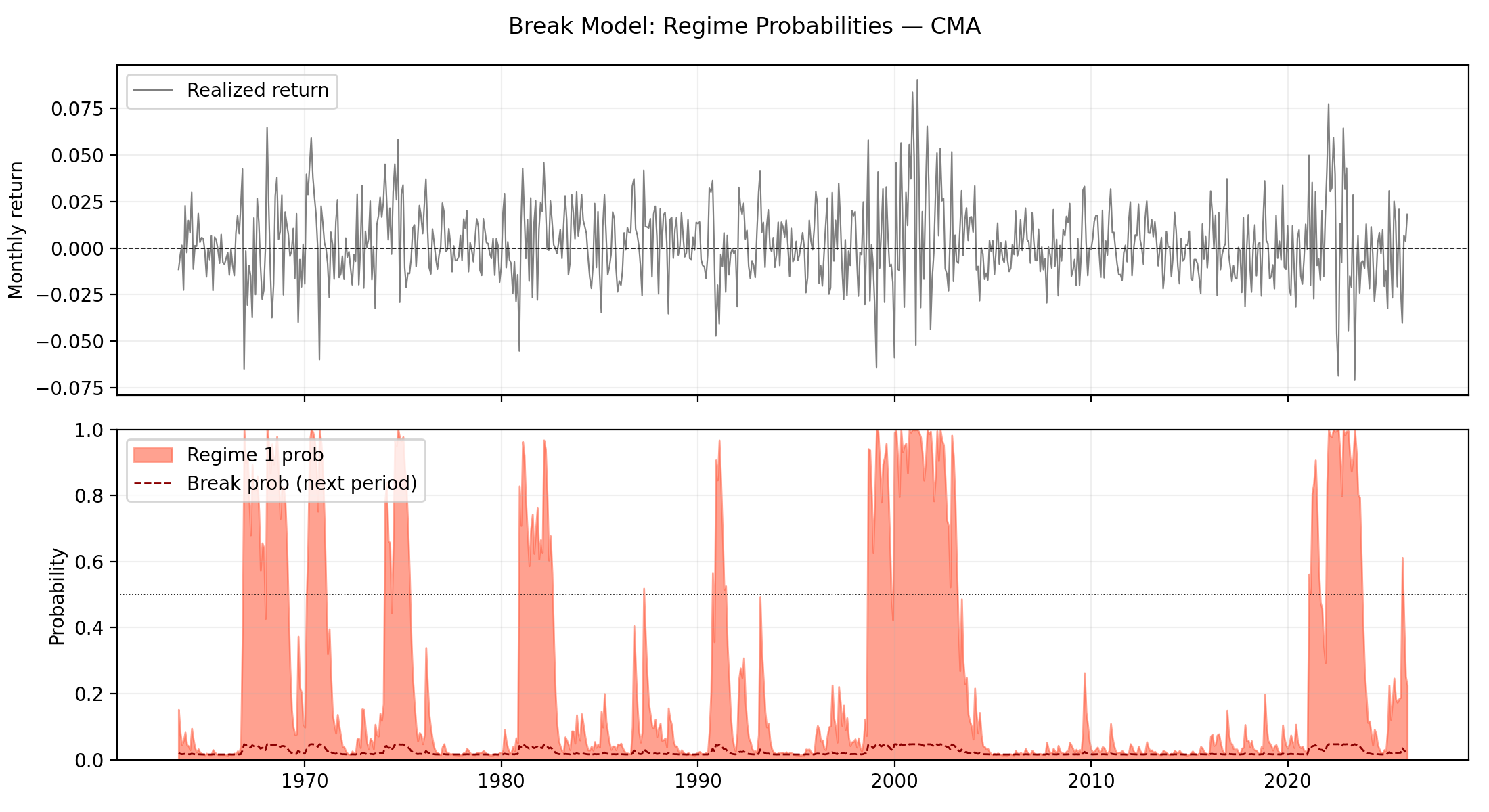}
\caption{CMA}
\end{subfigure}
\hfill
\begin{subfigure}{0.48\textwidth}
\includegraphics[width=\linewidth]{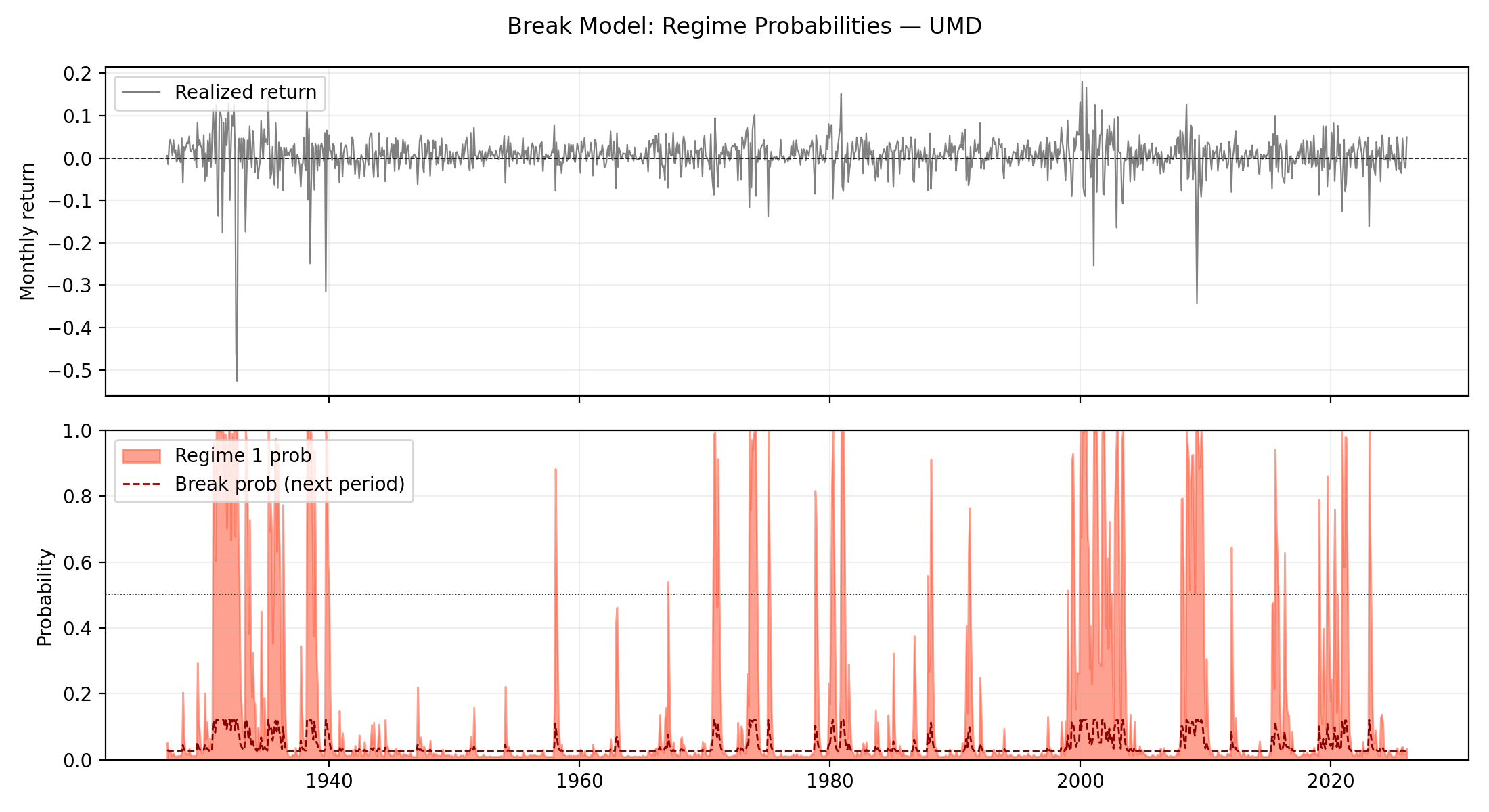}
\caption{UMD}
\end{subfigure}
\caption{Break-model state probabilities and next-period break probabilities.}
\label{fig:app-break-probs}
\end{figure}

\FloatBarrier

\clearpage

\subsection{Mislearning Series}

\begin{figure}[htbp]
\centering
\begin{subfigure}{0.48\textwidth}
\includegraphics[width=\linewidth]{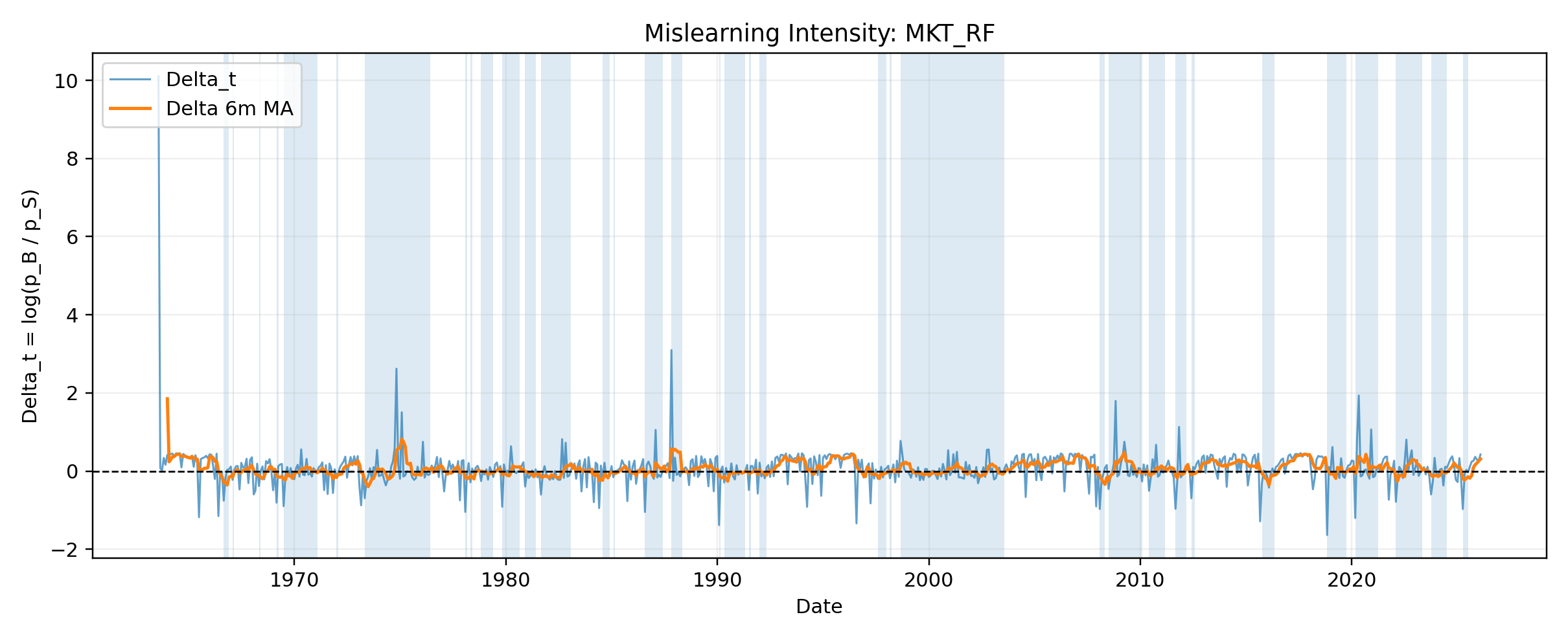}
\caption{MKT-RF}
\end{subfigure}
\hfill
\begin{subfigure}{0.48\textwidth}
\includegraphics[width=\linewidth]{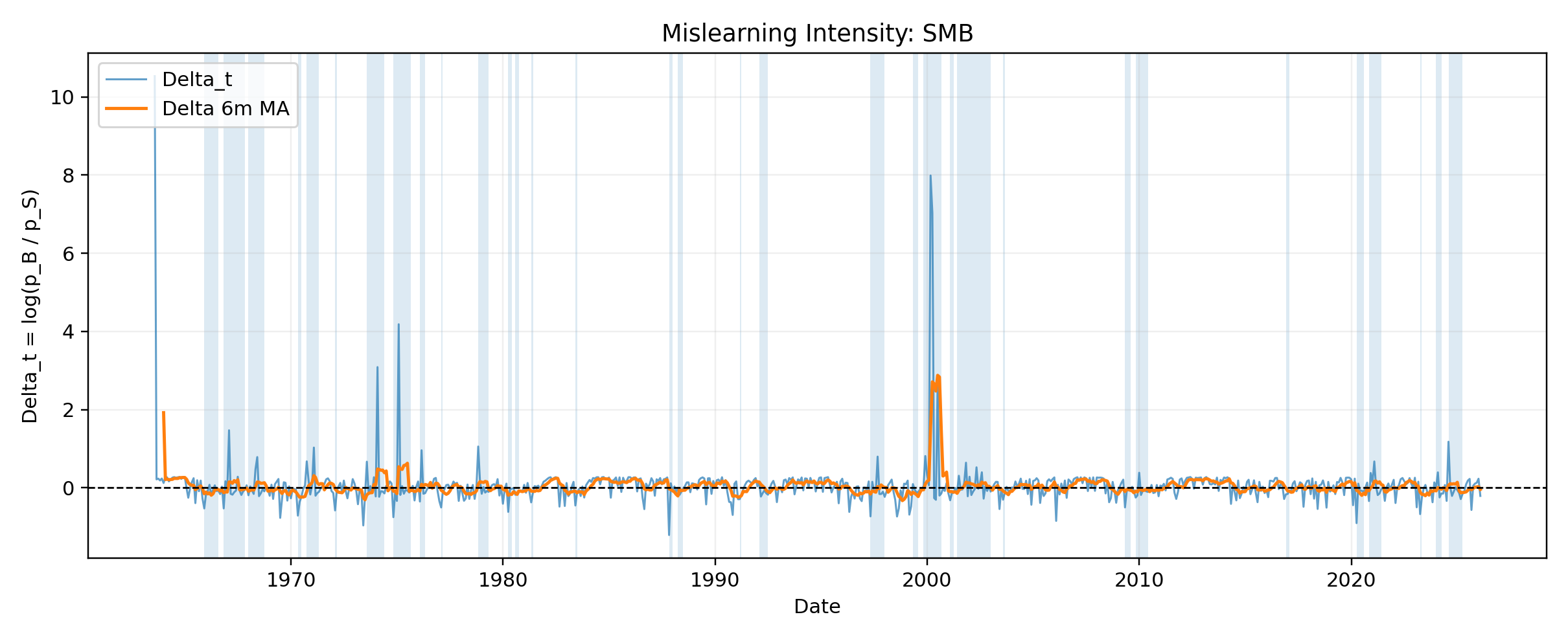}
\caption{SMB}
\end{subfigure}

\vspace{0.8em}

\begin{subfigure}{0.48\textwidth}
\includegraphics[width=\linewidth]{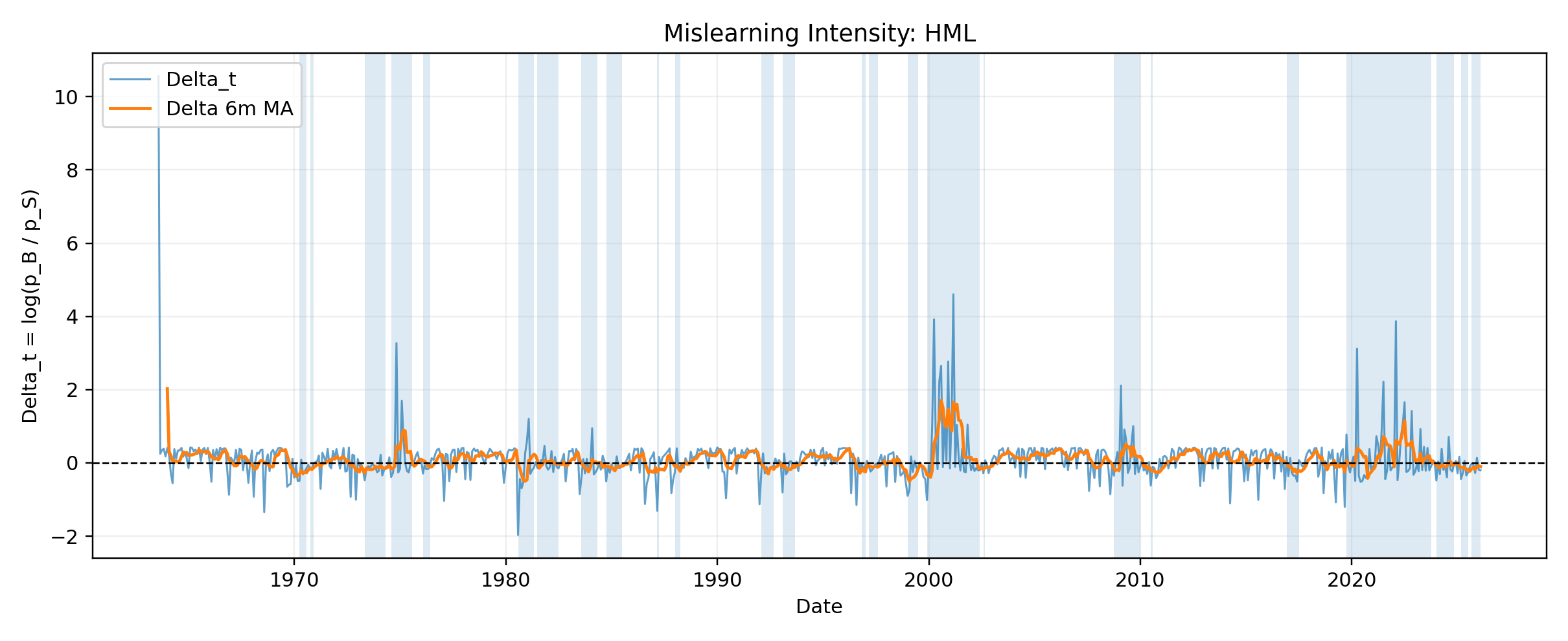}
\caption{HML}
\end{subfigure}
\hfill
\begin{subfigure}{0.48\textwidth}
\includegraphics[width=\linewidth]{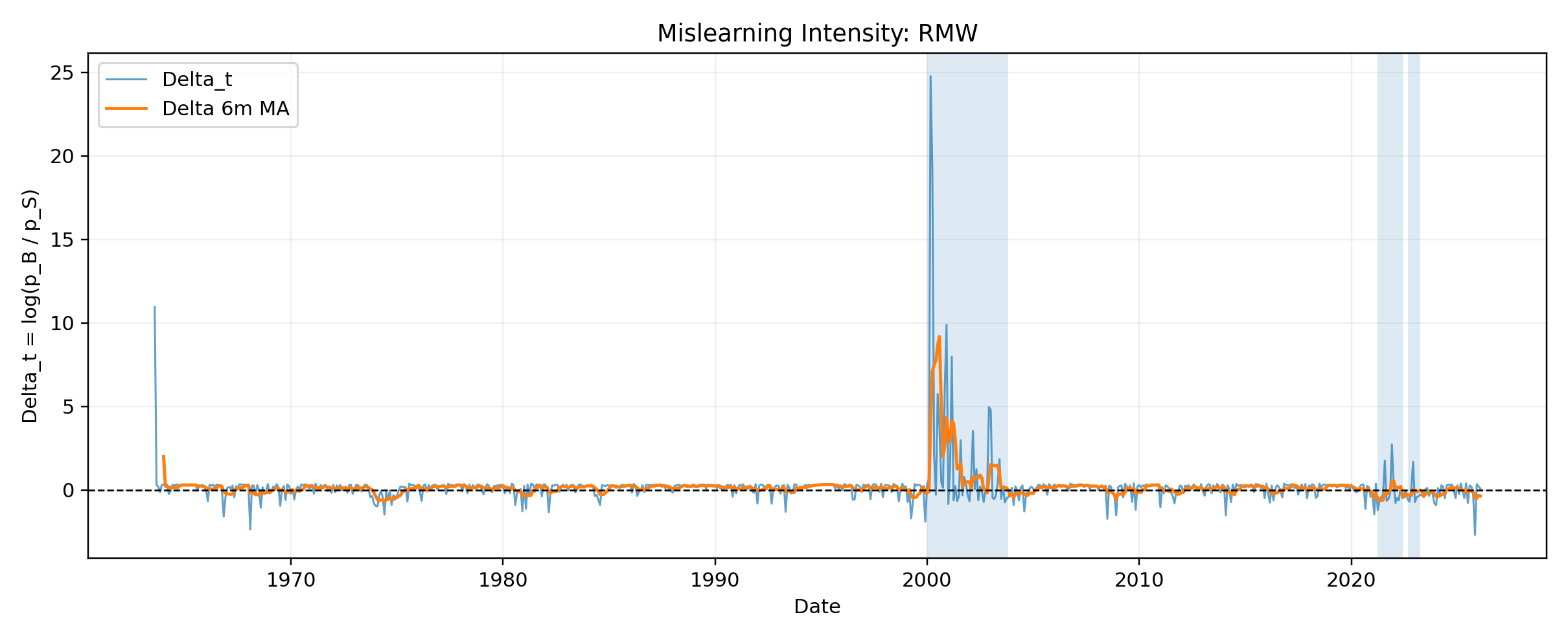}
\caption{RMW}
\end{subfigure}

\vspace{0.8em}

\begin{subfigure}{0.48\textwidth}
\includegraphics[width=\linewidth]{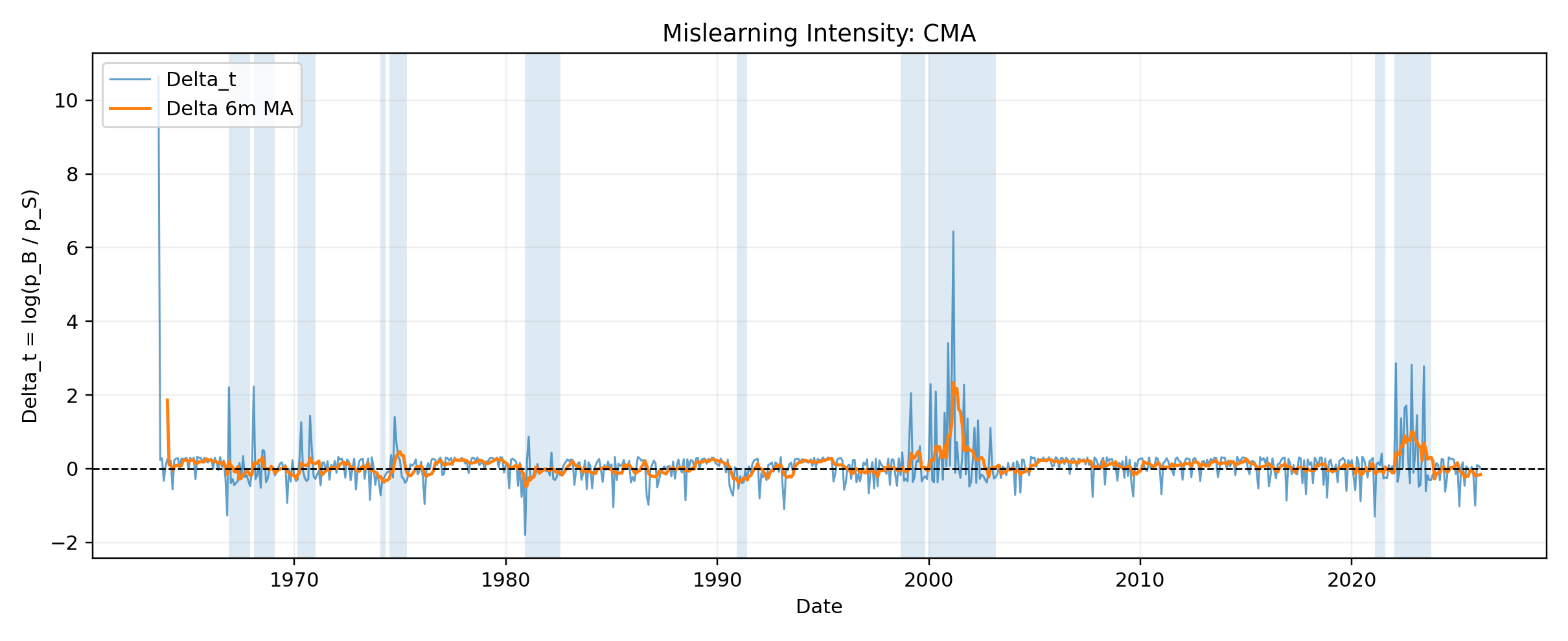}
\caption{CMA}
\end{subfigure}
\hfill
\begin{subfigure}{0.48\textwidth}
\includegraphics[width=\linewidth]{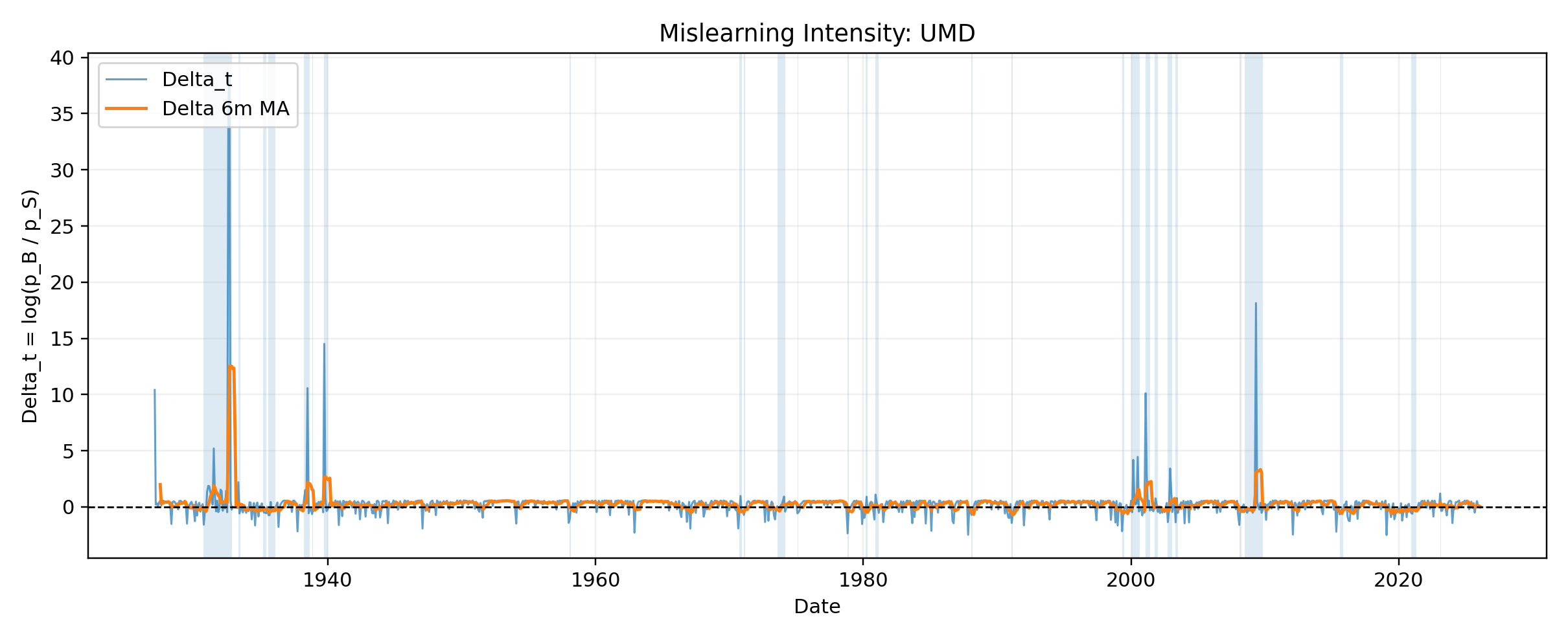}
\caption{UMD}
\end{subfigure}
\caption{Mislearning intensity $\Delta_t$ and six-month moving average by factor.}
\label{fig:app-delta-series}
\end{figure}

\FloatBarrier

\section{Full Predictive Regression Tables}
\label{app:predictive-full-tables}

To keep the main text readable, this appendix reports the complete baseline and controlled predictive regression outputs in consolidated form.

\subsection{Baseline Predictive Regressions (Full)}

\input{tabs/baseline_predictive_v1_full.tex}

\FloatBarrier

\subsection{Controlled Predictive Regressions (Full)}

\input{tabs/controlled_predictive_v1_full.tex}

\FloatBarrier

\section{Passive Outcome-Mapping Robustness}
\label{app:passive-robustness}

This appendix reports the stock-only robustness checks for the passive extension. The key outcome is the 12-month cumulative return. We report the validated lagged baseline, a month-fixed-effect interaction-only specification, a strictly one-sided detrended passive proxy, a specification that excludes March 2020 through June 2020, and a leave-one-year-out sign-stability diagnostic.

\input{tabs/passive_stockonly_robustness.tex}

\begin{figure}[htbp]
\centering
\includegraphics[width=0.82\textwidth]{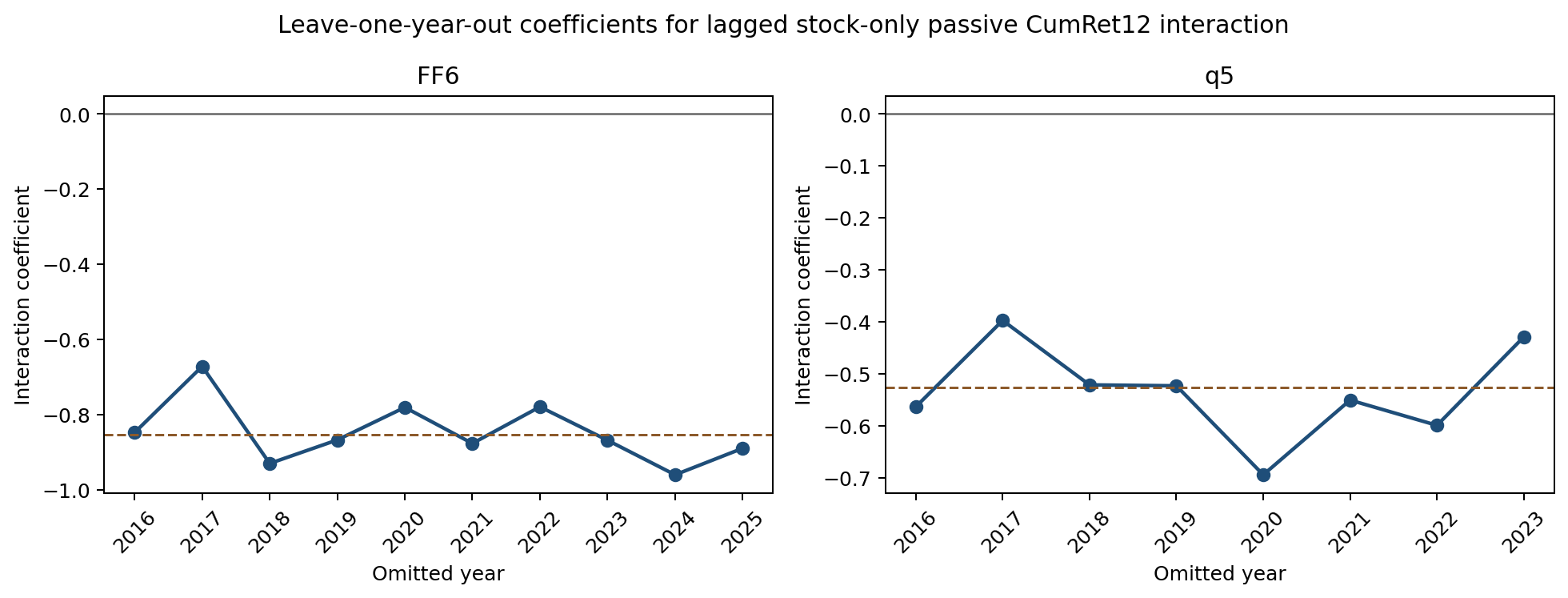}
\caption{Leave-one-year-out sign stability of the lagged stock-only passive cumulative-return interaction. Each point reports the interaction coefficient from the validated baseline after omitting all forecast-origin months in one calendar year. The horizontal line marks zero.}
\label{fig:passive-leaveoneout}
\end{figure}

Figure~\ref{fig:passive-leaveoneout} shows that the interaction remains negative in every leave-one-year-out exercise, ranging from -0.9592 to -0.6710 in FF6 and from -0.6945 to -0.3974 in q5.

\FloatBarrier

\section{Passive Proxy Construction and Timing Details}
\label{app:passive-proxy-details}

The passive extension uses only lagged aggregate passive ownership, $PassiveShare^{Total}_{t-1}$, as its main institutional proxy. The detrended passive series is retained solely as a robustness check using a strictly one-sided Hodrick--Prescott filter. Because publication-lag metadata are incomplete, same-month passive measures are not used as paper evidence. The purpose of this appendix is to document the stock-only proxy choice underlying the main-text extension.

\section{Additional Stock-Only Passive Support}
\label{app:passive-support}

This appendix reports narrower supporting evidence for the passive extension using the lagged stock proxy only. The table reports the 12-month Sharpe ratio and one downside-risk outcome in the benchmark factor systems. These results are supportive but secondary relative to the cumulative-return mapping in the main text.

\input{tabs/passive_stockonly_support.tex}
\section{FF6 Revised Cross-Factor Diagnostic}
\label{app:ff6-prop4-rank}

To place the FF6 and q5 heterogeneity analyses on equal footing, this appendix reports a revised FF6 rank-based diagnostic corresponding to the revised FF6 Proposition~4 table in the main text. The objective is to examine whether the stronger monotonic implication in Corollary~4.1 is supported in the data; that is, whether factors with higher break-proneness also rank higher in break-state mislearning severity and in the frequency of large mislearning spikes.

The FF6 evidence confirms the same general lesson as the q5 results: cross-factor heterogeneity is clearly present, but the mapping from break-proneness to break-state average mislearning is not strictly monotone. In other words, the data support heterogeneity, but not a simple one-dimensional ranking in which the most break-prone factors must always exhibit the highest average conditional mislearning.

\input{tabs/ff6_prop4_revised_rankcheck.tex}

\FloatBarrier

\section{Benchmark Non-Absorber Diagnostics}
\label{app:common-sample-passive}

This appendix reports the lagged benchmark break-onset interaction for the passive extension. The interaction is not negative and does not support a robust absorber mechanism. For that reason, the paper's institutional interpretation relies on the outcome-mapping evidence in the main text rather than on onset buffering.

\input{tabs/passive_stockonly_onset.tex}

\FloatBarrier

\section{Additional Model-Fit Tables}
\label{app:model-fit}

For completeness, this appendix reports supplementary model-fit outputs.

\input{tabs/stable_model_fit_v1.tex}

\input{tabs/break_model_fit_v1.tex}

\input{tabs/model_comparison_v1.tex}

\FloatBarrier
\section{Mathematical Proofs}
\label{app:proofs}

\subsection{Proof of Proposition 1: Slow Updating after Breaks}

Let $e_t = \hat{\lambda}_t - \lambda_t$ denote the investor's posterior mean error. Under the true data-generating process, the latent state evolves as $\lambda_{t+1} = A \lambda_t + \eta_{t+1} + J_{t+1}$. The investor, operating under the misspecified stable model, updates beliefs via the Kalman filter:
\[ \hat{\lambda}_{t+1} = A \hat{\lambda}_t + K_{t+1} (f_{t+1} - A \hat{\lambda}_t) \]
Substituting the observation equation $f_{t+1} = \lambda_{t+1} + u_{t+1}$ into the belief update yields:
\begin{align*}
e_{t+1} &= \hat{\lambda}_{t+1} - \lambda_{t+1} \\
&= A \hat{\lambda}_t + K_{t+1}(\lambda_{t+1} + u_{t+1} - A \hat{\lambda}_t) - \lambda_{t+1} \\
&= (I - K_{t+1})(A \hat{\lambda}_t - \lambda_{t+1}) + K_{t+1}u_{t+1}
\end{align*}
Since $\lambda_{t+1} = A\lambda_t + \eta_{t+1} + J_{t+1}$, we can rewrite the term in the parentheses as $A e_t - \eta_{t+1} - J_{t+1}$. This gives the exact error dynamics:
\[ e_{t+1} = (I - K_{t+1})A e_t - (I - K_{t+1})J_{t+1} - (I - K_{t+1})\eta_{t+1} + K_{t+1}u_{t+1} \]
Suppose a discrete structural break occurs at time $t^\star$, such that $J_{t^\star} \neq 0$. Conditioning on the realized break at $t^\star$ and on no additional breaks over the next $h$ periods, and taking expectations over the Gaussian innovations, yields the local post-break error path
\[
\mathbb{E}[e_{t^\star+h}\mid J_{t^\star},\, J_{t^\star+1}=\cdots=J_{t^\star+h}=0]
=
\big((I-K)A\big)^{h+1} e_{t^\star-1}
-
\big((I-K)A\big)^h (I-K)J_{t^\star},
\]
where $K$ represents the steady-state Kalman gain matrix. Assume the usual stabilizability and detectability conditions for the misspecified linear-Gaussian state-space system, so that the algebraic Riccati equation admits a unique stabilizing solution and the steady-state Kalman gain $K$ exists. We additionally assume
\[
\rho\!\left((I-K)A\right) < 1,
\]
where $\rho(\cdot)$ denotes the spectral radius. Under these standard filter-stability conditions, the post-break error dynamics are well defined and decay geometrically.

Under the misspecified belief system, the investor assumes a state innovation variance $\tilde{\Sigma}_\eta$ that is strictly smaller than the true variance ($\tilde{\Sigma}_\eta \ll \Sigma_\eta$). Under the standard monotonicity properties of the stabilizing Riccati solution, a smaller $\tilde{\Sigma}_\eta$ implies a smaller steady-state Kalman gain $K$. The initial error introduced by the jump, $-(I-K)J_{t^\star}$, is therefore larger because the filter places too little weight on the break-consistent innovation. Since $\rho((I-K)A) < 1$, this post-break error decays geometrically. A smaller $K$ implies weaker attenuation of the jump component and slower correction of the induced belief wedge. Hence, under the stability conditions above, the post-break bias remains persistent for multiple periods and decays more slowly when $\tilde{\Sigma}_\eta$ is smaller. \qed

\subsection{Additional Formal Results for Propositions 2--4}

To complete the theoretical argument, this subsection provides formal sufficient conditions for Propositions~2--4. The objective is not to claim that every empirical pattern must hold uniformly across all factor taxonomies, but rather to show that the model generates these implications under economically interpretable conditions.

\subsubsection*{Notation}

Let $m_{t|t-1}$ denote the stable model's one-step-ahead predictive mean for $f_t$, and let
\[
s_{S,t}^2
\]
denote the corresponding predictive variance. Under the break-aware model, the one-step-ahead predictive density is a two-component mixture with jump probability $p_t$, jump mean $\mu_J$, and jump variance increment $\sigma_J^2$. Define
\[
s_{B,t}^2 = s_{S,t}^2 + \sigma_J^2.
\]

\subsubsection*{Lemma 1 (Likelihood-ratio representation)}

Under the stable Gaussian predictive density
\[
p_S(f_t \mid \mathcal{F}_{t-1}) = \phi(f_t; m_{t|t-1}, s_{S,t}^2),
\]
and the break-aware mixture density
\[
p_B(f_t \mid \mathcal{F}_{t-1})
=
(1-p_t)\phi(f_t; m_{t|t-1}, s_{S,t}^2)
+
p_t\phi(f_t; m_{t|t-1}+\mu_J, s_{B,t}^2),
\]
the mislearning measure can be written as
\[
\Delta_t
=
\log\!\left[(1-p_t) + p_t \exp\bigl(g_t(f_t)\bigr)\right],
\]
where
\[
g_t(x)
=
\frac{1}{2}\log\!\left(\frac{s_{S,t}^2}{s_{B,t}^2}\right)
+
\frac{(x-m_{t|t-1})^2}{2s_{S,t}^2}
-
\frac{(x-m_{t|t-1}-\mu_J)^2}{2s_{B,t}^2}.
\]

\paragraph{Proof.}
By direct substitution,
\[
\frac{p_B(x \mid \mathcal{F}_{t-1})}{p_S(x \mid \mathcal{F}_{t-1})}
=
(1-p_t)
+
p_t
\frac{\phi(x; m_{t|t-1}+\mu_J, s_{B,t}^2)}
{\phi(x; m_{t|t-1}, s_{S,t}^2)}.
\]
Taking logs yields the stated expression with
\[
g_t(x)
=
\log
\frac{\phi(x; m_{t|t-1}+\mu_J, s_{B,t}^2)}
{\phi(x; m_{t|t-1}, s_{S,t}^2)}.
\]
Expanding the Gaussian densities gives the closed form above. \qed

\subsubsection*{Proposition 2: Formal Proof}

\paragraph{Claim.}
When realized returns are more consistent with the predictive density of the break model than that of the stable model, $\Delta_t$ rises. The increase is larger when the break is larger and when the stable model is more rigid.

\paragraph{Proof.}
By Lemma 1, $\Delta_t$ is strictly increasing in $g_t(f_t)$ whenever $0<p_t\le 1$, since
\[
\frac{\partial \Delta_t}{\partial g_t}
=
\frac{p_t e^{g_t}}{(1-p_t)+p_t e^{g_t}} > 0.
\]

To study a break-consistent realization, evaluate $g_t(x)$ at
\[
x = m_{t|t-1} + \mu_J.
\]
Then
\[
g_t(m_{t|t-1}+\mu_J)
=
\frac{1}{2}\log\!\left(\frac{s_{S,t}^2}{s_{B,t}^2}\right)
+
\frac{\mu_J^2}{2s_{S,t}^2},
\qquad
s_{B,t}^2=s_{S,t}^2+\sigma_J^2.
\]
The first term is negative because $s_{B,t}^2>s_{S,t}^2$, while the second term is positive and increasing in $|\mu_J|$. Hence, for sufficiently large $|\mu_J|$, we have
\[
g_t(m_{t|t-1}+\mu_J)>0,
\]
which implies $\Delta_t>0$.

Moreover, holding $s_{S,t}^2$ fixed,
\[
\frac{\partial g_t(m_{t|t-1}+\mu_J)}{\partial |\mu_J|}
=
\frac{|\mu_J|}{s_{S,t}^2}>0.
\]
Thus, along this canonical break realization, larger mean shifts increase $\Delta_t$.

Finally, writing $s:=s_{S,t}^2$, we have
\[
\frac{\partial g_t(m_{t|t-1}+\mu_J)}{\partial s}
=
\frac{\sigma_J^2}{2s(s+\sigma_J^2)}-\frac{\mu_J^2}{2s^2}.
\]

This derivative is strictly negative if and only if $\mu_J^2 > \sigma_J^2 \frac{s}{s+\sigma_J^2}$.
Therefore, $g_t$ decreases with $s_{S,t}^2$---and so a more rigid stable model magnifies the likelihood gap---whenever the jump magnitude is sufficiently large relative to the stable predictive variance. This establishes the proposition as a sufficient-condition result.
\qed

\subsubsection*{Lemma 2 (Subjective market-clearing condition under CARA--normal beliefs)}

Let
\[
m_t^S := \mathbb{E}_t^S[f_{t+1}]
\]
denote the subjective conditional mean of factor excess returns, and let $\Sigma_u$ denote the corresponding conditional covariance matrix. If net supply is $S_t$, then market clearing implies
\[
m_t^S = \gamma \Sigma_u S_t.
\]

\paragraph{Proof.}
The investor chooses holdings $x_t$ to maximize
\[
x_t^\top m_t^S - \frac{\gamma}{2}x_t^\top \Sigma_u x_t.
\]
The first-order condition is
\[
m_t^S - \gamma \Sigma_u x_t = 0,
\]
so
\[
x_t = \frac{1}{\gamma}\Sigma_u^{-1} m_t^S.
\]
Imposing market clearing, $x_t = S_t$, yields
\[
m_t^S = \gamma \Sigma_u S_t.
\]
\qed

\subsubsection*{Theorem 1 (Expected excess-return decomposition under misspecified beliefs)}
Let
\[
m_t^T := \mathbb{E}_t[f_{t+1}]
\]
denote the true conditional mean of factor excess returns, and define the belief wedge
\[
w_t := m_t^T - m_t^S.
\]
Then the true conditional expected excess return satisfies
\[
\mathbb{E}_t[f_{t+1}] = \gamma \Sigma_u S_t + w_t.
\]

\paragraph{Proof.}
From Lemma~2,
\[
m_t^S = \gamma \Sigma_u S_t.
\]
Hence
\[
\mathbb{E}_t[f_{t+1}]
= m_t^T
= m_t^S + (m_t^T - m_t^S)
= \gamma \Sigma_u S_t + w_t.
\]
\qed

\subsubsection*{Corollary 1 (Long-horizon uncertainty premium)}

Suppose there exists a horizon $h$ such that the expected cumulative correction of the belief wedge,
\[
C_{t,h} = \mathbb{E}_t\!\left[\sum_{j=0}^{h-1} w_{t+j}\right],
\]
is weakly increasing in $\Delta_t$. Then expected cumulative excess returns over horizon $h$ are weakly increasing in $\Delta_t$:
\[
\mathbb{E}_t\!\left[\sum_{j=1}^{h} f_{t+j}\right]
=
\mathbb{E}_t\!\left[\sum_{j=0}^{h-1} \gamma \Sigma_u S_{t+j}\right]
+
C_{t,h}.
\]
If the conditional variance of cumulative returns grows sufficiently slowly relative to the conditional mean, then the long-horizon Sharpe ratio is also weakly increasing in $\Delta_t$.

\paragraph{Interpretation.}
This is a sufficient-condition result. It formalizes the empirical Proposition~\ref{prop:uncertainty_premium}: when elevated $\Delta_t$ identifies states in which the future correction of the belief wedge is larger, long-horizon expected returns and Sharpe ratios rise with mislearning intensity.

\subsubsection*{Proposition 4: Formal Proof}

\paragraph{Claim.}
Let $B_{k,t}$ denote the break-state indicator for factor $k$, and define
\[
\pi_k = \Pr(B_{k,t}=1), \qquad
\mu_{1,k} = \mathbb{E}[\Delta_{k,t}\mid B_{k,t}=1], \qquad
\mu_{0,k} = \mathbb{E}[\Delta_{k,t}\mid B_{k,t}=0].
\]
Then
\[
\mathbb{E}[\Delta_{k,t}] = \pi_k \mu_{1,k} + (1-\pi_k)\mu_{0,k}.
\]
Moreover, for any fixed spike threshold $c$,
\[
\Pr(\Delta_{k,t}>c)
=
\pi_k q_{1,k}(c) + (1-\pi_k)q_{0,k}(c),
\]
where
\[
q_{1,k}(c)=\Pr(\Delta_{k,t}>c\mid B_{k,t}=1),
\qquad
q_{0,k}(c)=\Pr(\Delta_{k,t}>c\mid B_{k,t}=0).
\]

\paragraph{Proof.}
By the law of iterated expectations,
\[
\mathbb{E}[\Delta_{k,t}]
=
\mathbb{E}\!\left[\mathbb{E}[\Delta_{k,t}\mid B_{k,t}]\right]
=
\pi_k \mu_{1,k} + (1-\pi_k)\mu_{0,k}.
\]
Similarly, by the law of total probability,
\[
\Pr(\Delta_{k,t}>c)
=
\Pr(\Delta_{k,t}>c\mid B_{k,t}=1)\Pr(B_{k,t}=1)
+
\Pr(\Delta_{k,t}>c\mid B_{k,t}=0)\Pr(B_{k,t}=0),
\]
which gives
\[
\Pr(\Delta_{k,t}>c)
=
\pi_k q_{1,k}(c) + (1-\pi_k)q_{0,k}(c).
\]
Thus both unconditional average mislearning and unconditional spike frequency admit an exact decomposition into break-frequency and conditional-severity components. \qed

\subsubsection*{Corollary 4.1: Formal Proof}

\paragraph{Claim.}
Suppose the non-break component of mislearning is comparable across factors, so that
\[
\mu_{0,k}=\bar{\mu}_0,
\]
and define the break-state severity gap
\[
\delta_k := \mu_{1,k}-\mu_{0,k}.
\]
If $\delta_k\ge 0$ for all $k$ and $\delta_k$ is constant across factors or weakly increasing in $\pi_k$, then $\mathbb{E}[\Delta_{k,t}]$ is weakly increasing in $\pi_k$. Likewise, for any fixed threshold $c$, if
\[
q_{0,k}(c)=\bar q_0(c)
\]
and
\[
\eta_k(c):=q_{1,k}(c)-q_{0,k}(c)\ge 0
\]
is constant across factors or weakly increasing in $\pi_k$, then $\Pr(\Delta_{k,t}>c)$ is weakly increasing in $\pi_k$.

\paragraph{Proof.}
Under $\mu_{0,k}=\bar{\mu}_0$, Proposition~\ref{prop:cross_factor_decomposition} implies
\[
\mathbb{E}[\Delta_{k,t}]
=
\bar{\mu}_0 + \pi_k \delta_k.
\]
Now take two factors $k$ and $\ell$ such that $\pi_k \ge \pi_\ell$. If $\delta_k \ge \delta_\ell \ge 0$, then
\[
\pi_k\delta_k \ge \pi_\ell\delta_\ell,
\]
and therefore
\[
\mathbb{E}[\Delta_{k,t}] \ge \mathbb{E}[\Delta_{\ell,t}].
\]
Hence average mislearning is weakly increasing in break-proneness.

The same argument applies to spike frequency. Under $q_{0,k}(c)=\bar q_0(c)$, Proposition~\ref{prop:cross_factor_decomposition} implies
\[
\Pr(\Delta_{k,t}>c)
=
\bar q_0(c) + \pi_k \eta_k(c).
\]
If $\eta_k(c)\ge \eta_\ell(c)\ge 0$ whenever $\pi_k\ge \pi_\ell$, then
\[
\pi_k \eta_k(c) \ge \pi_\ell \eta_\ell(c),
\]
so $\Pr(\Delta_{k,t}>c)$ is weakly increasing in $\pi_k$. \qed

\subsubsection*{Interpretation for the empirical evidence}

The empirical evidence is most naturally interpreted through Proposition~\ref{prop:cross_factor_decomposition} together with the conditional implication in Corollary~\ref{cor:prop4_monotone}. The anomaly-level decomposition holds numerically by construction. The stronger monotonic implication does not emerge as a universal one-dimensional law in the full cross-section. Instead, reduced-form diagnostics indicate that IVOL predicts break-state conditional severity $\mu_{1,k}$ but does not predict break-proneness $\pi_k$. This makes IVOL a natural screening variable for identifying lower-friction environments in which break-state severity is more comparable across assets.

Consistent with this interpretation, IVOL-tertile validation shows that the original Prop~4 logic is most clearly visible in lower-friction environments. In the Low-IVOL anomaly subsample, both the relation between average mislearning and break-proneness and the relation between spike frequency and break-proneness are positive and economically clean. At the same time, the strongest average-mislearning slope does not occur exclusively in the Low-IVOL tertile. We therefore interpret the cross-sectional evidence as \emph{partial conditional support} for Corollary~\ref{cor:prop4_monotone}: the monotonic mapping becomes more visible when cross-anomaly severity heterogeneity is compressed, but it is not a universal law across the full anomaly universe.
\section{q-Factor Robustness Tables}
\label{app:q5-robustness-tables}

\subsection{Unrestricted Baseline}
\input{tabs/baseline_predictive_q5_robust.tex}

\FloatBarrier

\subsection{Common-Sample Baseline}
\input{tabs/baseline_predictive_q5_robust_common_sample.tex}

\FloatBarrier

\subsection{Controlled Specification}
\input{tabs/controlled_predictive_q5_robust.tex}

\FloatBarrier

\subsection{Rank-Based Diagnostic for Proposition~4}
\label{app:q5-prop4-rank}

To complement the revised q5 cross-factor heterogeneity table in the main text, this appendix reports a simple rank-based diagnostic. The purpose is to assess whether factors with higher break-proneness also rank higher in break-state mislearning severity and spike frequency. Because the q5 universe contains only five factors, this exercise should be interpreted as descriptive rather than as a high-powered cross-sectional statistical test.

The rank evidence confirms the mixed nature of the q5 results. The ordering of unconditional break-proneness and break-state conditional average mislearning is not monotone, whereas the ordering of break-proneness and pooled spike frequency is more positively aligned. This pattern reinforces the interpretation in Section~\ref{sec:q5-prop4}: q5 supports the existence of cross-factor heterogeneity, but not a clean one-to-one mapping from break-proneness to break-state mislearning severity or spike frequency.

\FloatBarrier
\input{tabs/q5_prop4_revised_rankcheck.tex}
\FloatBarrier

\section{Anomaly Family Classification and Model Diagnostics}
\label{app:anomaly-family}

This appendix documents the anomaly-family classification and associated model-diagnostic outputs used in the anomaly-universe analysis. Family assignments are based on transparent name-based rules with economically motivated exact-match overrides for ambiguous cases.

\input{tabs/anomaly_stage5diag_family_map_summary.tex}

\input{tabs/anomaly_stage5diag_fit_quality.tex}

\begin{figure}[htbp]
\centering
\includegraphics[width=0.72\textwidth]{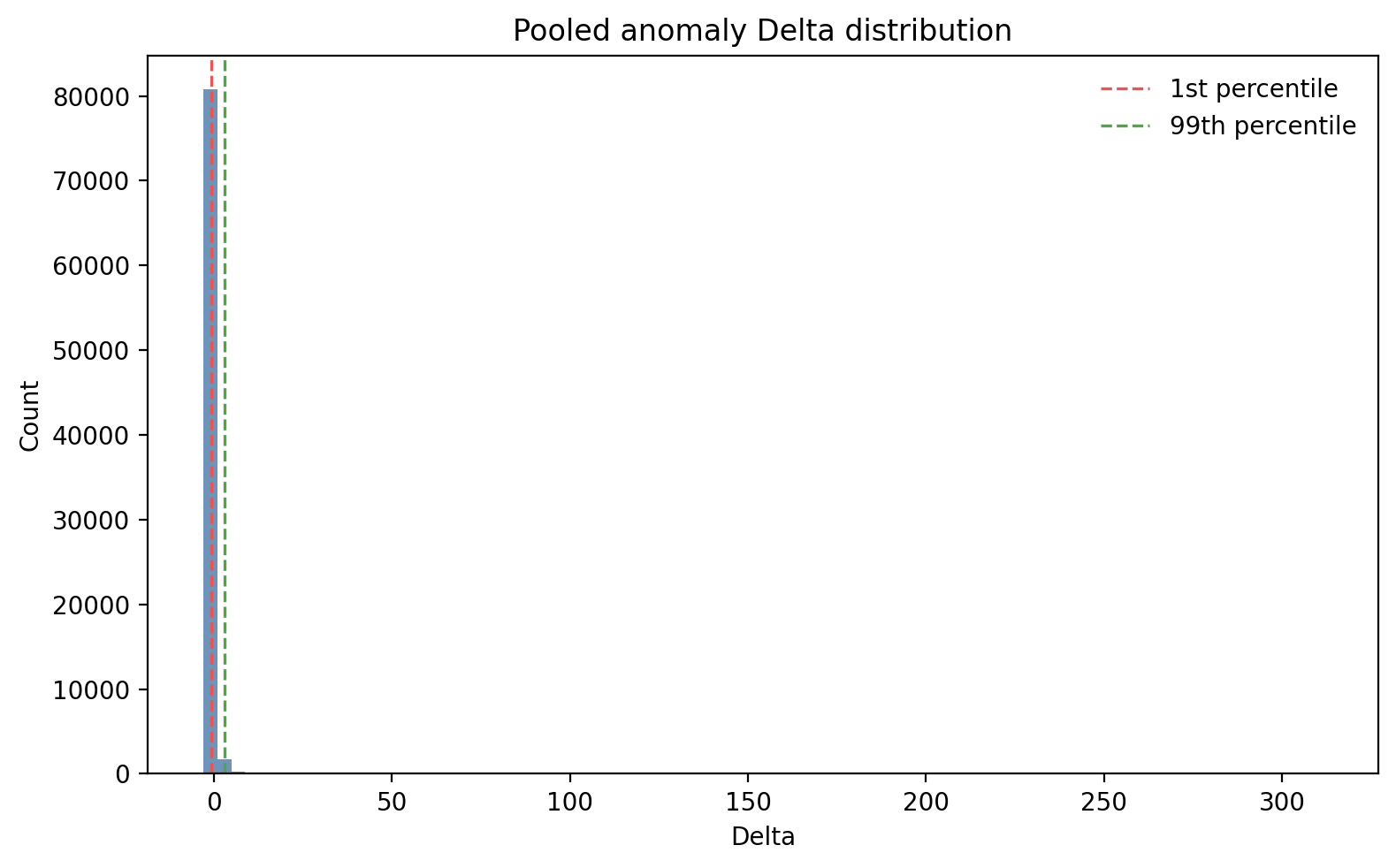}
\caption{Distribution of mislearning intensity $\Delta_t$ across the anomaly universe.}
\label{fig:app-anomaly-delta-hist}
\end{figure}

\FloatBarrier

\section{Reduced-Form Cross-Sectional Diagnostics for Proposition~\ref{prop:cross_factor_decomposition}}
\label{app:prop4-reducedform}

This appendix reports the reduced-form anomaly-level diagnostics underlying the empirical interpretation of Proposition~\ref{prop:cross_factor_decomposition} and Corollary~\ref{cor:prop4_monotone}. The decomposition identity holds numerically by construction. More importantly, IVOL is positively associated with break-state conditional severity $\mu_{1,k}$, while it does not predict break-proneness $\pi_k$. This motivates using IVOL as an ex-ante screening variable in the IVOL-tertile validation of the corollary.

\paragraph{Scope.}
All results in this section should be interpreted as reduced-form diagnostics rather than structural identification. In particular, these regressions are intended to document empirical regularities consistent with the decomposition in Proposition~\ref{prop:cross_factor_decomposition}, rather than to identify the underlying structural determinants of break frequency or break-state severity.

\input{tabs/prop4_decomp_summary.tex}

\FloatBarrier

\input{tabs/prop4_decomp_xsec.tex}

\FloatBarrier

\input{tabs/prop4_decomp_signsummary.tex}

\FloatBarrier

\section{IVOL-Screened Validation of Corollary~\ref{cor:prop4_monotone}}
\label{app:prop4-ivol}

This appendix reports a reduced-form anomaly-level validation of Corollary~\ref{cor:prop4_monotone}. The motivating fact is that IVOL predicts break-state conditional severity $\mu_{1,k}$ but does not predict break-proneness $\pi_k$. This makes IVOL a natural ex-ante screening variable for identifying cross-sectional environments in which break-state severity is more comparable across assets.

We sort the anomaly universe into IVOL tertiles and estimate within-tertile cross-sectional regressions of unconditional average mislearning and mislearning spike frequency on break-proneness. If the conditional monotonic logic of Corollary~\ref{cor:prop4_monotone} is operative, it should appear most clearly in the lower-friction, Low-IVOL subsample.

To further assess whether the conditions underlying Corollary~\ref{cor:prop4_monotone} are more likely to hold in lower-friction environments, we examine the cross-anomaly dispersion of break-state conditional mislearning. The standard deviation of $\mu_{1,k}$ increases substantially across IVOL groups, rising from approximately 0.10 in the Low-IVOL tertile to above 0.30 in the High-IVOL tertile. A similar, though less pronounced, increase is observed for the non-break component $\mu_{0,k}$ and for unconditional average mislearning.

This pattern indicates that cross-anomaly heterogeneity in break-state severity is substantially compressed in low-IVOL environments and strongly amplified in high-IVOL environments. Because the monotonic implication in Corollary~\ref{cor:prop4_monotone} requires that break-state severity (and thus the severity gap $\delta_k = \mu_{1,k} - \mu_{0,k}$) be comparable across factors, this dispersion pattern provides a direct structural explanation for why the monotonic relation between break-proneness and average mislearning is more visible in the Low-IVOL subsample but not in the full cross-section.

\input{tabs/prop4_tertile_regressions.tex}

\input{tabs/prop4_tertile_descriptives.tex}

\begin{figure}[htbp]
\centering
\includegraphics[width=0.82\textwidth]{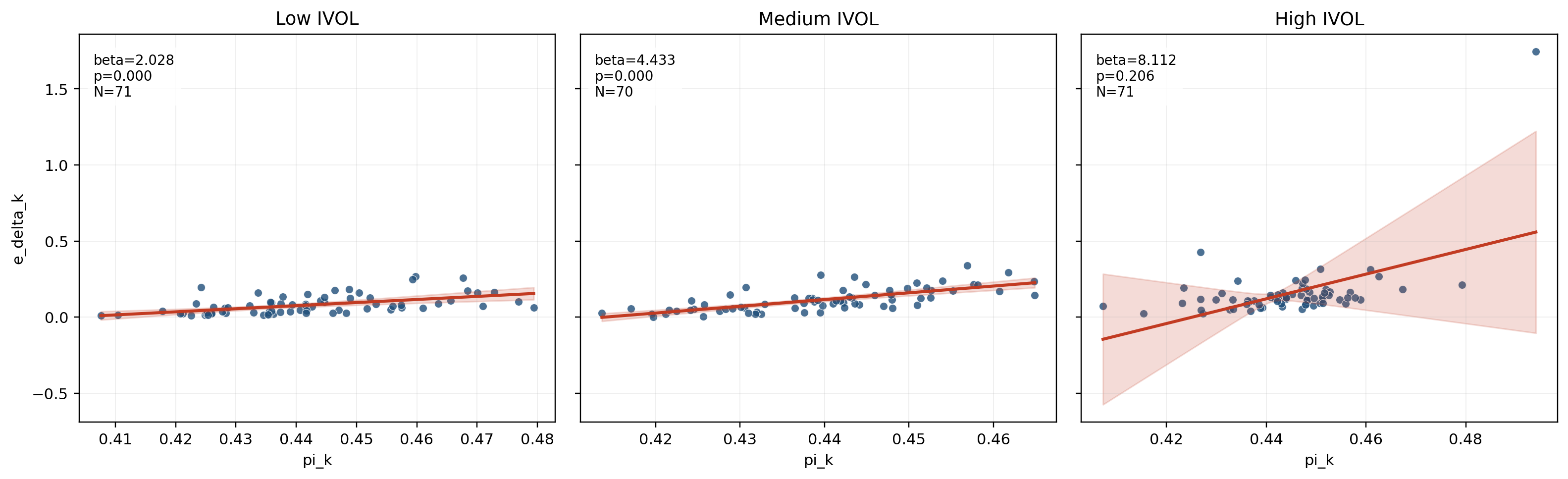}
\caption{IVOL-tertile slopes for unconditional average mislearning on break-proneness.}
\label{fig:app-prop4-tertile-edelta}
\end{figure}

\FloatBarrier

\begin{figure}[htbp]
\centering
\includegraphics[width=0.82\textwidth]{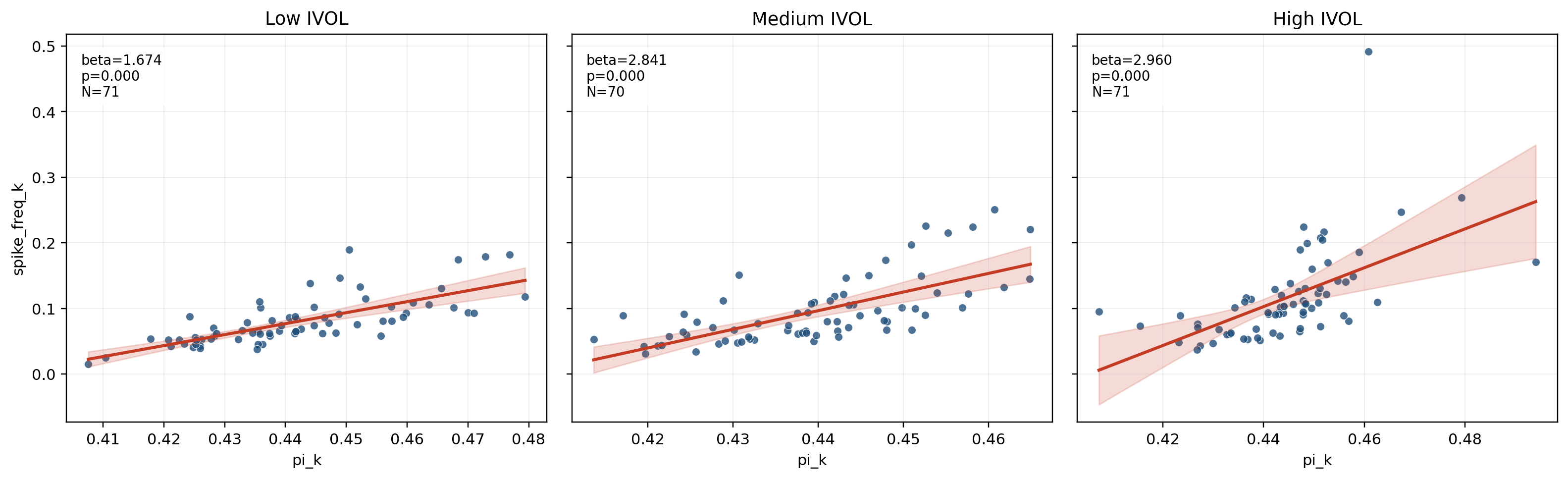}
\caption{IVOL-tertile slopes for mislearning spike frequency on break-proneness.}
\label{fig:app-prop4-tertile-spike}
\end{figure}

\FloatBarrier

\section{Additional Anomaly-Universe Predictive Diagnostics}
\label{app:anomaly-predictive}

This appendix reports additional predictive diagnostics for the anomaly universe. In particular, it shows that the weak pooled 12-month future-Sharpe result is not driven by the choice of standard-error estimator or clustering scheme. It also reports supplementary family-level outcome summaries and extreme-value robustness checks that are referenced in the main-text anomaly section.

\input{tabs/anomaly_stage5diag_alt_inference.tex}

\FloatBarrier

\input{tabs/anomaly_stage5diag_alt_family_summary.tex}

\FloatBarrier

\input{tabs/anomaly_stage5diag_fit_extremes.tex}

\FloatBarrier

\end{document}

%% file: tabs/ff6_prop4_revised_main.tex
\begin{table}[htbp]
\centering
\small
\setlength{\tabcolsep}{4pt}
\caption{FF6 factors: break diagnostics and predictive slopes}
\label{tab:ff6_prop4_revised}
\begin{tabular}{l r r r r c r r}
\toprule
Factor 
& Pr(Break) 
& Break share 
& $\Delta$ (break) 
& Spike freq 
& \shortstack[c]{Sharpe coef.\\($p$, 12m)} 
& Obs. 
& Break mths \\
\midrule
MKT (RF) & 0.4878 & 0.4660 & -0.0237 & 0.0360 & 0.2887 (0.0005) & 751 & 350 \\
SMB      & 0.3423 & 0.2716 &  0.0570 & 0.0253 & -0.0843 (0.2074) & 751 & 204 \\
HML      & 0.3664 & 0.3409 &  0.0578 & 0.0466 & 0.2333 (0.0044) & 751 & 256 \\
RMW      & 0.1019 & 0.0985 &  1.2098 & 0.0293 & 0.0710 (0.0028) & 751 & 74 \\
CMA      & 0.2436 & 0.2210 &  0.1994 & 0.0453 & -0.0049 (0.9517) & 751 & 166 \\
UMD      & 0.1569 & 0.1312 &  0.7763 & 0.3011 & -0.0057 (0.8031) & 1189 & 156 \\
\bottomrule
\end{tabular}
\end{table}

%% file: tabs/baseline_predictive_v1.tex
\begin{table}[htbp]
\centering
\footnotesize
\setlength{\tabcolsep}{4pt}
\caption{Baseline pooled predictive regressions of future factor performance on mislearning intensity $\Delta_t$.}
\label{tab:baseline-predictive}
\begin{tabular}{l c c c}
\toprule
Outcome & Horizon & Coef. (SE, $t$, $p$) & Obs. ($R^2$) \\
\midrule

\multicolumn{4}{l}{\textit{Panel A. Horizon = 3 months}} \\
\midrule
Sharpe & 3 & 0.0102 (0.0476, 0.21, 0.831) & 4926 (0.0108) \\
Cum. return & 3 & -0.0012 (0.0027, -0.42, 0.673) & 4926 (0.0076) \\
Volatility & 3 & 0.0061 (0.0037, 1.64, 0.100) & 4926 (0.1260) \\
Downside vol. & 3 & -0.0006 (0.0004, -1.28, 0.200) & 4926 (0.0154) \\
Max DD & 3 & -0.0001 (0.0004, -0.24, 0.808) & 4926 (0.0258) \\

\midrule
\multicolumn{4}{l}{\textit{Panel B. Horizon = 6 months}} \\
\midrule
Sharpe & 6 & 0.0226 (0.0251, 0.90, 0.368) & 4908 (0.0263) \\
Cum. return & 6 & 0.0002 (0.0032, 0.06, 0.951) & 4908 (0.0135) \\
Volatility & 6 & 0.0043 (0.0028, 1.56, 0.120) & 4908 (0.1631) \\
Downside vol. & 6 & 0.0039 (0.0036, 1.07, 0.285) & 4908 (0.0555) \\
Max DD & 6 & -0.0004 (0.0007, -0.64, 0.524) & 4908 (0.0495) \\

\midrule
\multicolumn{4}{l}{\textit{Panel C. Horizon = 12 months}} \\
\midrule
Sharpe & 12 & 0.0288 (0.0219, 1.31, 0.189) & 4872 (0.0382) \\
Cum. return & 12 & 0.0040 (0.0049, 0.82, 0.415) & 4872 (0.0255) \\
Volatility & 12 & 0.0036 (0.0019, 1.89, 0.059) & 4872 (0.2091) \\
Downside vol. & 12 & 0.0033 (0.0025, 1.33, 0.185) & 4872 (0.1216) \\
Max DD & 12 & 0.0000 (0.0013, -0.06, 0.955) & 4872 (0.0799) \\

\bottomrule
\end{tabular}
\end{table}

%% file: tabs/controlled_predictive_v1.tex
\begin{table}[htbp]
\centering
\footnotesize
\setlength{\tabcolsep}{4pt}
\caption{Controlled pooled predictive regressions with additional risk controls.}
\label{tab:controlled-predictive}
\begin{tabular}{l c c c}
\toprule
Outcome & Horizon & Coef. (SE, $t$, $p$) & Obs. ($R^2$) \\
\midrule

\multicolumn{4}{l}{\textit{Panel A. Horizon = 3 months}} \\
\midrule
Sharpe & 3 & 0.0769 (0.0887, 0.87, 0.386) & 2580 (0.0187) \\
Cum. return & 3 & 0.0024 (0.0019, 1.23, 0.218) & 2580 (0.0236) \\
Volatility & 3 & 0.0021 (0.0014, 1.45, 0.148) & 2580 (0.2887) \\
Downside vol. & 3 & -0.0008 (0.0011, -0.79, 0.428) & 2580 (0.0413) \\
Max DD & 3 & 0.0001 (0.0007, 0.16, 0.876) & 2580 (0.0786) \\

\midrule
\multicolumn{4}{l}{\textit{Panel B. Horizon = 6 months}} \\
\midrule
Sharpe & 6 & 0.0521 (0.0438, 1.19, 0.234) & 2562 (0.0489) \\
Cum. return & 6 & 0.0034 (0.0031, 1.09, 0.274) & 2562 (0.0403) \\
Volatility & 6 & 0.0012 (0.0018, 0.66, 0.510) & 2562 (0.3916) \\
Downside vol. & 6 & -0.0005 (0.0015, -0.30, 0.761) & 2562 (0.1336) \\
Max DD & 6 & -0.0006 (0.0011, -0.55, 0.580) & 2562 (0.1322) \\

\midrule
\multicolumn{4}{l}{\textit{Panel C. Horizon = 12 months}} \\
\midrule
Sharpe & 12 & 0.0801 (0.0288, 2.78, 0.006) & 2526 (0.0735) \\
Cum. return & 12 & 0.0145 (0.0048, 3.00, 0.003) & 2526 (0.0789) \\
Volatility & 12 & 0.0009 (0.0017, 0.51, 0.608) & 2526 (0.4163) \\
Downside vol. & 12 & -0.0017 (0.0015, -1.16, 0.248) & 2526 (0.2734) \\
Max DD & 12 & -0.0040 (0.0016, -2.52, 0.012) & 2526 (0.1594) \\

\bottomrule
\end{tabular}
\end{table}

%% file: tabs/q5_robust_summary.tex
\begin{table}[htbp]
\centering
\small
\setlength{\tabcolsep}{4pt}
\caption{Pooled Predictive Coefficient Signs: q5 Robustness Check}
\label{tab:q5-summary}
\begin{tabular}{p{3.2cm}c c c c}
\toprule
Specification & Horizon & $\Delta_t\!\to\!$ Sharpe & $\Delta_t\!\to\!$ Cum. Return & $\Delta_t\!\to\!$ Volatility \\
\midrule
\textbf{Baseline} & 3m  & $-$ (ns)             & $-$ (ns)              & $+$ (ns) \\
                   & 6m  & $-$ (ns)             & $\approx 0$          & $+$ (ns) \\
                   & 12m & $+$ ($p\approx0.20$) & $+$ ($p\approx0.06$) & $+$ (ns) \\
\midrule
\textbf{Common-sample baseline} & 3m  & $-$ (ns)             & $-$ (ns)              & $+$ (ns) \\
                                 & 6m  & $\approx 0$         & $\approx 0$          & $+$ (ns) \\
                                 & 12m & $+$ ($p\approx0.10$) & $+$ ($p\approx0.10$) & $++$ ($p\approx0.01$) \\
\midrule
\textbf{Controlled} & 3m  & $-$ (ns)             & $-$ (ns)              & $+$ (ns) \\
                     & 6m  & $+$ (ns)             & $-$ (ns)              & $+$ (ns) \\
                     & 12m & $+$ ($p\approx0.10$) & $+$ (ns)              & $\approx 0$ \\
\bottomrule
\end{tabular}

\vspace{0.4em}
\footnotesize
\textit{Notes}: Signs report coefficient direction; parentheses report significance. 
``Common-sample baseline'' uses the controlled-regression sample without controls; ``ns'' denotes statistical insignificance.
\end{table}

%% file: tabs/q5_prop4_revised_main.tex
\begin{table}[htbp]
\centering
\small
\setlength{\tabcolsep}{4pt}
\caption{q5 factors: break diagnostics and predictive slopes}
\label{tab:q5_prop4_revised}
\begin{tabular}{l r r r r c r r}
\toprule
Factor 
& Pr(Break) 
& Break share 
& $\Delta$ (break) 
& Spike freq 
& \shortstack[c]{Sharpe coef.\\($p$, 12m)} 
& Obs. 
& Break mths \\
\midrule
MKT & 0.4146 & 0.3635 & -0.0016 & 0.0603 & 0.4286 (0.078) & 696 & 253 \\
ME  & 0.0364 & 0.0172 & 2.9783 & 0.0086 & 0.0174 (0.428) & 696 & 12 \\
IA  & 0.1753 & 0.1595 & 0.3764 & 0.0560 & 0.1103 (0.210) & 696 & 111 \\
ROE & 0.2101 & 0.1782 & 0.3072 & 0.2601 & 0.0230 (0.770) & 696 & 124 \\
EG  & 0.3426 & 0.3233 & 0.1152 & 0.1149 & 0.0616 (0.732) & 696 & 225 \\
\bottomrule
\end{tabular}
\end{table}

%% file: tabs/anomaly_stage5diag_family_regressions.tex
\begin{table}[htbp]
\centering
\footnotesize
\setlength{\tabcolsep}{3pt}
\caption{Anomaly family heterogeneity in 12-month predictive regressions}
\label{tab:anomaly-family-regressions}
\begin{tabular}{>{\raggedright\arraybackslash}p{3.1cm} c c c c c}
\toprule
Family & $N$ & Sharpe (base) & Sharpe (ctrl) & CumRet (ctrl) & Vol (ctrl) \\
\midrule
Value & 18 & 0.0101 (0.281) & -0.0042 (0.578) & 0.0029 (0.160) & 0.0062 (0.000) \\
Momentum & 33 & 0.0052 (0.636) & 0.0176 (0.072) & 0.0091 (0.001) & 0.0058 (0.031) \\
Investment & 16 & -0.0218 (0.002) & -0.0256 (0.003) & -0.0036 (0.001) & 0.0027 (0.019) \\
Profit./quality & 10 & 0.0564 (0.000) & 0.0588 (0.000) & 0.0235 (0.000) & 0.0062 (0.000) \\
Accruals/acct. & 18 & -0.0083 (0.631) & -0.0077 (0.664) & 0.0035 (0.125) & 0.0026 (0.008) \\
Beta/risk/vol. & 35 & -0.0024 (0.601) & 0.0000 (0.999) & -0.0022 (0.594) & 0.0037 (0.020) \\
Growth/issuance & 32 & -0.0045 (0.087) & -0.0002 (0.926) & 0.0004 (0.502) & 0.0011 (0.092) \\
\shortstack[l]{Reversal/seasonality/\\microstr.} & 33 & -0.0142 (0.472) & -0.0474 (0.072) & 0.0001 (0.987) & 0.0019 (0.327) \\
Other & 17 & -0.0040 (0.094) & -0.0079 (0.006) & -0.0008 (0.084) & 0.0022 (0.004) \\
\bottomrule
\end{tabular}
\end{table}

%% file: tabs/anomaly_stage5diag_alt_outcomes.tex
\begin{table}[htbp]
\centering
\footnotesize
\setlength{\tabcolsep}{3pt}
\caption{Alternative outcome definitions in pooled anomaly predictive regressions}
\label{tab:anomaly-alt-outcomes}
\begin{tabular}{l c c c}
\toprule
Outcome & Coef. ($p$) & $N$ \\
\midrule

\multicolumn{3}{l}{\textit{Panel A. Baseline, all months}} \\
\midrule
Sharpe (12m) & -0.0025 (0.234) & 158,038 \\
Cum. return (12m) & 0.0022 (0.167) & 158,038 \\
Volatility (12m) & 0.0048 (0.001) & 158,038 \\
Downside semivol. (12m) & 0.0031 (0.001) & 158,038 \\
Drawdown (12m) & 0.0032 (0.001) & 158,038 \\
Volatility ratio (12m) & 0.0046 (0.266) & 158,038 \\

\midrule
\multicolumn{3}{l}{\textit{Baseline, break state ($p>0.5$)}} \\
\midrule
Cum. return (12m) & 0.0013 (0.402) & 27,761 \\

\midrule
\multicolumn{3}{l}{\textit{Panel B. Controlled (lagged), all months}} \\
\midrule
Sharpe (12m) & -0.0012 (0.587) & 82,887 \\
Cum. return (12m) & 0.0010 (0.496) & 82,887 \\
Volatility (12m) & 0.0027 (0.004) & 82,887 \\
Downside semivol. (12m) & 0.0019 (0.003) & 82,887 \\
Drawdown (12m) & 0.0018 (0.002) & 82,887 \\
Volatility ratio (12m) & 0.0118 (0.013) & 82,887 \\

\midrule
\multicolumn{3}{l}{\textit{Controlled (lagged), break state ($p>0.5$)}} \\
\midrule
Cum. return (12m) & 0.0007 (0.636) & 17,427 \\

\bottomrule
\end{tabular}
\end{table}

%% file: tabs/passive_stockonly_main.tex
\begin{table}[htbp]
\centering
\caption{Lagged stock-only passive ownership extension: future 12-month cumulative-return mapping}
\label{tab:passive-stockonly}
\begin{tabular}{lrrrr}
\toprule
System & Interaction coef. & SE & $p$ & $N$ \\
\midrule
FF6 & -0.8535 & 0.1653 & 0.0000 & 648 \\
q5 & -0.5262 & 0.2570 & 0.0406 & 475 \\
\bottomrule
\end{tabular}
\end{table}

%% file: tabs/baseline_predictive_v1_full.tex
\begingroup
\footnotesize
\setlength{\tabcolsep}{3.5pt}

\begin{center}
\begin{longtable}{p{1.7cm} p{1.3cm} p{2.2cm} c r r r r r r}
\caption{Baseline predictive regressions (full).}
\label{tab:baseline_predictive_full_panel} \\
\toprule
Sample & Factor & Outcome & Ctrl. & Coef. & SE & $t$ & $p$ & Obs. & $R^2$ \\
\midrule
\endfirsthead

\multicolumn{10}{l}{\textit{Table \thetable\ (continued)}} \\
\toprule
Sample & Factor & Outcome & Ctrl. & Coef. & SE & $t$ & $p$ & Obs. & $R^2$ \\
\midrule
\endhead

\midrule
\multicolumn{10}{r}{\textit{Continued on next page}} \\
\endfoot

\bottomrule
\endlastfoot

\multicolumn{10}{l}{\textit{Panel A. Horizon $h=3$}} \\
\midrule
Pooled & All & Sharpe & No & 0.0102 & 0.0476 & 0.2141 & 0.8305 & 4,926 & 0.0108 \\
Factor & CMA & Sharpe & No & -0.1330 & 0.2611 & -0.5095 & 0.6104 & 748 & <0.0001 \\
Factor & HML & Sharpe & No & 0.3213 & 0.2526 & 1.2720 & 0.2034 & 748 & 0.0010 \\
Factor & MKT-RF & Sharpe & No & 0.8178 & 0.5327 & 1.5351 & 0.1247 & 748 & 0.0032 \\
Factor & RMW & Sharpe & No & 0.1758 & 0.1040 & 1.6911 & 0.0908 & 748 & 0.0008 \\
Factor & SMB & Sharpe & No & -0.6297 & 0.2037 & -3.0906 & 0.0020 & 748 & 0.0023 \\
Factor & UMD & Sharpe & No & -0.0601 & 0.0567 & -1.0601 & 0.2891 & 1,186 & 0.0003 \\

Pooled & All & CumRet & No & -0.0012 & 0.0027 & -0.4216 & 0.6733 & 4,926 & 0.0076 \\
Factor & CMA & CumRet & No & 0.0020 & 0.0029 & 0.6766 & 0.4986 & 748 & 0.0010 \\
Factor & HML & CumRet & No & 0.0058 & 0.0039 & 1.4835 & 0.1379 & 748 & 0.0042 \\
Factor & MKT-RF & CumRet & No & 0.0054 & 0.0039 & 1.3579 & 0.1745 & 748 & 0.0013 \\
Factor & RMW & CumRet & No & 0.0059 & 0.0014 & 4.1135 & <0.0001 & 748 & 0.0394 \\
Factor & SMB & CumRet & No & -0.0084 & 0.0044 & -1.9072 & 0.0565 & 748 & 0.0092 \\
Factor & UMD & CumRet & No & -0.0045 & 0.0026 & -1.7316 & 0.0833 & 1,186 & 0.0103 \\

Pooled & All & Volatility & No & 0.0061 & 0.0037 & 1.6431 & 0.1004 & 4,926 & 0.1260 \\
Factor & CMA & Volatility & No & 0.0049 & 0.0034 & 1.4459 & 0.1482 & 748 & 0.0104 \\
Factor & HML & Volatility & No & 0.0032 & 0.0043 & 0.7382 & 0.4604 & 748 & 0.0020 \\
Factor & MKT-RF & Volatility & No & -0.0087 & 0.0063 & -1.3953 & 0.1629 & 748 & 0.0045 \\
Factor & RMW & Volatility & No & 0.0062 & 0.0017 & 3.6413 & 0.0003 & 748 & 0.0487 \\
Factor & SMB & Volatility & No & 0.0025 & 0.0063 & 0.4021 & 0.6876 & 748 & 0.0011 \\
Factor & UMD & Volatility & No & 0.0074 & 0.0039 & 1.9205 & 0.0548 & 1,186 & 0.0213 \\

Pooled & All & Downside vol. & No & -0.0006 & 0.0004 & -1.2828 & 0.1996 & 4,926 & 0.0154 \\
Factor & CMA & Downside vol. & No & -0.0007 & 0.0014 & -0.5180 & 0.6045 & 748 & 0.0007 \\
Factor & HML & Downside vol. & No & -0.0024 & 0.0012 & -2.0199 & 0.0434 & 748 & 0.0033 \\
Factor & MKT-RF & Downside vol. & No & -0.0051 & 0.0022 & -2.2799 & 0.0226 & 748 & 0.0037 \\
Factor & RMW & Downside vol. & No & -0.0009 & 0.0004 & -2.2128 & 0.0269 & 748 & 0.0049 \\
Factor & SMB & Downside vol. & No & 0.0034 & 0.0037 & 0.9317 & 0.3515 & 748 & 0.0067 \\
Factor & UMD & Downside vol. & No & -0.0003 & 0.0006 & -0.5601 & 0.5754 & 1,186 & 0.0002 \\

Pooled & All & Max DD & No & -0.0001 & 0.0004 & -0.2434 & 0.8077 & 4,926 & 0.0258 \\
Factor & CMA & Max DD & No & 0.0012 & 0.0010 & 1.2445 & 0.2133 & 748 & 0.0025 \\
Factor & HML & Max DD & No & -0.0006 & 0.0016 & -0.3553 & 0.7224 & 748 & 0.0002 \\
Factor & MKT-RF & Max DD & No & -0.0047 & 0.0024 & -1.9184 & 0.0551 & 748 & 0.0037 \\
Factor & RMW & Max DD & No & 0.0008 & 0.0008 & 1.0604 & 0.2889 & 748 & 0.0034 \\
Factor & SMB & Max DD & No & 0.0012 & 0.0018 & 0.6456 & 0.5185 & 748 & 0.0010 \\
Factor & UMD & Max DD & No & -0.0004 & 0.0004 & -0.9278 & 0.3535 & 1,186 & 0.0002 \\

Pooled & All & Failure & No & -0.0008 & 0.0060 & -0.1319 & 0.8951 & 4,926 & <0.0001 \\
Factor & CMA & Failure & No & -0.0175 & 0.0123 & -1.4201 & 0.1556 & 748 & 0.0014 \\
Factor & HML & Failure & No & -0.0281 & 0.0163 & -1.7249 & 0.0845 & 748 & 0.0037 \\
Factor & MKT-RF & Failure & No & -0.0391 & 0.0208 & -1.8809 & 0.0600 & 748 & 0.0047 \\
Factor & RMW & Failure & No & -0.0108 & 0.0048 & -2.2453 & 0.0247 & 748 & 0.0026 \\
Factor & SMB & Failure & No & 0.0087 & 0.0218 & 0.3972 & 0.6912 & 748 & 0.0003 \\
Factor & UMD & Failure & No & 0.0078 & 0.0059 & 1.3240 & 0.1855 & 1,186 & 0.0023 \\

\midrule
\multicolumn{10}{l}{\textit{Panel B. Horizon $h=6$}} \\
\midrule
Pooled & All & Sharpe & No & 0.0226 & 0.0251 & 0.9002 & 0.3680 & 4,908 & 0.0263 \\
Factor & CMA & Sharpe & No & -0.0445 & 0.1089 & -0.4083 & 0.6830 & 745 & 0.0001 \\
Factor & HML & Sharpe & No & 0.1044 & 0.1398 & 0.7467 & 0.4552 & 745 & 0.0009 \\
Factor & MKT-RF & Sharpe & No & 0.2904 & 0.1422 & 2.0419 & 0.0412 & 745 & 0.0042 \\
Factor & RMW & Sharpe & No & 0.0695 & 0.0308 & 2.2568 & 0.0240 & 745 & 0.0018 \\
Factor & SMB & Sharpe & No & -0.2293 & 0.1320 & -1.7374 & 0.0823 & 745 & 0.0031 \\
Factor & UMD & Sharpe & No & 0.0087 & 0.0284 & 0.3049 & 0.7605 & 1,183 & 0.0001 \\

Pooled & All & CumRet & No & 0.0002 & 0.0032 & 0.0612 & 0.9512 & 4,908 & 0.0135 \\
Factor & CMA & CumRet & No & 0.0041 & 0.0043 & 0.9444 & 0.3450 & 745 & 0.0019 \\
Factor & HML & CumRet & No & 0.0113 & 0.0082 & 1.3727 & 0.1699 & 745 & 0.0069 \\
Factor & MKT-RF & CumRet & No & 0.0097 & 0.0063 & 1.5516 & 0.1207 & 745 & 0.0020 \\
Factor & RMW & CumRet & No & 0.0069 & 0.0018 & 3.9250 & <0.0001 & 745 & 0.0267 \\
Factor & SMB & CumRet & No & -0.0081 & 0.0036 & -2.2579 & 0.0240 & 745 & 0.0042 \\
Factor & UMD & CumRet & No & -0.0034 & 0.0026 & -1.3068 & 0.1913 & 1,183 & 0.0032 \\

Pooled & All & Volatility & No & 0.0043 & 0.0028 & 1.5568 & 0.1195 & 4,908 & 0.1631 \\
Factor & CMA & Volatility & No & 0.0053 & 0.0033 & 1.5865 & 0.1126 & 745 & 0.0139 \\
Factor & HML & Volatility & No & 0.0006 & 0.0045 & 0.1315 & 0.8954 & 745 & 0.0001 \\
Factor & MKT-RF & Volatility & No & -0.0124 & 0.0052 & -2.3992 & 0.0164 & 745 & 0.0116 \\
Factor & RMW & Volatility & No & 0.0067 & 0.0015 & 4.4528 & <0.0001 & 745 & 0.0554 \\
Factor & SMB & Volatility & No & 0.0018 & 0.0068 & 0.2597 & 0.7951 & 745 & 0.0006 \\
Factor & UMD & Volatility & No & 0.0047 & 0.0026 & 1.8409 & 0.0656 & 1,183 & 0.0079 \\

Pooled & All & Downside vol. & No & 0.0039 & 0.0036 & 1.0687 & 0.2852 & 4,908 & 0.0555 \\
Factor & CMA & Downside vol. & No & 0.0017 & 0.0012 & 1.4473 & 0.1478 & 745 & 0.0029 \\
Factor & HML & Downside vol. & No & 0.0005 & 0.0021 & 0.2401 & 0.8102 & 745 & 0.0001 \\
Factor & MKT-RF & Downside vol. & No & -0.0083 & 0.0034 & -2.4327 & 0.0150 & 745 & 0.0068 \\
Factor & RMW & Downside vol. & No & 0.0019 & 0.0009 & 2.0224 & 0.0431 & 745 & 0.0110 \\
Factor & SMB & Downside vol. & No & 0.0032 & 0.0041 & 0.7820 & 0.4342 & 745 & 0.0045 \\
Factor & UMD & Downside vol. & No & 0.0058 & 0.0036 & 1.6149 & 0.1063 & 1,183 & 0.0182 \\

Pooled & All & Max DD & No & -0.0004 & 0.0007 & -0.6373 & 0.5239 & 4,908 & 0.0495 \\
Factor & CMA & Max DD & No & 0.0018 & 0.0015 & 1.2173 & 0.2235 & 745 & 0.0024 \\
Factor & HML & Max DD & No & -0.0017 & 0.0022 & -0.7758 & 0.4378 & 745 & 0.0007 \\
Factor & MKT-RF & Max DD & No & -0.0095 & 0.0037 & -2.5575 & 0.0105 & 745 & 0.0065 \\
Factor & RMW & Max DD & No & 0.0015 & 0.0008 & 1.9712 & 0.0487 & 745 & 0.0044 \\
Factor & SMB & Max DD & No & -0.0006 & 0.0018 & -0.3626 & 0.7169 & 745 & 0.0001 \\
Factor & UMD & Max DD & No & -0.0008 & 0.0008 & -1.0441 & 0.2964 & 1,183 & 0.0004 \\

Pooled & All & Failure & No & 0.0012 & 0.0057 & 0.2062 & 0.8367 & 4,908 & <0.0001 \\
Factor & CMA & Failure & No & -0.0058 & 0.0134 & -0.4328 & 0.6652 & 745 & 0.0002 \\
Factor & HML & Failure & No & -0.0025 & 0.0137 & -0.1798 & 0.8573 & 745 & <0.0001 \\
Factor & MKT-RF & Failure & No & -0.0287 & 0.0156 & -1.8380 & 0.0661 & 745 & 0.0025 \\
Factor & RMW & Failure & No & -0.0085 & 0.0040 & -2.1284 & 0.0333 & 745 & 0.0016 \\
Factor & SMB & Failure & No & 0.0069 & 0.0216 & 0.3207 & 0.7484 & 745 & 0.0002 \\
Factor & UMD & Failure & No & 0.0068 & 0.0053 & 1.2788 & 0.2010 & 1,183 & 0.0017 \\

\midrule
\multicolumn{10}{l}{\textit{Panel C. Horizon $h=12$}} \\
\midrule
Pooled & All & Sharpe & No & 0.0288 & 0.0219 & 1.3134 & 0.1890 & 4,872 & 0.0382 \\
Factor & CMA & Sharpe & No & -0.0049 & 0.0814 & -0.0606 & 0.9517 & 739 & <0.0001 \\
Factor & HML & Sharpe & No & 0.2333 & 0.0819 & 2.8468 & 0.0044 & 739 & 0.0121 \\
Factor & MKT-RF & Sharpe & No & 0.2887 & 0.0834 & 3.4619 & 0.0005 & 739 & 0.0136 \\
Factor & RMW & Sharpe & No & 0.0710 & 0.0237 & 2.9919 & 0.0028 & 739 & 0.0062 \\
Factor & SMB & Sharpe & No & -0.0843 & 0.0669 & -1.2607 & 0.2074 & 739 & 0.0012 \\
Factor & UMD & Sharpe & No & -0.0057 & 0.0230 & -0.2494 & 0.8031 & 1,177 & 0.0001 \\

Pooled & All & CumRet & No & 0.0040 & 0.0049 & 0.8152 & 0.4150 & 4,872 & 0.0255 \\
Factor & CMA & CumRet & No & 0.0039 & 0.0079 & 0.4937 & 0.6215 & 739 & 0.0008 \\
Factor & HML & CumRet & No & 0.0294 & 0.0177 & 1.6599 & 0.0969 & 739 & 0.0208 \\
Factor & MKT-RF & CumRet & No & 0.0233 & 0.0123 & 1.8898 & 0.0588 & 739 & 0.0056 \\
Factor & RMW & CumRet & No & 0.0207 & 0.0042 & 4.9228 & <0.0001 & 739 & 0.0990 \\
Factor & SMB & CumRet & No & -0.0070 & 0.0042 & -1.6602 & 0.0969 & 739 & 0.0014 \\
Factor & UMD & CumRet & No & -0.0045 & 0.0036 & -1.2438 & 0.2136 & 1,177 & 0.0028 \\

Pooled & All & Volatility & No & 0.0036 & 0.0019 & 1.8861 & 0.0593 & 4,872 & 0.2091 \\
Factor & CMA & Volatility & No & 0.0036 & 0.0033 & 1.0933 & 0.2742 & 739 & 0.0074 \\
Factor & HML & Volatility & No & 0.0017 & 0.0048 & 0.3604 & 0.7185 & 739 & 0.0008 \\
Factor & MKT-RF & Volatility & No & -0.0155 & 0.0044 & -3.5000 & 0.0005 & 739 & 0.0238 \\
Factor & RMW & Volatility & No & 0.0062 & 0.0017 & 3.6041 & 0.0003 & 739 & 0.0490 \\
Factor & SMB & Volatility & No & -0.0008 & 0.0051 & -0.1570 & 0.8752 & 739 & 0.0002 \\
Factor & UMD & Volatility & No & 0.0041 & 0.0021 & 1.9183 & 0.0551 & 1,177 & 0.0065 \\

Pooled & All & Downside vol. & No & 0.0033 & 0.0025 & 1.3251 & 0.1851 & 4,872 & 0.1216 \\
Factor & CMA & Downside vol. & No & 0.0014 & 0.0014 & 1.0026 & 0.3160 & 739 & 0.0025 \\
Factor & HML & Downside vol. & No & -0.0017 & 0.0023 & -0.7445 & 0.4566 & 739 & 0.0015 \\
Factor & MKT-RF & Downside vol. & No & -0.0122 & 0.0038 & -3.2089 & 0.0013 & 739 & 0.0156 \\
Factor & RMW & Downside vol. & No & 0.0011 & 0.0007 & 1.4168 & 0.1565 & 739 & 0.0032 \\
Factor & SMB & Downside vol. & No & 0.0012 & 0.0035 & 0.3442 & 0.7307 & 739 & 0.0007 \\
Factor & UMD & Downside vol. & No & 0.0057 & 0.0028 & 2.0087 & 0.0446 & 1,177 & 0.0110 \\

Pooled & All & Max DD & No & -0.0001 & 0.0013 & -0.0564 & 0.9550 & 4,872 & 0.0799 \\
Factor & CMA & Max DD & No & 0.0018 & 0.0027 & 0.6863 & 0.4925 & 739 & 0.0014 \\
Factor & HML & Max DD & No & -0.0055 & 0.0032 & -1.7406 & 0.0817 & 739 & 0.0038 \\
Factor & MKT-RF & Max DD & No & -0.0201 & 0.0072 & -2.8116 & 0.0049 & 739 & 0.0152 \\
Factor & RMW & Max DD & No & 0.0005 & 0.0011 & 0.4465 & 0.6553 & 739 & 0.0002 \\
Factor & SMB & Max DD & No & -0.0039 & 0.0018 & -2.1621 & 0.0306 & 739 & 0.0025 \\
Factor & UMD & Max DD & No & 0.0013 & 0.0017 & 0.7661 & 0.4436 & 1,177 & 0.0006 \\

Pooled & All & Failure & No & 0.0003 & 0.0058 & 0.0458 & 0.9635 & 4,872 & <0.0001 \\
Factor & CMA & Failure & No & 0.0045 & 0.0189 & 0.2370 & 0.8127 & 739 & 0.0001 \\
Factor & HML & Failure & No & -0.0107 & 0.0176 & -0.6077 & 0.5434 & 739 & 0.0005 \\
Factor & MKT-RF & Failure & No & -0.0580 & 0.0297 & -1.9500 & 0.0512 & 739 & 0.0103 \\
Factor & RMW & Failure & No & -0.0076 & 0.0062 & -1.2169 & 0.2236 & 739 & 0.0013 \\
Factor & SMB & Failure & No & -0.0147 & 0.0086 & -1.7223 & 0.0850 & 739 & 0.0010 \\
Factor & UMD & Failure & No & 0.0080 & 0.0047 & 1.6853 & 0.0919 & 1,177 & 0.0024 \\

\end{longtable}
\end{center}
\endgroup

%% file: tabs/controlled_predictive_v1_full.tex
\begingroup
\footnotesize
\setlength{\tabcolsep}{3.5pt}

\begin{center}
\begin{longtable}{p{1.7cm} p{1.3cm} p{2.2cm} c r r r r r r}
\caption{Controlled predictive regressions (full).}
\label{tab:controlled_predictive_full_panel} \\
\toprule
Sample & Factor & Outcome & Ctrl. & Coef. & SE & $t$ & $p$ & Obs. & $R^2$ \\
\midrule
\endfirsthead

\multicolumn{10}{l}{\textit{Table \thetable\ (continued)}} \\
\toprule
Sample & Factor & Outcome & Ctrl. & Coef. & SE & $t$ & $p$ & Obs. & $R^2$ \\
\midrule
\endhead

\midrule
\multicolumn{10}{r}{\textit{Continued on next page}} \\
\endfoot

\bottomrule
\endlastfoot

\multicolumn{10}{l}{\textit{Panel A. Horizon $h=3$}} \\
\midrule
Pooled & All & Sharpe & Yes & 0.0769 & 0.0887 & 0.8668 & 0.3861 & 2,580 & 0.0187 \\
Factor & CMA & Sharpe & Yes & 0.2754 & 0.3370 & 0.8173 & 0.4138 & 430 & 0.0028 \\
Factor & HML & Sharpe & Yes & 0.5259 & 0.4061 & 1.2951 & 0.1953 & 430 & 0.0445 \\
Factor & MKT-RF & Sharpe & Yes & 2.2505 & 1.4062 & 1.6004 & 0.1095 & 430 & 0.0213 \\
Factor & RMW & Sharpe & Yes & 0.1278 & 0.1014 & 1.2599 & 0.2077 & 430 & 0.0025 \\
Factor & SMB & Sharpe & Yes & -0.3665 & 0.3860 & -0.9494 & 0.3424 & 430 & 0.0152 \\
Factor & UMD & Sharpe & Yes & -0.2604 & 0.2896 & -0.8992 & 0.3686 & 430 & 0.0081 \\

Pooled & All & CumRet & Yes & 0.0024 & 0.0019 & 1.2321 & 0.2179 & 2,580 & 0.0236 \\
Factor & CMA & CumRet & Yes & 0.0053 & 0.0037 & 1.4460 & 0.1482 & 430 & 0.0363 \\
Factor & HML & CumRet & Yes & 0.0084 & 0.0052 & 1.6115 & 0.1071 & 430 & 0.1087 \\
Factor & MKT-RF & CumRet & Yes & 0.0112 & 0.0137 & 0.8198 & 0.4123 & 430 & 0.0249 \\
Factor & RMW & CumRet & Yes & 0.0039 & 0.0023 & 1.7130 & 0.0867 & 430 & 0.1015 \\
Factor & SMB & CumRet & Yes & -0.0131 & 0.0068 & -1.9327 & 0.0533 & 430 & 0.0491 \\
Factor & UMD & CumRet & Yes & -0.0029 & 0.0040 & -0.7180 & 0.4727 & 430 & 0.0689 \\

Pooled & All & Volatility & Yes & 0.0021 & 0.0014 & 1.4468 & 0.1479 & 2,580 & 0.2887 \\
Factor & CMA & Volatility & Yes & 0.0047 & 0.0032 & 1.4600 & 0.1443 & 430 & 0.3053 \\
Factor & HML & Volatility & Yes & 0.0017 & 0.0044 & 0.3965 & 0.6917 & 430 & 0.2784 \\
Factor & MKT-RF & Volatility & Yes & -0.0263 & 0.0108 & -2.4336 & 0.0150 & 430 & 0.2041 \\
Factor & RMW & Volatility & Yes & 0.0010 & 0.0010 & 1.0013 & 0.3167 & 430 & 0.3679 \\
Factor & SMB & Volatility & Yes & 0.0109 & 0.0034 & 3.1904 & 0.0014 & 430 & 0.0771 \\
Factor & UMD & Volatility & Yes & 0.0006 & 0.0028 & 0.2172 & 0.8280 & 430 & 0.2809 \\

Pooled & All & Downside vol. & Yes & -0.0008 & 0.0011 & -0.7921 & 0.4283 & 2,580 & 0.0413 \\
Factor & CMA & Downside vol. & Yes & -0.0041 & 0.0018 & -2.2740 & 0.0230 & 430 & 0.0584 \\
Factor & HML & Downside vol. & Yes & -0.0055 & 0.0022 & -2.4654 & 0.0137 & 430 & 0.0353 \\
Factor & MKT-RF & Downside vol. & Yes & -0.0072 & 0.0050 & -1.4256 & 0.1540 & 430 & 0.0100 \\
Factor & RMW & Downside vol. & Yes & -0.0012 & 0.0006 & -1.9692 & 0.0489 & 430 & 0.0220 \\
Factor & SMB & Downside vol. & Yes & 0.0079 & 0.0041 & 1.9367 & 0.0528 & 430 & 0.0476 \\
Factor & UMD & Downside vol. & Yes & 0.0007 & 0.0021 & 0.3439 & 0.7309 & 430 & 0.1529 \\

Pooled & All & Max DD & Yes & 0.0001 & 0.0007 & 0.1563 & 0.8758 & 2,580 & 0.0786 \\
Factor & CMA & Max DD & Yes & 0.0000 & 0.0016 & 0.0019 & 0.9985 & 430 & 0.0663 \\
Factor & HML & Max DD & Yes & -0.0005 & 0.0024 & -0.1896 & 0.8496 & 430 & 0.0754 \\
Factor & MKT-RF & Max DD & Yes & -0.0059 & 0.0054 & -1.0899 & 0.2758 & 430 & 0.0216 \\
Factor & RMW & Max DD & Yes & 0.0001 & 0.0008 & 0.0913 & 0.9273 & 430 & 0.0562 \\
Factor & SMB & Max DD & Yes & 0.0031 & 0.0019 & 1.6935 & 0.0904 & 430 & 0.0129 \\
Factor & UMD & Max DD & Yes & 0.0000 & 0.0018 & 0.0134 & 0.9893 & 430 & 0.1737 \\

Pooled & All & Failure & Yes & -0.0094 & 0.0085 & -1.1004 & 0.2711 & 2,580 & 0.0355 \\
Factor & CMA & Failure & Yes & -0.0310 & 0.0231 & -1.3408 & 0.1800 & 430 & 0.0291 \\
Factor & HML & Failure & Yes & -0.0516 & 0.0231 & -2.2315 & 0.0256 & 430 & 0.0800 \\
Factor & MKT-RF & Failure & Yes & -0.0373 & 0.0452 & -0.8243 & 0.4098 & 430 & 0.0237 \\
Factor & RMW & Failure & Yes & -0.0130 & 0.0061 & -2.1073 & 0.0351 & 430 & 0.0233 \\
Factor & SMB & Failure & Yes & 0.0285 & 0.0313 & 0.9101 & 0.3628 & 430 & 0.0078 \\
Factor & UMD & Failure & Yes & 0.0058 & 0.0168 & 0.3465 & 0.7290 & 430 & 0.1548 \\

\midrule
\multicolumn{10}{l}{\textit{Panel B. Horizon $h=6$}} \\
\midrule
Pooled & All & Sharpe & Yes & 0.0521 & 0.0438 & 1.1905 & 0.2338 & 2,562 & 0.0489 \\
Factor & CMA & Sharpe & Yes & 0.0959 & 0.1676 & 0.5721 & 0.5672 & 427 & 0.0172 \\
Factor & HML & Sharpe & Yes & 0.1807 & 0.2180 & 0.8289 & 0.4072 & 427 & 0.0387 \\
Factor & MKT-RF & Sharpe & Yes & 0.4228 & 0.3150 & 1.3423 & 0.1795 & 427 & 0.0047 \\
Factor & RMW & Sharpe & Yes & 0.0423 & 0.0375 & 1.1295 & 0.2587 & 427 & 0.0106 \\
Factor & SMB & Sharpe & Yes & -0.1302 & 0.1401 & -0.9289 & 0.3529 & 427 & 0.0407 \\
Factor & UMD & Sharpe & Yes & 0.0061 & 0.0818 & 0.0746 & 0.9405 & 427 & 0.0552 \\

Pooled & All & CumRet & Yes & 0.0034 & 0.0031 & 1.0929 & 0.2744 & 2,562 & 0.0403 \\
Factor & CMA & CumRet & Yes & 0.0095 & 0.0048 & 1.9739 & 0.0484 & 427 & 0.0632 \\
Factor & HML & CumRet & Yes & 0.0221 & 0.0103 & 2.1456 & 0.0319 & 427 & 0.1257 \\
Factor & MKT-RF & CumRet & Yes & 0.0133 & 0.0160 & 0.8300 & 0.4065 & 427 & 0.0391 \\
Factor & RMW & CumRet & Yes & 0.0018 & 0.0018 & 1.0107 & 0.3122 & 427 & 0.1573 \\
Factor & SMB & CumRet & Yes & -0.0135 & 0.0062 & -2.1832 & 0.0290 & 427 & 0.0774 \\
Factor & UMD & CumRet & Yes & -0.0028 & 0.0066 & -0.4242 & 0.6714 & 427 & 0.1131 \\

Pooled & All & Volatility & Yes & 0.0012 & 0.0018 & 0.6587 & 0.5101 & 2,562 & 0.3916 \\
Factor & CMA & Volatility & Yes & 0.0042 & 0.0024 & 1.7466 & 0.0807 & 427 & 0.4414 \\
Factor & HML & Volatility & Yes & -0.0029 & 0.0043 & -0.6648 & 0.5062 & 427 & 0.3561 \\
Factor & MKT-RF & Volatility & Yes & -0.0162 & 0.0073 & -2.2343 & 0.0255 & 427 & 0.2195 \\
Factor & RMW & Volatility & Yes & 0.0017 & 0.0011 & 1.5043 & 0.1325 & 427 & 0.3922 \\
Factor & SMB & Volatility & Yes & 0.0105 & 0.0034 & 3.0954 & 0.0020 & 427 & 0.1001 \\
Factor & UMD & Volatility & Yes & -0.0035 & 0.0024 & -1.4698 & 0.1416 & 427 & 0.4211 \\

Pooled & All & Downside vol. & Yes & -0.0005 & 0.0015 & -0.3037 & 0.7614 & 2,562 & 0.1336 \\
Factor & CMA & Downside vol. & Yes & -0.0012 & 0.0018 & -0.6513 & 0.5148 & 427 & 0.1745 \\
Factor & HML & Downside vol. & Yes & -0.0010 & 0.0031 & -0.3161 & 0.7519 & 427 & 0.0760 \\
Factor & MKT-RF & Downside vol. & Yes & -0.0134 & 0.0066 & -2.0349 & 0.0419 & 427 & 0.0181 \\
Factor & RMW & Downside vol. & Yes & 0.0013 & 0.0008 & 1.6692 & 0.0951 & 427 & 0.0876 \\
Factor & SMB & Downside vol. & Yes & 0.0093 & 0.0028 & 3.3143 & 0.0009 & 427 & 0.0582 \\
Factor & UMD & Downside vol. & Yes & -0.0044 & 0.0030 & -1.4780 & 0.1394 & 427 & 0.3101 \\

Pooled & All & Max DD & Yes & -0.0006 & 0.0011 & -0.5528 & 0.5804 & 2,562 & 0.1322 \\
Factor & CMA & Max DD & Yes & -0.0009 & 0.0022 & -0.4263 & 0.6699 & 427 & 0.1094 \\
Factor & HML & Max DD & Yes & -0.0062 & 0.0038 & -1.6234 & 0.1045 & 427 & 0.0686 \\
Factor & MKT-RF & Max DD & Yes & -0.0085 & 0.0077 & -1.1092 & 0.2673 & 427 & 0.0285 \\
Factor & RMW & Max DD & Yes & 0.0005 & 0.0007 & 0.7020 & 0.4827 & 427 & 0.0897 \\
Factor & SMB & Max DD & Yes & 0.0008 & 0.0023 & 0.3443 & 0.7306 & 427 & 0.0072 \\
Factor & UMD & Max DD & Yes & -0.0008 & 0.0021 & -0.3877 & 0.6982 & 427 & 0.3474 \\

Pooled & All & Failure & Yes & -0.0087 & 0.0081 & -1.0800 & 0.2801 & 2,562 & 0.0334 \\
Factor & CMA & Failure & Yes & -0.0304 & 0.0255 & -1.1924 & 0.2331 & 427 & 0.0173 \\
Factor & HML & Failure & Yes & -0.0194 & 0.0245 & -0.7942 & 0.4271 & 427 & 0.0559 \\
Factor & MKT-RF & Failure & Yes & -0.0365 & 0.0302 & -1.2102 & 0.2262 & 427 & 0.0072 \\
Factor & RMW & Failure & Yes & -0.0081 & 0.0051 & -1.5920 & 0.1114 & 427 & 0.0331 \\
Factor & SMB & Failure & Yes & 0.0223 & 0.0298 & 0.7480 & 0.4545 & 427 & 0.0085 \\
Factor & UMD & Failure & Yes & -0.0045 & 0.0194 & -0.2321 & 0.8165 & 427 & 0.1529 \\

\midrule
\multicolumn{10}{l}{\textit{Panel C. Horizon $h=12$}} \\
\midrule
Pooled & All & Sharpe & Yes & 0.0801 & 0.0288 & 2.7774 & 0.0055 & 2,526 & 0.0735 \\
Factor & CMA & Sharpe & Yes & 0.0611 & 0.1015 & 0.6024 & 0.5469 & 421 & 0.0616 \\
Factor & HML & Sharpe & Yes & 0.3762 & 0.1476 & 2.5494 & 0.0108 & 421 & 0.0435 \\
Factor & MKT-RF & Sharpe & Yes & 0.4028 & 0.2243 & 1.7961 & 0.0725 & 421 & 0.0161 \\
Factor & RMW & Sharpe & Yes & 0.0616 & 0.0269 & 2.2888 & 0.0221 & 421 & 0.0178 \\
Factor & SMB & Sharpe & Yes & -0.0747 & 0.0726 & -1.0279 & 0.3040 & 421 & 0.0855 \\
Factor & UMD & Sharpe & Yes & 0.0303 & 0.0455 & 0.6664 & 0.5051 & 421 & 0.1321 \\

Pooled & All & CumRet & Yes & 0.0145 & 0.0048 & 3.0026 & 0.0027 & 2,526 & 0.0789 \\
Factor & CMA & CumRet & Yes & 0.0073 & 0.0076 & 0.9610 & 0.3366 & 421 & 0.1123 \\
Factor & HML & CumRet & Yes & 0.0550 & 0.0224 & 2.4608 & 0.0139 & 421 & 0.1475 \\
Factor & MKT-RF & CumRet & Yes & 0.0583 & 0.0304 & 1.9191 & 0.0550 & 421 & 0.0528 \\
Factor & RMW & CumRet & Yes & 0.0136 & 0.0038 & 3.5825 & 0.0003 & 421 & 0.2473 \\
Factor & SMB & CumRet & Yes & -0.0127 & 0.0065 & -1.9466 & 0.0516 & 421 & 0.1609 \\
Factor & UMD & CumRet & Yes & -0.0002 & 0.0071 & -0.0268 & 0.9786 & 421 & 0.1314 \\

Pooled & All & Volatility & Yes & 0.0009 & 0.0017 & 0.5124 & 0.6084 & 2,526 & 0.4163 \\
Factor & CMA & Volatility & Yes & 0.0026 & 0.0018 & 1.4253 & 0.1541 & 421 & 0.4377 \\
Factor & HML & Volatility & Yes & -0.0005 & 0.0042 & -0.1284 & 0.8978 & 421 & 0.3158 \\
Factor & MKT-RF & Volatility & Yes & -0.0222 & 0.0069 & -3.2269 & 0.0013 & 421 & 0.2424 \\
Factor & RMW & Volatility & Yes & 0.0018 & 0.0011 & 1.5658 & 0.1174 & 421 & 0.3457 \\
Factor & SMB & Volatility & Yes & 0.0035 & 0.0032 & 1.1021 & 0.2704 & 421 & 0.1021 \\
Factor & UMD & Volatility & Yes & -0.0023 & 0.0027 & -0.8506 & 0.3950 & 421 & 0.3652 \\

Pooled & All & Downside vol. & Yes & -0.0017 & 0.0015 & -1.1555 & 0.2479 & 2,526 & 0.2734 \\
Factor & CMA & Downside vol. & Yes & 0.0002 & 0.0013 & 0.1581 & 0.8744 & 421 & 0.2902 \\
Factor & HML & Downside vol. & Yes & -0.0033 & 0.0026 & -1.2569 & 0.2088 & 421 & 0.2221 \\
Factor & MKT-RF & Downside vol. & Yes & -0.0209 & 0.0071 & -2.9253 & 0.0034 & 421 & 0.0978 \\
Factor & RMW & Downside vol. & Yes & -0.0002 & 0.0008 & -0.1968 & 0.8440 & 421 & 0.0952 \\
Factor & SMB & Downside vol. & Yes & 0.0061 & 0.0019 & 3.1976 & 0.0014 & 421 & 0.0814 \\
Factor & UMD & Downside vol. & Yes & -0.0040 & 0.0033 & -1.2213 & 0.2220 & 421 & 0.3445 \\

Pooled & All & Max DD & Yes & -0.0040 & 0.0016 & -2.5172 & 0.0118 & 2,526 & 0.1594 \\
Factor & CMA & Max DD & Yes & -0.0011 & 0.0026 & -0.4240 & 0.6716 & 421 & 0.1221 \\
Factor & HML & Max DD & Yes & -0.0147 & 0.0061 & -2.4209 & 0.0155 & 421 & 0.0524 \\
Factor & MKT-RF & Max DD & Yes & -0.0349 & 0.0167 & -2.0957 & 0.0361 & 421 & 0.0689 \\
Factor & RMW & Max DD & Yes & -0.0012 & 0.0012 & -1.0689 & 0.2851 & 421 & 0.1072 \\
Factor & SMB & Max DD & Yes & -0.0030 & 0.0033 & -0.9139 & 0.3608 & 421 & 0.0051 \\
Factor & UMD & Max DD & Yes & -0.0046 & 0.0030 & -1.5652 & 0.1175 & 421 & 0.3751 \\

Pooled & All & Failure & Yes & -0.0090 & 0.0082 & -1.1004 & 0.2711 & 2,526 & 0.0227 \\
Factor & CMA & Failure & Yes & -0.0141 & 0.0293 & -0.4801 & 0.6312 & 421 & 0.0477 \\
Factor & HML & Failure & Yes & -0.0299 & 0.0360 & -0.8296 & 0.4068 & 421 & 0.0186 \\
Factor & MKT-RF & Failure & Yes & -0.1010 & 0.0503 & -2.0081 & 0.0446 & 421 & 0.0255 \\
Factor & RMW & Failure & Yes & -0.0056 & 0.0042 & -1.3301 & 0.1835 & 421 & 0.0398 \\
Factor & SMB & Failure & Yes & -0.0249 & 0.0164 & -1.5175 & 0.1291 & 421 & 0.0136 \\
Factor & UMD & Failure & Yes & 0.0056 & 0.0159 & 0.3492 & 0.7269 & 421 & 0.2431 \\

\end{longtable}
\end{center}
\endgroup

%% file: tabs/passive_stockonly_robustness.tex
\begin{table}[htbp]
\centering
\caption{Appendix passive robustness checks for cumulative-return mapping}
\label{tab:passive-robustness}
\begin{tabular}{llrrrr}
\toprule
Specification & Proxy & coef. & SE & $p$ & $N$ \\
\midrule
\multicolumn{6}{l}{\textbf{Panel A: FF6 System}} \\
\midrule
Validated lagged baseline & $PassiveShareTotal_{t-1}$ & -0.8535 & 0.1653 & 0.0000 & 648 \\
Month-FE interaction only & $PassiveShareTotal_{t-1}$ & -0.8627 & 0.1932 & 0.0000 & 648 \\
Lagged detrended share & $PassiveShareDetrended_{t-1}$ & -3.9159 & 2.7188 & 0.1498 & 438 \\
Exclude 2020-03 to 2020-06 & $PassiveShareTotal_{t-1}$ & -0.7966 & 0.1633 & 0.0000 & 624 \\
\midrule
\multicolumn{6}{l}{\textbf{Panel B: q5 System}} \\
\midrule
Validated lagged baseline & $PassiveShareTotal_{t-1}$ & -0.5262 & 0.2570 & 0.0406 & 475 \\
Month-FE interaction only & $PassiveShareTotal_{t-1}$ & -0.5518 & 0.3125 & 0.0774 & 475 \\
Lagged detrended share & $PassiveShareDetrended_{t-1}$ & -0.3751 & 1.6774 & 0.8231 & 300 \\
Exclude 2020-03 to 2020-06 & $PassiveShareTotal_{t-1}$ & -0.5858 & 0.2289 & 0.0105 & 455 \\
\bottomrule
\end{tabular}
\end{table}

%% file: tabs/passive_stockonly_support.tex
\begin{table}[htbp]
\centering
\caption{Sharpe and downside semivolatility mapping}
\label{tab:passive-support}
\begin{tabular}{llrrrr}
\toprule
System & Outcome & Interaction coef. & SE & $p$ & $N$ \\
\midrule
FF6 & Future 12m Sharpe & -9.9401 & 2.9235 & 0.0007 & 648 \\
FF6 & Future 12m downside semivolatility & 0.0805 & 0.0633 & 0.2040 & 648 \\
q5 & Future 12m Sharpe & -6.8701 & 4.4476 & 0.1224 & 475 \\
q5 & Future 12m downside semivolatility & 0.0787 & 0.0519 & 0.1290 & 475 \\
\bottomrule
\end{tabular}
\end{table}

%% file: tabs/ff6_prop4_revised_rankcheck.tex
\begin{table}[htbp]
\centering
\footnotesize
\setlength{\tabcolsep}{4pt}
\caption{FF6 factors: rank comparison across break diagnostics and predictive slopes}
\label{tab:ff6_prop4_rankcheck}
\begin{tabular}{l r r r r r}
\toprule
Factor 
& Rank (Pr(Break)) 
& Rank (Break share) 
& Rank ($\Delta$ break) 
& Rank (Spike freq.) 
& Rank (Sharpe coef.) \\
\midrule
MKT & 1 & 1 & 6 & 4 & 1 \\
SMB & 3 & 3 & 5 & 6 & 6 \\
HML & 2 & 2 & 4 & 2 & 2 \\
RMW & 6 & 6 & 1 & 5 & 3 \\
CMA & 4 & 4 & 3 & 3 & 4 \\
UMD & 5 & 5 & 2 & 1 & 5 \\
\bottomrule
\end{tabular}
\end{table}

%% file: tabs/passive_stockonly_onset.tex
\begin{table}[htbp]
\centering
\caption{Appendix lagged break-onset interaction benchmark}
\label{tab:common-sample-passive}
\begin{tabular}{lrrrr}
\toprule
System & Interaction coef. & SE & p-value & N \\
\midrule
FF6 & 0.9507 & 0.7205 & 0.1870 & 720 \\
q5 & 0.8068 & 1.1692 & 0.4902 & 535 \\
\bottomrule
\end{tabular}
\end{table}

%% file: tabs/stable_model_fit_v1.tex
\begin{table}[htbp]
\centering
\footnotesize
\setlength{\tabcolsep}{4pt}
\caption{Stable-model fit diagnostics.}
\label{tab:app-stable-fit}
\begin{tabular}{l c c c r r r r r r}
\toprule
Factor & Start & End & Obs. & LogLik & AIC & BIC & $\rho$ & $\sigma_u$ & $\sigma_\eta$ \\
\midrule
MKT & 1963--07 & 2026--01 & 751  & 1255.79 & -2505.58 & -2491.72 & 0.0587 & 0.0013 & 0.0449 \\
SMB & 1963--07 & 2026--01 & 751  & 1551.30 & -3096.61 & -3082.74 & 0.3680 & 0.0277 & 0.0114 \\
HML & 1963--07 & 2026--01 & 751  & 1572.75 & -3139.51 & -3125.64 & 0.4923 & 0.0244 & 0.0150 \\
RMW & 1963--07 & 2026--01 & 751  & 1787.94 & -3569.88 & -3556.02 & 0.1602 & <0.0001 & 0.0221 \\
CMA & 1963--07 & 2026--01 & 751  & 1842.33 & -3678.65 & -3664.79 & 0.4695 & 0.0173 & 0.0101 \\
UMD & 1927--01 & 2026--01 & 1189 & 1937.79 & -3869.58 & -3854.34 & 0.0849 & 0.0001 & 0.0470 \\
\bottomrule
\end{tabular}
\end{table}

%% file: tabs/break_model_fit_v1.tex
\begin{table}[htbp]
\centering
\footnotesize
\setlength{\tabcolsep}{3pt}
\caption{Break-model fit diagnostics.}
\label{tab:app-break-fit}
\begin{tabular}{l c c c r r r r r r r r r}
\toprule
Factor & Start & End & Obs. & LogLik & AIC & BIC 
& Mean$_0$ & Mean$_1$ & SD$_0$ & SD$_1$ & $p_{00}$ & $p_{11}$ \\
\midrule
MKT & 1963--07 & 2026--01 & 751  & 1314.32 & -2616.64 & -2588.91 
& 0.0109 & 0.0011 & 0.0281 & 0.0558 & 0.9489 & 0.9475 \\

SMB & 1963--07 & 2026--01 & 751  & 1596.43 & -3180.86 & -3153.13 
& -0.0009 & 0.0067 & 0.0219 & 0.0409 & 0.9536 & 0.9136 \\

HML & 1963--07 & 2026--01 & 751  & 1649.15 & -3286.30 & -3258.57 
& 0.0002 & 0.0074 & 0.0187 & 0.0417 & 0.9673 & 0.9459 \\

RMW & 1963--07 & 2026--01 & 751  & 1943.01 & -3874.03 & -3846.30 
& 0.0020 & 0.0084 & 0.0158 & 0.0512 & 0.9964 & 0.9656 \\

CMA & 1963--07 & 2026--01 & 751  & 1915.91 & -3819.82 & -3792.09 
& 0.0009 & 0.0072 & 0.0150 & 0.0317 & 0.9851 & 0.9534 \\

UMD & 1927--01 & 2026--01 & 1189 & 2277.97 & -4543.94 & -4513.46 
& 0.0081 & -0.0037 & 0.0265 & 0.0972 & 0.9758 & 0.8792 \\
\bottomrule
\end{tabular}
\end{table}

%% file: tabs/model_comparison_v1.tex
\begin{table}[htbp]
\centering
\footnotesize
\setlength{\tabcolsep}{4pt}
\caption{Model-comparison diagnostics between the stable and break-aware specifications.}
\label{tab:app-model-comparison}
\begin{tabular}{l c ccc ccc ccc r r}
\toprule
& & \multicolumn{3}{c}{Stable} & \multicolumn{3}{c}{Markov} & \multicolumn{3}{c}{$\Delta$ (Markov $-$ Stable)} & \\
\cmidrule(lr){3-5} \cmidrule(lr){6-8} \cmidrule(lr){9-11}
Factor & Obs. 
& LogLik & AIC & BIC 
& LogLik & AIC & BIC 
& $\Delta$LL & $\Delta$AIC & $\Delta$BIC 
& Params \\
\midrule
UMD & 1189 
& 1937.79 & -3869.58 & -3854.34 
& 2277.97 & -4543.94 & -4513.46 
& 340.18 & 674.36 & 659.12 
& 3 / 6 \\

RMW & 751 
& 1787.94 & -3569.88 & -3556.02 
& 1943.01 & -3874.03 & -3846.30 
& 155.07 & 304.14 & 290.28 
& 3 / 6 \\

HML & 751 
& 1572.75 & -3139.51 & -3125.64 
& 1649.15 & -3286.30 & -3258.57 
& 76.40 & 146.79 & 132.93 
& 3 / 6 \\

CMA & 751 
& 1842.33 & -3678.65 & -3664.79 
& 1915.91 & -3819.82 & -3792.09 
& 73.58 & 141.17 & 127.30 
& 3 / 6 \\

MKT & 751 
& 1255.79 & -2505.58 & -2491.72 
& 1314.32 & -2616.64 & -2588.91 
& 58.53 & 111.06 & 97.19 
& 3 / 6 \\

SMB & 751 
& 1551.30 & -3096.61 & -3082.74 
& 1596.43 & -3180.86 & -3153.13 
& 45.13 & 84.25 & 70.39 
& 3 / 6 \\

\bottomrule
\end{tabular}
\end{table}

%% file: tabs/baseline_predictive_q5_robust.tex
\begingroup
\footnotesize
\setlength{\tabcolsep}{3.5pt}

\begin{center}
\begin{longtable}{p{1.5cm} p{1.1cm} p{2.0cm} c r r r r r r}
\caption{Baseline predictive results.}
\label{tab:baseline_q5_robust} \\
\toprule
Sample & Factor & Outcome & Ctrl. & Coef. & SE & $t$ & $p$ & Obs. & $R^2$ \\
\midrule
\endfirsthead

\multicolumn{10}{l}{\textit{Table \thetable\ (continued)}} \\
\toprule
Sample & Factor & Outcome & Ctrl. & Coef. & SE & $t$ & $p$ & Obs. & $R^2$ \\
\midrule
\endhead

\midrule
\multicolumn{10}{r}{\textit{Continued on next page}} \\
\endfoot

\bottomrule
\endlastfoot

\multicolumn{10}{l}{\textit{Panel A. Horizon $h=3$}} \\
\midrule
Pooled & All & Sharpe & No & -0.1485 & 0.1617 & -0.9185 & 0.3583 & 3,465 & 0.0108 \\
Factor & EG & Sharpe & No & -0.7998 & 0.6339 & -1.2616 & 0.2071 & 693 & 0.0007 \\
Factor & IA & Sharpe & No & -0.0633 & 0.3974 & -0.1592 & 0.8735 & 693 & 0.0000 \\
Factor & ME & Sharpe & No & -0.3646 & 0.1680 & -2.1696 & 0.0300 & 693 & 0.0015 \\
Factor & MKT & Sharpe & No & 0.9802 & 0.4553 & 2.1530 & 0.0313 & 693 & 0.0045 \\
Factor & ROE & Sharpe & No & 0.0197 & 0.2491 & 0.0790 & 0.9370 & 693 & 0.0000 \\

Pooled & All & CumRet & No & -0.0017 & 0.0026 & -0.6607 & 0.5088 & 3,465 & 0.0114 \\
Factor & EG & CumRet & No & -0.0046 & 0.0027 & -1.7004 & 0.0891 & 693 & 0.0053 \\
Factor & IA & CumRet & No & 0.0030 & 0.0040 & 0.7375 & 0.4608 & 693 & 0.0028 \\
Factor & ME & CumRet & No & -0.0081 & 0.0026 & -3.0613 & 0.0022 & 693 & 0.0168 \\
Factor & MKT & CumRet & No & 0.0071 & 0.0050 & 1.4397 & 0.1500 & 693 & 0.0018 \\
Factor & ROE & CumRet & No & 0.0009 & 0.0028 & 0.3220 & 0.7475 & 693 & 0.0002 \\

Pooled & All & Volatility & No & 0.0026 & 0.0021 & 1.2022 & 0.2293 & 3,465 & 0.1925 \\
Factor & EG & Volatility & No & 0.0051 & 0.0036 & 1.4182 & 0.1561 & 693 & 0.0087 \\
Factor & IA & Volatility & No & 0.0078 & 0.0043 & 1.8340 & 0.0667 & 693 & 0.0292 \\
Factor & ME & Volatility & No & 0.0035 & 0.0030 & 1.1572 & 0.2472 & 693 & 0.0038 \\
Factor & MKT & Volatility & No & -0.0099 & 0.0066 & -1.4977 & 0.1342 & 693 & 0.0047 \\
Factor & ROE & Volatility & No & 0.0005 & 0.0027 & 0.1853 & 0.8530 & 693 & 0.0001 \\

Pooled & All & Downside vol. & No & 0.0006 & 0.0015 & 0.4230 & 0.6723 & 3,465 & 0.0436 \\
Factor & EG & Downside vol. & No & 0.0012 & 0.0016 & 0.7255 & 0.4681 & 693 & 0.0023 \\
Factor & IA & Downside vol. & No & 0.0003 & 0.0012 & 0.2764 & 0.7823 & 693 & 0.0002 \\
Factor & ME & Downside vol. & No & 0.0038 & 0.0022 & 1.7671 & 0.0772 & 693 & 0.0160 \\
Factor & MKT & Downside vol. & No & -0.0048 & 0.0023 & -2.1036 & 0.0354 & 693 & 0.0026 \\
Factor & ROE & Downside vol. & No & -0.0013 & 0.0009 & -1.4766 & 0.1398 & 693 & 0.0016 \\

Pooled & All & Max DD & No & 0.0015 & 0.0009 & 1.5568 & 0.1195 & 3,465 & 0.0776 \\
Factor & EG & Max DD & No & 0.0037 & 0.0021 & 1.8022 & 0.0715 & 693 & 0.0226 \\
Factor & IA & Max DD & No & 0.0022 & 0.0011 & 1.8744 & 0.0609 & 693 & 0.0096 \\
Factor & ME & Max DD & No & 0.0020 & 0.0013 & 1.5137 & 0.1301 & 693 & 0.0056 \\
Factor & MKT & Max DD & No & -0.0050 & 0.0023 & -2.2056 & 0.0274 & 693 & 0.0035 \\
Factor & ROE & Max DD & No & 0.0013 & 0.0015 & 0.8497 & 0.3955 & 693 & 0.0018 \\

Pooled & All & Failure & No & 0.0078 & 0.0112 & 0.6977 & 0.4854 & 3,465 & 0.0003 \\
Factor & EG & Failure & No & 0.0096 & 0.0219 & 0.4387 & 0.6609 & 693 & 0.0004 \\
Factor & IA & Failure & No & 0.0055 & 0.0256 & 0.2135 & 0.8310 & 693 & 0.0002 \\
Factor & ME & Failure & No & 0.0259 & 0.0131 & 1.9799 & 0.0477 & 693 & 0.0056 \\
Factor & MKT & Failure & No & -0.0366 & 0.0218 & -1.6812 & 0.0927 & 693 & 0.0034 \\
Factor & ROE & Failure & No & 0.0029 & 0.0186 & 0.1542 & 0.8775 & 693 & 0.0001 \\

\midrule
\multicolumn{10}{l}{\textit{Panel B. Horizon $h=6$}} \\
\midrule
Pooled & All & Sharpe & No & -0.0196 & 0.0473 & -0.4142 & 0.6788 & 3,450 & 0.0536 \\
Factor & EG & Sharpe & No & -0.1596 & 0.1355 & -1.1783 & 0.2387 & 690 & 0.0014 \\
Factor & IA & Sharpe & No & -0.1842 & 0.1576 & -1.1688 & 0.2425 & 690 & 0.0022 \\
Factor & ME & Sharpe & No & 0.0180 & 0.0789 & 0.2276 & 0.8200 & 690 & 0.0000 \\
Factor & MKT & Sharpe & No & 0.3194 & 0.1542 & 2.0708 & 0.0384 & 690 & 0.0044 \\
Factor & ROE & Sharpe & No & 0.0211 & 0.0859 & 0.2457 & 0.8059 & 690 & 0.0000 \\

Pooled & All & CumRet & No & 0.0004 & 0.0023 & 0.1972 & 0.8437 & 3,450 & 0.0220 \\
Factor & EG & CumRet & No & -0.0057 & 0.0034 & -1.6549 & 0.0979 & 690 & 0.0037 \\
Factor & IA & CumRet & No & 0.0030 & 0.0042 & 0.7214 & 0.4707 & 690 & 0.0013 \\
Factor & ME & CumRet & No & -0.0028 & 0.0037 & -0.7527 & 0.4516 & 690 & 0.0010 \\
Factor & MKT & CumRet & No & 0.0136 & 0.0077 & 1.7638 & 0.0778 & 690 & 0.0031 \\
Factor & ROE & CumRet & No & 0.0013 & 0.0028 & 0.4653 & 0.6417 & 690 & 0.0003 \\

Pooled & All & Volatility & No & 0.0033 & 0.0020 & 1.6439 & 0.1002 & 3,450 & 0.2846 \\
Factor & EG & Volatility & No & 0.0021 & 0.0034 & 0.6154 & 0.5383 & 690 & 0.0016 \\
Factor & IA & Volatility & No & 0.0097 & 0.0035 & 2.7660 & 0.0057 & 690 & 0.0497 \\
Factor & ME & Volatility & No & 0.0059 & 0.0026 & 2.3018 & 0.0213 & 690 & 0.0126 \\
Factor & MKT & Volatility & No & -0.0104 & 0.0056 & -1.8438 & 0.0652 & 690 & 0.0068 \\
Factor & ROE & Volatility & No & 0.0011 & 0.0023 & 0.4847 & 0.6279 & 690 & 0.0004 \\

Pooled & All & Downside vol. & No & 0.0007 & 0.0016 & 0.4500 & 0.6527 & 3,450 & 0.1025 \\
Factor & EG & Downside vol. & No & 0.0027 & 0.0024 & 1.1453 & 0.2521 & 690 & 0.0056 \\
Factor & IA & Downside vol. & No & 0.0023 & 0.0016 & 1.4396 & 0.1500 & 690 & 0.0074 \\
Factor & ME & Downside vol. & No & 0.0029 & 0.0018 & 1.5872 & 0.1125 & 690 & 0.0068 \\
Factor & MKT & Downside vol. & No & -0.0090 & 0.0034 & -2.6216 & 0.0088 & 690 & 0.0064 \\
Factor & ROE & Downside vol. & No & -0.0008 & 0.0017 & -0.4619 & 0.6441 & 690 & 0.0003 \\

Pooled & All & Max DD & No & 0.0007 & 0.0010 & 0.6724 & 0.5013 & 3,450 & 0.1409 \\
Factor & EG & Max DD & No & 0.0021 & 0.0020 & 1.0768 & 0.2816 & 690 & 0.0029 \\
Factor & IA & Max DD & No & 0.0037 & 0.0017 & 2.2036 & 0.0276 & 690 & 0.0125 \\
Factor & ME & Max DD & No & 0.0005 & 0.0013 & 0.3620 & 0.7174 & 690 & 0.0001 \\
Factor & MKT & Max DD & No & -0.0086 & 0.0045 & -1.9295 & 0.0537 & 690 & 0.0043 \\
Factor & ROE & Max DD & No & 0.0011 & 0.0018 & 0.5785 & 0.5629 & 690 & 0.0005 \\

Pooled & All & Failure & No & 0.0079 & 0.0113 & 0.7020 & 0.4827 & 3,450 & 0.0004 \\
Factor & EG & Failure & No & 0.0032 & 0.0257 & 0.1238 & 0.9015 & 690 & 0.0000 \\
Factor & IA & Failure & No & -0.0130 & 0.0125 & -1.0388 & 0.2989 & 690 & 0.0009 \\
Factor & ME & Failure & No & 0.0263 & 0.0119 & 2.2131 & 0.0269 & 690 & 0.0059 \\
Factor & MKT & Failure & No & -0.0336 & 0.0187 & -1.7955 & 0.0726 & 690 & 0.0029 \\
Factor & ROE & Failure & No & 0.0193 & 0.0183 & 1.0568 & 0.2906 & 690 & 0.0027 \\

\midrule
\multicolumn{10}{l}{\textit{Panel C. Horizon $h=12$}} \\
\midrule
Pooled & All & Sharpe & No & 0.0372 & 0.0288 & 1.2936 & 0.1958 & 3,420 & 0.1214 \\
Factor & EG & Sharpe & No & -0.0370 & 0.1191 & -0.3109 & 0.7559 & 684 & 0.0002 \\
Factor & IA & Sharpe & No & -0.0372 & 0.0804 & -0.4624 & 0.6438 & 684 & 0.0004 \\
Factor & ME & Sharpe & No & 0.0630 & 0.0602 & 1.0472 & 0.2950 & 684 & 0.0014 \\
Factor & MKT & Sharpe & No & 0.2187 & 0.1263 & 1.7311 & 0.0834 & 684 & 0.0065 \\
Factor & ROE & Sharpe & No & 0.0418 & 0.0543 & 0.7700 & 0.4413 & 684 & 0.0007 \\

Pooled & All & CumRet & No & 0.0056 & 0.0029 & 1.9059 & 0.0567 & 3,420 & 0.0448 \\
Factor & EG & CumRet & No & -0.0072 & 0.0084 & -0.8546 & 0.3928 & 684 & 0.0025 \\
Factor & IA & CumRet & No & 0.0060 & 0.0078 & 0.7675 & 0.4428 & 684 & 0.0023 \\
Factor & ME & CumRet & No & 0.0042 & 0.0049 & 0.8513 & 0.3946 & 684 & 0.0011 \\
Factor & MKT & CumRet & No & 0.0261 & 0.0165 & 1.5795 & 0.1142 & 684 & 0.0056 \\
Factor & ROE & CumRet & No & 0.0072 & 0.0046 & 1.5785 & 0.1145 & 684 & 0.0040 \\

Pooled & All & Volatility & No & 0.0022 & 0.0017 & 1.2854 & 0.1987 & 3,420 & 0.3683 \\
Factor & EG & Volatility & No & 0.0024 & 0.0037 & 0.6675 & 0.5044 & 684 & 0.0025 \\
Factor & IA & Volatility & No & 0.0082 & 0.0029 & 2.8525 & 0.0043 & 684 & 0.0408 \\
Factor & ME & Volatility & No & 0.0032 & 0.0018 & 1.8385 & 0.0660 & 684 & 0.0052 \\
Factor & MKT & Volatility & No & -0.0133 & 0.0053 & -2.5020 & 0.0123 & 684 & 0.0146 \\
Factor & ROE & Volatility & No & 0.0016 & 0.0031 & 0.5197 & 0.6032 & 684 & 0.0010 \\

Pooled & All & Downside vol. & No & 0.0010 & 0.0014 & 0.7012 & 0.4832 & 3,420 & 0.1976 \\
Factor & EG & Downside vol. & No & 0.0023 & 0.0025 & 0.9068 & 0.3645 & 684 & 0.0033 \\
Factor & IA & Downside vol. & No & 0.0041 & 0.0017 & 2.3454 & 0.0190 & 684 & 0.0247 \\
Factor & ME & Downside vol. & No & 0.0029 & 0.0016 & 1.8638 & 0.0624 & 684 & 0.0078 \\
Factor & MKT & Downside vol. & No & -0.0115 & 0.0046 & -2.4822 & 0.0131 & 684 & 0.0110 \\
Factor & ROE & Downside vol. & No & 0.0001 & 0.0026 & 0.0222 & 0.9823 & 684 & 0.0000 \\

Pooled & All & Max DD & No & -0.0011 & 0.0013 & -0.8450 & 0.3981 & 3,420 & 0.2151 \\
Factor & EG & Max DD & No & 0.0046 & 0.0040 & 1.1450 & 0.2522 & 684 & 0.0071 \\
Factor & IA & Max DD & No & 0.0051 & 0.0032 & 1.5582 & 0.1192 & 684 & 0.0129 \\
Factor & ME & Max DD & No & -0.0019 & 0.0015 & -1.2863 & 0.1983 & 684 & 0.0012 \\
Factor & MKT & Max DD & No & -0.0222 & 0.0086 & -2.5805 & 0.0099 & 684 & 0.0149 \\
Factor & ROE & Max DD & No & -0.0007 & 0.0030 & -0.2198 & 0.8260 & 684 & 0.0001 \\

Pooled & All & Failure & No & -0.0081 & 0.0072 & -1.1222 & 0.2618 & 3,420 & 0.0004 \\
Factor & EG & Failure & No & -0.0002 & 0.0277 & -0.0084 & 0.9933 & 684 & 0.0000 \\
Factor & IA & Failure & No & 0.0089 & 0.0190 & 0.4677 & 0.6400 & 684 & 0.0004 \\
Factor & ME & Failure & No & -0.0087 & 0.0035 & -2.5045 & 0.0123 & 684 & 0.0007 \\
Factor & MKT & Failure & No & -0.0698 & 0.0335 & -2.0830 & 0.0373 & 684 & 0.0124 \\
Factor & ROE & Failure & No & -0.0031 & 0.0167 & -0.1861 & 0.8523 & 684 & 0.0001 \\

\end{longtable}
\end{center}
\endgroup

%% file: tabs/baseline_predictive_q5_robust_common_sample.tex
\begingroup
\footnotesize
\setlength{\tabcolsep}{3.5pt}

\begin{center}
\begin{longtable}{p{1.5cm} p{1.1cm} p{2.0cm} c r r r r r r}
\caption{Baseline predictive results (common sample, q5 robustness).}
\label{tab:baseline_q5_robust_common_sample} \\
\toprule
Sample & Factor & Outcome & Ctrl. & Coef. & SE & $t$ & $p$ & Obs. & $R^2$ \\
\midrule
\endfirsthead

\multicolumn{10}{l}{\textit{Table \thetable\ (continued)}} \\
\toprule
Sample & Factor & Outcome & Ctrl. & Coef. & SE & $t$ & $p$ & Obs. & $R^2$ \\
\midrule
\endhead

\midrule
\multicolumn{10}{r}{\textit{Continued on next page}} \\
\endfoot

\bottomrule
\endlastfoot

\multicolumn{10}{l}{\textit{Panel A. Horizon $h=3$}} \\
\midrule
Pooled & All & Sharpe & No & -0.0125 & 0.2427 & -0.0516 & 0.9589 & 2,085 & 0.0099 \\
Factor & EG & Sharpe & No & -0.9489 & 1.0793 & -0.8792 & 0.3793 & 417 & 0.0006 \\
Factor & IA & Sharpe & No & 0.7272 & 0.3929 & 1.8510 & 0.0642 & 417 & 0.0066 \\
Factor & ME & Sharpe & No & -0.3330 & 0.1025 & -3.2483 & 0.0012 & 417 & 0.0019 \\
Factor & MKT & Sharpe & No & 2.0549 & 1.1513 & 1.7847 & 0.0743 & 417 & 0.0119 \\
Factor & ROE & Sharpe & No & 0.0063 & 0.3068 & 0.0204 & 0.9837 & 417 & 0.0000 \\

Pooled & All & CumRet & No & -0.0023 & 0.0037 & -0.6273 & 0.5305 & 2,085 & 0.0168 \\
Factor & EG & CumRet & No & -0.0055 & 0.0045 & -1.2194 & 0.2227 & 417 & 0.0057 \\
Factor & IA & CumRet & No & 0.0092 & 0.0036 & 2.5614 & 0.0104 & 417 & 0.0263 \\
Factor & ME & CumRet & No & -0.0096 & 0.0017 & -5.6455 & 0.0000 & 417 & 0.0328 \\
Factor & MKT & CumRet & No & 0.0071 & 0.0134 & 0.5301 & 0.5960 & 417 & 0.0010 \\
Factor & ROE & CumRet & No & -0.0001 & 0.0043 & -0.0315 & 0.9749 & 417 & 0.0000 \\

Pooled & All & Volatility & No & 0.0046 & 0.0024 & 1.8986 & 0.0576 & 2,085 & 0.1477 \\
Factor & EG & Volatility & No & 0.0104 & 0.0046 & 2.2357 & 0.0254 & 417 & 0.0274 \\
Factor & IA & Volatility & No & 0.0117 & 0.0049 & 2.3821 & 0.0172 & 417 & 0.0574 \\
Factor & ME & Volatility & No & 0.0057 & 0.0021 & 2.7454 & 0.0060 & 417 & 0.0125 \\
Factor & MKT & Volatility & No & -0.0415 & 0.0118 & -3.5241 & 0.0004 & 417 & 0.0430 \\
Factor & ROE & Volatility & No & 0.0018 & 0.0043 & 0.4109 & 0.6812 & 417 & 0.0008 \\

Pooled & All & Downside vol. & No & 0.0012 & 0.0021 & 0.5460 & 0.5850 & 2,085 & 0.0253 \\
Factor & EG & Downside vol. & No & 0.0020 & 0.0028 & 0.7240 & 0.4691 & 417 & 0.0048 \\
Factor & IA & Downside vol. & No & -0.0010 & 0.0014 & -0.7323 & 0.4640 & 417 & 0.0020 \\
Factor & ME & Downside vol. & No & 0.0052 & 0.0014 & 3.7073 & 0.0002 & 417 & 0.0390 \\
Factor & MKT & Downside vol. & No & -0.0099 & 0.0057 & -1.7436 & 0.0812 & 417 & 0.0077 \\
Factor & ROE & Downside vol. & No & -0.0022 & 0.0015 & -1.4123 & 0.1579 & 417 & 0.0034 \\

Pooled & All & Max DD & No & 0.0023 & 0.0011 & 1.9734 & 0.0485 & 2,085 & 0.0498 \\
Factor & EG & Max DD & No & 0.0057 & 0.0033 & 1.7325 & 0.0832 & 417 & 0.0382 \\
Factor & IA & Max DD & No & 0.0014 & 0.0015 & 0.8943 & 0.3711 & 417 & 0.0037 \\
Factor & ME & Max DD & No & 0.0029 & 0.0007 & 4.0618 & 0.0000 & 417 & 0.0148 \\
Factor & MKT & Max DD & No & -0.0085 & 0.0060 & -1.4154 & 0.1570 & 417 & 0.0053 \\
Factor & ROE & Max DD & No & 0.0020 & 0.0022 & 0.9079 & 0.3640 & 417 & 0.0033 \\

Pooled & All & Failure & No & 0.0082 & 0.0170 & 0.4840 & 0.6284 & 2,085 & 0.0051 \\
Factor & EG & Failure & No & 0.0186 & 0.0367 & 0.5061 & 0.6128 & 417 & 0.0010 \\
Factor & IA & Failure & No & -0.0362 & 0.0154 & -2.3504 & 0.0188 & 417 & 0.0062 \\
Factor & ME & Failure & No & 0.0340 & 0.0092 & 3.7123 & 0.0002 & 417 & 0.0137 \\
Factor & MKT & Failure & No & -0.0531 & 0.0539 & -0.9859 & 0.3242 & 417 & 0.0042 \\
Factor & ROE & Failure & No & 0.0100 & 0.0292 & 0.3408 & 0.7332 & 417 & 0.0007 \\

\midrule
\multicolumn{10}{l}{\textit{Panel B. Horizon $h=6$}} \\
\midrule
Pooled & All & Sharpe & No & 0.0247 & 0.0565 & 0.4379 & 0.6614 & 2,070 & 0.0561 \\
Factor & EG & Sharpe & No & -0.0324 & 0.1843 & -0.1757 & 0.8605 & 414 & 0.0001 \\
Factor & IA & Sharpe & No & 0.1137 & 0.1053 & 1.0794 & 0.2804 & 414 & 0.0014 \\
Factor & ME & Sharpe & No & -0.0313 & 0.0291 & -1.0736 & 0.2830 & 414 & 0.0003 \\
Factor & MKT & Sharpe & No & 0.4722 & 0.3372 & 1.4002 & 0.1615 & 414 & 0.0049 \\
Factor & ROE & Sharpe & No & -0.0117 & 0.1032 & -0.1132 & 0.9098 & 414 & 0.0000 \\

Pooled & All & CumRet & No & 0.0002 & 0.0032 & 0.0602 & 0.9520 & 2,070 & 0.0310 \\
Factor & EG & CumRet & No & -0.0035 & 0.0058 & -0.6057 & 0.5447 & 414 & 0.0011 \\
Factor & IA & CumRet & No & 0.0106 & 0.0038 & 2.7672 & 0.0057 & 414 & 0.0156 \\
Factor & ME & CumRet & No & -0.0053 & 0.0015 & -3.6520 & 0.0003 & 414 & 0.0055 \\
Factor & MKT & CumRet & No & 0.0053 & 0.0166 & 0.3204 & 0.7486 & 414 & 0.0003 \\
Factor & ROE & CumRet & No & 0.0014 & 0.0045 & 0.3058 & 0.7597 & 414 & 0.0002 \\

Pooled & All & Volatility & No & 0.0062 & 0.0018 & 3.4750 & 0.0005 & 2,070 & 0.2169 \\
Factor & EG & Volatility & No & 0.0065 & 0.0043 & 1.5033 & 0.1328 & 414 & 0.0118 \\
Factor & IA & Volatility & No & 0.0134 & 0.0028 & 4.8476 & 0.0000 & 414 & 0.0814 \\
Factor & ME & Volatility & No & 0.0081 & 0.0009 & 9.4554 & 0.0000 & 414 & 0.0301 \\
Factor & MKT & Volatility & No & -0.0291 & 0.0115 & -2.5324 & 0.0113 & 414 & 0.0270 \\
Factor & ROE & Volatility & No & 0.0027 & 0.0033 & 0.8081 & 0.4191 & 414 & 0.0018 \\

Pooled & All & Downside vol. & No & 0.0020 & 0.0018 & 1.1060 & 0.2687 & 2,070 & 0.0624 \\
Factor & EG & Downside vol. & No & 0.0045 & 0.0034 & 1.3104 & 0.1901 & 414 & 0.0117 \\
Factor & IA & Downside vol. & No & 0.0026 & 0.0021 & 1.2240 & 0.2209 & 414 & 0.0106 \\
Factor & ME & Downside vol. & No & 0.0040 & 0.0011 & 3.6457 & 0.0003 & 414 & 0.0198 \\
Factor & MKT & Downside vol. & No & -0.0153 & 0.0073 & -2.0979 & 0.0359 & 414 & 0.0120 \\
Factor & ROE & Downside vol. & No & -0.0002 & 0.0028 & -0.0615 & 0.9510 & 414 & 0.0000 \\

Pooled & All & Max DD & No & 0.0017 & 0.0011 & 1.5006 & 0.1335 & 2,070 & 0.0848 \\
Factor & EG & Max DD & No & 0.0032 & 0.0031 & 1.0283 & 0.3038 & 414 & 0.0046 \\
Factor & IA & Max DD & No & 0.0036 & 0.0022 & 1.6732 & 0.0943 & 414 & 0.0112 \\
Factor & ME & Max DD & No & 0.0014 & 0.0006 & 2.4893 & 0.0128 & 414 & 0.0015 \\
Factor & MKT & Max DD & No & -0.0128 & 0.0097 & -1.3143 & 0.1887 & 414 & 0.0049 \\
Factor & ROE & Max DD & No & 0.0024 & 0.0025 & 0.9273 & 0.3538 & 414 & 0.0018 \\

Pooled & All & Failure & No & 0.0121 & 0.0149 & 0.8147 & 0.4152 & 2,070 & 0.0086 \\
Factor & EG & Failure & No & 0.0085 & 0.0423 & 0.2008 & 0.8408 & 414 & 0.0002 \\
Factor & IA & Failure & No & -0.0276 & 0.0177 & -1.5584 & 0.1191 & 414 & 0.0031 \\
Factor & ME & Failure & No & 0.0341 & 0.0079 & 4.3161 & 0.0000 & 414 & 0.0135 \\
Factor & MKT & Failure & No & -0.0496 & 0.0352 & -1.4092 & 0.1588 & 414 & 0.0038 \\
Factor & ROE & Failure & No & 0.0241 & 0.0280 & 0.8594 & 0.3901 & 414 & 0.0035 \\

\midrule
\multicolumn{10}{l}{\textit{Panel C. Horizon $h=12$}} \\
\midrule
Pooled & All & Sharpe & No & 0.0612 & 0.0429 & 1.4276 & 0.1534 & 2,040 & 0.0991 \\
Factor & EG & Sharpe & No & 0.0616 & 0.1798 & 0.3427 & 0.7318 & 408 & 0.0006 \\
Factor & IA & Sharpe & No & 0.1103 & 0.0879 & 1.2539 & 0.2099 & 408 & 0.0042 \\
Factor & ME & Sharpe & No & 0.0174 & 0.0220 & 0.7923 & 0.4282 & 408 & 0.0002 \\
Factor & MKT & Sharpe & No & 0.4286 & 0.2430 & 1.7637 & 0.0778 & 408 & 0.0120 \\
Factor & ROE & Sharpe & No & 0.0230 & 0.0785 & 0.2929 & 0.7696 & 408 & 0.0002 \\

Pooled & All & CumRet & No & 0.0082 & 0.0048 & 1.7085 & 0.0875 & 2,040 & 0.0618 \\
Factor & EG & CumRet & No & -0.0009 & 0.0149 & -0.0579 & 0.9538 & 408 & 0.0000 \\
Factor & IA & CumRet & No & 0.0181 & 0.0085 & 2.1284 & 0.0333 & 408 & 0.0209 \\
Factor & ME & CumRet & No & 0.0004 & 0.0025 & 0.1752 & 0.8609 & 408 & 0.0000 \\
Factor & MKT & CumRet & No & 0.0487 & 0.0335 & 1.4550 & 0.1457 & 408 & 0.0103 \\
Factor & ROE & CumRet & No & 0.0100 & 0.0074 & 1.3496 & 0.1771 & 408 & 0.0057 \\

Pooled & All & Volatility & No & 0.0044 & 0.0016 & 2.8202 & 0.0048 & 2,040 & 0.2629 \\
Factor & EG & Volatility & No & 0.0081 & 0.0034 & 2.3729 & 0.0177 & 408 & 0.0201 \\
Factor & IA & Volatility & No & 0.0111 & 0.0022 & 5.0450 & 0.0000 & 408 & 0.0611 \\
Factor & ME & Volatility & No & 0.0043 & 0.0010 & 4.4615 & 0.0000 & 408 & 0.0112 \\
Factor & MKT & Volatility & No & -0.0333 & 0.0095 & -3.5016 & 0.0005 & 408 & 0.0436 \\
Factor & ROE & Volatility & No & 0.0034 & 0.0046 & 0.7408 & 0.4588 & 408 & 0.0031 \\

Pooled & All & Downside vol. & No & 0.0027 & 0.0014 & 1.8776 & 0.0604 & 2,040 & 0.1440 \\
Factor & EG & Downside vol. & No & 0.0061 & 0.0025 & 2.4429 & 0.0146 & 408 & 0.0181 \\
Factor & IA & Downside vol. & No & 0.0054 & 0.0016 & 3.3313 & 0.0009 & 408 & 0.0517 \\
Factor & ME & Downside vol. & No & 0.0040 & 0.0007 & 5.6719 & 0.0000 & 408 & 0.0243 \\
Factor & MKT & Downside vol. & No & -0.0262 & 0.0066 & -3.9686 & 0.0001 & 408 & 0.0362 \\
Factor & ROE & Downside vol. & No & 0.0015 & 0.0040 & 0.3684 & 0.7126 & 408 & 0.0008 \\

Pooled & All & Max DD & No & -0.0002 & 0.0016 & -0.1383 & 0.8900 & 2,040 & 0.1202 \\
Factor & EG & Max DD & No & 0.0089 & 0.0055 & 1.6016 & 0.1092 & 408 & 0.0183 \\
Factor & IA & Max DD & No & 0.0050 & 0.0042 & 1.2133 & 0.2250 & 408 & 0.0119 \\
Factor & ME & Max DD & No & -0.0011 & 0.0008 & -1.2663 & 0.2054 & 408 & 0.0005 \\
Factor & MKT & Max DD & No & -0.0440 & 0.0176 & -2.5022 & 0.0123 & 408 & 0.0298 \\
Factor & ROE & Max DD & No & -0.0005 & 0.0049 & -0.1008 & 0.9197 & 408 & 0.0000 \\

Pooled & All & Failure & No & -0.0103 & 0.0107 & -0.9595 & 0.3373 & 2,040 & 0.0135 \\
Factor & EG & Failure & No & -0.0001 & 0.0467 & -0.0021 & 0.9983 & 408 & 0.0000 \\
Factor & IA & Failure & No & 0.0043 & 0.0267 & 0.1614 & 0.8718 & 408 & 0.0001 \\
Factor & ME & Failure & No & -0.0074 & 0.0038 & -1.9697 & 0.0489 & 408 & 0.0008 \\
Factor & MKT & Failure & No & -0.1258 & 0.0545 & -2.3083 & 0.0210 & 408 & 0.0248 \\
Factor & ROE & Failure & No & -0.0127 & 0.0288 & -0.4407 & 0.6594 & 408 & 0.0008 \\

\end{longtable}
\end{center}
\endgroup

%% file: tabs/controlled_predictive_q5_robust.tex
\begingroup
\footnotesize
\setlength{\tabcolsep}{3.5pt}

\begin{center}
\begin{longtable}{p{1.5cm} p{1.1cm} p{2.0cm} c r r r r r r}
\caption{Controlled predictive results.}
\label{tab:controlled_q5_robust} \\
\toprule
Sample & Factor & Outcome & Ctrl. & Coef. & SE & $t$ & $p$ & Obs. & $R^2$ \\
\midrule
\endfirsthead

\multicolumn{10}{l}{\textit{Table \thetable\ (continued)}} \\
\toprule
Sample & Factor & Outcome & Ctrl. & Coef. & SE & $t$ & $p$ & Obs. & $R^2$ \\
\midrule
\endhead

\midrule
\multicolumn{10}{r}{\textit{Continued on next page}} \\
\endfoot

\bottomrule
\endlastfoot

\multicolumn{10}{l}{\textit{Panel A. Horizon $h=3$}} \\
\midrule
Pooled & All & Sharpe & Yes & -0.1367 & 0.3336 & -0.4098 & 0.6819 & 2,085 & 0.0119 \\
Factor & EG & Sharpe & Yes & -1.5406 & 2.1266 & -0.7244 & 0.4688 & 417 & 0.0126 \\
Factor & IA & Sharpe & Yes & 0.7620 & 0.3962 & 1.9231 & 0.0545 & 417 & 0.0068 \\
Factor & ME & Sharpe & Yes & -0.7432 & 0.3356 & -2.2148 & 0.0268 & 417 & 0.0629 \\
Factor & MKT & Sharpe & Yes & 2.2465 & 1.4468 & 1.5527 & 0.1205 & 417 & 0.0195 \\
Factor & ROE & Sharpe & Yes & 0.1527 & 0.3240 & 0.4714 & 0.6374 & 417 & 0.0313 \\

Pooled & All & CumRet & Yes & -0.0036 & 0.0036 & -0.9859 & 0.3242 & 2,085 & 0.0258 \\
Factor & EG & CumRet & Yes & -0.0043 & 0.0042 & -1.0289 & 0.3035 & 417 & 0.0159 \\
Factor & IA & CumRet & Yes & 0.0070 & 0.0037 & 1.9204 & 0.0548 & 417 & 0.0417 \\
Factor & ME & CumRet & Yes & -0.0120 & 0.0018 & -6.8503 & 0.0000 & 417 & 0.0815 \\
Factor & MKT & CumRet & Yes & 0.0126 & 0.0161 & 0.7829 & 0.4337 & 417 & 0.0245 \\
Factor & ROE & CumRet & Yes & -0.0006 & 0.0043 & -0.1429 & 0.8864 & 417 & 0.0572 \\

Pooled & All & Volatility & Yes & 0.0014 & 0.0020 & 0.6928 & 0.4885 & 2,085 & 0.2812 \\
Factor & EG & Volatility & Yes & 0.0048 & 0.0041 & 1.1504 & 0.2500 & 417 & 0.2201 \\
Factor & IA & Volatility & Yes & 0.0041 & 0.0041 & 0.9900 & 0.3222 & 417 & 0.3280 \\
Factor & ME & Volatility & Yes & 0.0043 & 0.0020 & 2.1900 & 0.0285 & 417 & 0.0563 \\
Factor & MKT & Volatility & Yes & -0.0294 & 0.0112 & -2.6313 & 0.0085 & 417 & 0.2063 \\
Factor & ROE & Volatility & Yes & -0.0034 & 0.0035 & -0.9724 & 0.3309 & 417 & 0.3275 \\

Pooled & All & Downside vol. & Yes & 0.0008 & 0.0022 & 0.3607 & 0.7183 & 2,085 & 0.0441 \\
Factor & EG & Downside vol. & Yes & 0.0002 & 0.0023 & 0.0917 & 0.9269 & 417 & 0.0995 \\
Factor & IA & Downside vol. & Yes & -0.0024 & 0.0014 & -1.6524 & 0.0985 & 417 & 0.0535 \\
Factor & ME & Downside vol. & Yes & 0.0053 & 0.0014 & 3.7444 & 0.0002 & 417 & 0.0427 \\
Factor & MKT & Downside vol. & Yes & -0.0092 & 0.0056 & -1.6494 & 0.0991 & 417 & 0.0121 \\
Factor & ROE & Downside vol. & Yes & -0.0030 & 0.0015 & -1.9305 & 0.0536 & 417 & 0.1434 \\

Pooled & All & Max DD & Yes & 0.0015 & 0.0011 & 1.3566 & 0.1749 & 2,085 & 0.0843 \\
Factor & EG & Max DD & Yes & 0.0035 & 0.0029 & 1.2241 & 0.2209 & 417 & 0.1688 \\
Factor & IA & Max DD & Yes & -0.0004 & 0.0015 & -0.2812 & 0.7786 & 417 & 0.0813 \\
Factor & ME & Max DD & Yes & 0.0030 & 0.0007 & 4.1433 & 0.0000 & 417 & 0.0153 \\
Factor & MKT & Max DD & Yes & -0.0062 & 0.0059 & -1.0495 & 0.2940 & 417 & 0.0221 \\
Factor & ROE & Max DD & Yes & 0.0006 & 0.0022 & 0.2512 & 0.8017 & 417 & 0.1568 \\

Pooled & All & Failure & Yes & 0.0019 & 0.0172 & 0.1117 & 0.9110 & 2,085 & 0.0279 \\
Factor & EG & Failure & Yes & -0.0056 & 0.0322 & -0.1731 & 0.8626 & 417 & 0.0422 \\
Factor & IA & Failure & Yes & -0.0512 & 0.0194 & -2.6403 & 0.0083 & 417 & 0.0302 \\
Factor & ME & Failure & Yes & 0.0382 & 0.0093 & 4.1153 & 0.0000 & 417 & 0.0172 \\
Factor & MKT & Failure & Yes & -0.0358 & 0.0528 & -0.6771 & 0.4983 & 417 & 0.0237 \\
Factor & ROE & Failure & Yes & -0.0042 & 0.0278 & -0.1504 & 0.8804 & 417 & 0.1249 \\

\midrule
\multicolumn{10}{l}{\textit{Panel B. Horizon $h=6$}} \\
\midrule
Pooled & All & Sharpe & Yes & 0.0334 & 0.0575 & 0.5808 & 0.5614 & 2,070 & 0.0600 \\
Factor & EG & Sharpe & Yes & 0.1804 & 0.1692 & 1.0665 & 0.2862 & 414 & 0.0791 \\
Factor & IA & Sharpe & Yes & 0.0763 & 0.1369 & 0.5571 & 0.5775 & 414 & 0.0073 \\
Factor & ME & Sharpe & Yes & -0.1417 & 0.0468 & -3.0253 & 0.0025 & 414 & 0.0775 \\
Factor & MKT & Sharpe & Yes & 0.4679 & 0.3117 & 1.5010 & 0.1333 & 414 & 0.0055 \\
Factor & ROE & Sharpe & Yes & 0.0657 & 0.1061 & 0.6191 & 0.5359 & 414 & 0.0918 \\

Pooled & All & CumRet & Yes & -0.0023 & 0.0031 & -0.7326 & 0.4638 & 2,070 & 0.0475 \\
Factor & EG & CumRet & Yes & -0.0016 & 0.0054 & -0.2955 & 0.7676 & 414 & 0.0190 \\
Factor & IA & CumRet & Yes & 0.0058 & 0.0046 & 1.2421 & 0.2142 & 414 & 0.0477 \\
Factor & ME & CumRet & Yes & -0.0100 & 0.0020 & -5.1077 & 0.0000 & 414 & 0.1172 \\
Factor & MKT & CumRet & Yes & 0.0155 & 0.0177 & 0.8784 & 0.3797 & 414 & 0.0389 \\
Factor & ROE & CumRet & Yes & 0.0001 & 0.0039 & 0.0314 & 0.9750 & 414 & 0.1199 \\

Pooled & All & Volatility & Yes & 0.0023 & 0.0017 & 1.3538 & 0.1758 & 2,070 & 0.3866 \\
Factor & EG & Volatility & Yes & -0.0008 & 0.0033 & -0.2443 & 0.8070 & 414 & 0.3575 \\
Factor & IA & Volatility & Yes & 0.0050 & 0.0024 & 2.1200 & 0.0340 & 414 & 0.4335 \\
Factor & ME & Volatility & Yes & 0.0069 & 0.0013 & 5.1564 & 0.0000 & 414 & 0.0730 \\
Factor & MKT & Volatility & Yes & -0.0168 & 0.0075 & -2.2499 & 0.0245 & 414 & 0.2241 \\
Factor & ROE & Volatility & Yes & -0.0033 & 0.0028 & -1.1937 & 0.2326 & 414 & 0.4509 \\

Pooled & All & Downside vol. & Yes & 0.0009 & 0.0018 & 0.4714 & 0.6374 & 2,070 & 0.1205 \\
Factor & EG & Downside vol. & Yes & 0.0004 & 0.0029 & 0.1469 & 0.8832 & 414 & 0.2370 \\
Factor & IA & Downside vol. & Yes & -0.0008 & 0.0015 & -0.5270 & 0.5982 & 414 & 0.2071 \\
Factor & ME & Downside vol. & Yes & 0.0045 & 0.0011 & 3.9777 & 0.0001 & 414 & 0.0229 \\
Factor & MKT & Downside vol. & Yes & -0.0137 & 0.0070 & -1.9495 & 0.0512 & 414 & 0.0179 \\
Factor & ROE & Downside vol. & Yes & -0.0023 & 0.0021 & -1.1103 & 0.2669 & 414 & 0.3298 \\

Pooled & All & Max DD & Yes & 0.0001 & 0.0011 & 0.1063 & 0.9153 & 2,070 & 0.1401 \\
Factor & EG & Max DD & Yes & -0.0016 & 0.0026 & -0.6139 & 0.5393 & 414 & 0.2407 \\
Factor & IA & Max DD & Yes & 0.0000 & 0.0025 & 0.0180 & 0.9857 & 414 & 0.1294 \\
Factor & ME & Max DD & Yes & 0.0018 & 0.0009 & 1.9897 & 0.0466 & 414 & 0.0036 \\
Factor & MKT & Max DD & Yes & -0.0084 & 0.0086 & -0.9759 & 0.3291 & 414 & 0.0289 \\
Factor & ROE & Max DD & Yes & -0.0006 & 0.0021 & -0.2598 & 0.7950 & 414 & 0.2918 \\

Pooled & All & Failure & Yes & 0.0071 & 0.0151 & 0.4700 & 0.6383 & 2,070 & 0.0314 \\
Factor & EG & Failure & Yes & -0.0145 & 0.0421 & -0.3440 & 0.7308 & 414 & 0.0418 \\
Factor & IA & Failure & Yes & -0.0440 & 0.0192 & -2.2974 & 0.0216 & 414 & 0.0143 \\
Factor & ME & Failure & Yes & 0.0401 & 0.0083 & 4.8260 & 0.0000 & 414 & 0.0194 \\
Factor & MKT & Failure & Yes & -0.0420 & 0.0343 & -1.2221 & 0.2217 & 414 & 0.0073 \\
Factor & ROE & Failure & Yes & 0.0121 & 0.0242 & 0.4989 & 0.6179 & 414 & 0.2204 \\

\midrule
\multicolumn{10}{l}{\textit{Panel C. Horizon $h=12$}} \\
\midrule
Pooled & All & Sharpe & Yes & 0.0703 & 0.0430 & 1.6330 & 0.1025 & 2,040 & 0.1051 \\
Factor & EG & Sharpe & Yes & 0.2367 & 0.1499 & 1.5793 & 0.1143 & 408 & 0.1399 \\
Factor & IA & Sharpe & Yes & 0.0810 & 0.0799 & 1.0146 & 0.3103 & 408 & 0.0359 \\
Factor & ME & Sharpe & Yes & -0.0543 & 0.0336 & -1.6179 & 0.1057 & 408 & 0.1293 \\
Factor & MKT & Sharpe & Yes & 0.4305 & 0.2300 & 1.8721 & 0.0612 & 408 & 0.0166 \\
Factor & ROE & Sharpe & Yes & 0.0857 & 0.0686 & 1.2481 & 0.2120 & 408 & 0.1973 \\

Pooled & All & CumRet & Yes & 0.0038 & 0.0045 & 0.8439 & 0.3987 & 2,040 & 0.0844 \\
Factor & EG & CumRet & Yes & 0.0020 & 0.0116 & 0.1722 & 0.8633 & 408 & 0.0345 \\
Factor & IA & CumRet & Yes & 0.0081 & 0.0062 & 1.3037 & 0.1923 & 408 & 0.0931 \\
Factor & ME & CumRet & Yes & -0.0076 & 0.0032 & -2.3931 & 0.0167 & 408 & 0.2015 \\
Factor & MKT & CumRet & Yes & 0.0648 & 0.0310 & 2.0884 & 0.0368 & 408 & 0.0546 \\
Factor & ROE & CumRet & Yes & 0.0078 & 0.0064 & 1.2181 & 0.2232 & 408 & 0.1567 \\

Pooled & All & Volatility & Yes & 0.0008 & 0.0013 & 0.6302 & 0.5285 & 2,040 & 0.4376 \\
Factor & EG & Volatility & Yes & 0.0012 & 0.0023 & 0.5355 & 0.5923 & 408 & 0.3526 \\
Factor & IA & Volatility & Yes & 0.0030 & 0.0021 & 1.4484 & 0.1475 & 408 & 0.4209 \\
Factor & ME & Volatility & Yes & 0.0032 & 0.0011 & 2.9241 & 0.0035 & 408 & 0.0867 \\
Factor & MKT & Volatility & Yes & -0.0220 & 0.0070 & -3.1549 & 0.0016 & 408 & 0.2385 \\
Factor & ROE & Volatility & Yes & -0.0018 & 0.0027 & -0.6787 & 0.4973 & 408 & 0.4193 \\

Pooled & All & Downside vol. & Yes & 0.0010 & 0.0014 & 0.7335 & 0.4632 & 2,040 & 0.2490 \\
Factor & EG & Downside vol. & Yes & 0.0018 & 0.0020 & 0.8987 & 0.3688 & 408 & 0.2306 \\
Factor & IA & Downside vol. & Yes & 0.0009 & 0.0013 & 0.7286 & 0.4663 & 408 & 0.4128 \\
Factor & ME & Downside vol. & Yes & 0.0040 & 0.0007 & 5.4046 & 0.0000 & 408 & 0.0485 \\
Factor & MKT & Downside vol. & Yes & -0.0213 & 0.0071 & -3.0019 & 0.0027 & 408 & 0.0966 \\
Factor & ROE & Downside vol. & Yes & -0.0015 & 0.0023 & -0.6442 & 0.5194 & 408 & 0.3400 \\

Pooled & All & Max DD & Yes & -0.0024 & 0.0014 & -1.6691 & 0.0951 & 2,040 & 0.1937 \\
Factor & EG & Max DD & Yes & 0.0026 & 0.0044 & 0.5872 & 0.5571 & 408 & 0.2581 \\
Factor & IA & Max DD & Yes & -0.0006 & 0.0030 & -0.1913 & 0.8483 & 408 & 0.1541 \\
Factor & ME & Max DD & Yes & -0.0005 & 0.0016 & -0.2864 & 0.7746 & 408 & 0.0033 \\
Factor & MKT & Max DD & Yes & -0.0364 & 0.0173 & -2.0997 & 0.0358 & 408 & 0.0676 \\
Factor & ROE & Max DD & Yes & -0.0041 & 0.0030 & -1.3505 & 0.1768 & 408 & 0.3737 \\

Pooled & All & Failure & Yes & -0.0160 & 0.0106 & -1.5113 & 0.1307 & 2,040 & 0.0360 \\
Factor & EG & Failure & Yes & -0.0251 & 0.0366 & -0.6859 & 0.4928 & 408 & 0.0822 \\
Factor & IA & Failure & Yes & -0.0157 & 0.0197 & -0.7973 & 0.4253 & 408 & 0.0458 \\
Factor & ME & Failure & Yes & -0.0082 & 0.0074 & -1.1115 & 0.2664 & 408 & 0.0016 \\
Factor & MKT & Failure & Yes & -0.1179 & 0.0540 & -2.1828 & 0.0291 & 408 & 0.0297 \\
Factor & ROE & Failure & Yes & -0.0204 & 0.0214 & -0.9518 & 0.3412 & 408 & 0.2246 \\

\end{longtable}
\end{center}
\endgroup

%% file: tabs/q5_prop4_revised_rankcheck.tex
\begin{table}[htbp]
\centering
\footnotesize
\setlength{\tabcolsep}{4pt}
\caption{q5 factors: rank comparison across break diagnostics and predictive slopes}
\label{tab:q5_prop4_rankcheck}
\begin{tabular}{l r r r r r}
\toprule
Factor 
& Rank (Pr(Break)) 
& Rank (Break share) 
& Rank ($\Delta$ break) 
& Rank (Spike freq.) 
& Rank (Sharpe coef.) \\
\midrule
MKT & 1 & 1 & 5 & 3 & 1 \\
ME  & 5 & 5 & 1 & 5 & 5 \\
IA  & 4 & 4 & 2 & 4 & 2 \\
ROE & 3 & 3 & 3 & 1 & 4 \\
EG  & 2 & 2 & 4 & 2 & 3 \\
\bottomrule
\end{tabular}
\end{table}

%% file: tabs/anomaly_stage5diag_family_map_summary.tex
\begin{tabular}{lr}
\toprule
Family & Anomaly Count \\
\midrule
Value & 18 \\
Momentum & 33 \\
Investment & 16 \\
Profitability \& Quality & 10 \\
Accruals \& Accounting & 18 \\
Beta \& Risk \& Volatility & 35 \\
Growth \& Issuance & 32 \\
Reversal  \& Seasonality \& Microstructure & 33 \\
Other & 17 \\
\bottomrule
\end{tabular}

%% file: tabs/anomaly_stage5diag_fit_quality.tex
\begingroup
\footnotesize
\setlength{\tabcolsep}{2pt}
\renewcommand{\arraystretch}{0.95}
\begin{longtable}{lrrrcrrrr}
\caption{Per-anomaly model fit and Delta quality diagnostics}\label{tab:anomaly_stage5diag_fit_quality}\\
\toprule
Anomaly & N\_Obs & Sample & BreakFit & DegBreak & DeltaStd & ExtRatio & DeltaAbn & Problem \\
\midrule
\endfirsthead
\toprule
Anomaly & Obs. & Sample & Break fit & Deg. break & $\Delta$ Std. & Ext. ratio & $\Delta$ Abn. & Problem \\
\midrule
\endhead
\bottomrule
\endfoot
Beta & 1102 & long & 1 & 0 & 0.4625 & 6.71 & 1 & 1 \\
BidAskSpread & 1127 & long & 1 & 0 & 1.2308 & 5.85 & 1 & 1 \\
DivInit & 1127 & long & 1 & 1 & 1.0689 & 3.25 & 1 & 1 \\
DivYieldST & 1121 & long & 1 & 0 & 0.6644 & 4.31 & 1 & 1 \\
DolVol & 1125 & long & 1 & 0 & 1.0136 & 3.19 & 1 & 1 \\
IndMom & 1122 & long & 1 & 0 & 1.7040 & 17.44 & 1 & 1 \\
IndRetBig & 1127 & long & 1 & 0 & 1.4322 & 26.05 & 1 & 1 \\
IntMom & 1116 & long & 1 & 0 & 0.7846 & 5.96 & 1 & 1 \\
MRreversal & 1110 & long & 1 & 0 & 0.7245 & 3.14 & 1 & 1 \\
Mom12m & 1116 & long & 1 & 0 & 0.9105 & 5.56 & 1 & 1 \\
Mom6m & 1122 & long & 1 & 0 & 1.2259 & 8.92 & 1 & 1 \\
MomOffSeason & 1115 & long & 1 & 0 & 0.6836 & 8.92 & 1 & 1 \\
MomOffSeason16YrPlus & 947 & long & 1 & 0 & 0.4718 & 11.07 & 1 & 1 \\
MomRev & 1091 & long & 1 & 0 & 1.1699 & 10.51 & 1 & 1 \\
MomSeason16YrPlus & 936 & long & 1 & 0 & 0.4794 & 13.33 & 1 & 1 \\
MomVol & 1104 & long & 1 & 0 & 0.8315 & 7.05 & 1 & 1 \\
Price & 1128 & long & 1 & 0 & 0.9959 & 4.65 & 1 & 1 \\
PriceDelayRsq & 1105 & long & 1 & 0 & 0.4232 & 3.15 & 1 & 1 \\
PriceDelaySlope & 1105 & long & 1 & 0 & 0.3862 & 3.62 & 1 & 1 \\
ReturnSkew & 1127 & long & 1 & 0 & 0.9293 & 9.77 & 1 & 1 \\
ReturnSkew3F & 1121 & long & 1 & 0 & 0.9491 & 7.78 & 1 & 1 \\
STreversal & 1127 & long & 1 & 0 & 1.6210 & 10.99 & 1 & 1 \\
ShareIss1Y & 1110 & long & 1 & 0 & 1.8995 & 48.17 & 1 & 1 \\
Size & 1122 & long & 1 & 0 & 0.9637 & 4.20 & 1 & 1 \\
Spinoff & 1128 & long & 1 & 0 & 1.0463 & 3.69 & 1 & 1 \\
TrendFactor & 1126 & long & 1 & 0 & 0.9003 & 11.21 & 1 & 1 \\
sinAlgo & 1128 & long & 1 & 0 & 0.8900 & 2.23 & 1 & 1 \\
std\_turn & 1104 & long & 1 & 0 & 0.8403 & 12.29 & 1 & 1 \\
AM & 822 & medium & 1 & 0 & 1.2121 & 50.51 & 1 & 1 \\
AssetGrowth & 810 & medium & 1 & 0 & 1.1712 & 26.62 & 1 & 1 \\
BPEBM & 690 & medium & 1 & 0 & 1.4680 & 19.24 & 1 & 1 \\
BookLeverage & 822 & medium & 1 & 0 & 2.8596 & 79.43 & 1 & 1 \\
CashProd & 828 & medium & 1 & 0 & 1.0897 & 23.61 & 1 & 1 \\
CoskewACX & 689 & medium & 1 & 0 & 1.0842 & 24.00 & 1 & 1 \\
DelNetFin & 810 & medium & 1 & 0 & 0.5479 & 11.63 & 1 & 1 \\
EarningsSurprise & 679 & medium & 1 & 0 & 0.6387 & 6.64 & 1 & 1 \\
GrAdExp & 654 & medium & 1 & 1 & 0.1947 & 1.84 & 0 & 1 \\
GrSaleToGrOverhead & 810 & medium & 1 & 0 & 0.6162 & 17.52 & 1 & 1 \\
Herf & 827 & medium & 1 & 0 & 2.5591 & 43.96 & 1 & 1 \\
HerfAsset & 817 & medium & 1 & 0 & 4.0911 & 33.29 & 1 & 1 \\
HerfBE & 817 & medium & 1 & 0 & 3.7951 & 31.07 & 1 & 1 \\
IntanCFP & 768 & medium & 1 & 0 & 1.4764 & 19.55 & 1 & 1 \\
Leverage & 822 & medium & 1 & 0 & 1.8669 & 62.76 & 1 & 1 \\
NOA & 684 & medium & 1 & 0 & 1.1776 & 14.50 & 1 & 1 \\
OPLeverage & 822 & medium & 1 & 0 & 1.0841 & 32.83 & 1 & 1 \\
OperProf & 678 & medium & 1 & 0 & 2.0252 & 72.84 & 1 & 1 \\
RD & 822 & medium & 1 & 0 & 2.6547 & 51.54 & 1 & 1 \\
RoE & 702 & medium & 1 & 0 & 1.4277 & 21.51 & 1 & 1 \\
SurpriseRD & 814 & medium & 1 & 0 & 1.5327 & 22.12 & 1 & 1 \\
Tax & 822 & medium & 1 & 0 & 1.1565 & 23.86 & 1 & 1 \\
TotalAccruals & 810 & medium & 1 & 0 & 1.6630 & 52.30 & 1 & 1 \\
roaq & 642 & medium & 1 & 0 & 0.7741 & 9.66 & 1 & 1 \\
tang & 828 & medium & 1 & 0 & 3.5144 & 30.01 & 1 & 1 \\
AnalystRevision & 526 & short & 1 & 0 & 1.0483 & 12.83 & 1 & 1 \\
AnalystValue & 522 & short & 1 & 0 & 1.8583 & 70.89 & 1 & 1 \\
AnnouncementReturn & 580 & short & 1 & 0 & 1.3870 & 13.36 & 1 & 1 \\
Cash & 579 & short & 1 & 0 & 2.9501 & 55.59 & 1 & 1 \\
DelDRC & 234 & short & 1 & 1 & 0.1088 & 1.15 & 0 & 1 \\
IO\_ShortInterest & 481 & short & 1 & 0 & 9.6629 & 29.49 & 1 & 1 \\
OScore & 576 & short & 1 & 0 & 1.3320 & 31.35 & 1 & 1 \\
OptionVolume1 & 271 & short & 1 & 1 & 0.2469 & 2.06 & 0 & 1 \\
OrderBacklogChg & 570 & short & 1 & 0 & 0.4601 & 8.50 & 1 & 1 \\
PatentsRD & 387 & short & 1 & 0 & 16.7522 & 164.54 & 1 & 1 \\
RDcap & 474 & short & 1 & 0 & 1.2518 & 61.14 & 1 & 1 \\
REV6 & 520 & short & 1 & 0 & 1.6280 & 29.17 & 1 & 1 \\
XFIN & 570 & short & 1 & 0 & 1.8907 & 64.53 & 1 & 1 \\
dVolPut & 263 & short & 1 & 1 & 0.2510 & 4.17 & 0 & 1 \\
sfe & 513 & short & 1 & 0 & 1.7392 & 31.87 & 1 & 1 \\
BetaFP & 1101 & long & 1 & 0 & 0.2898 & 4.24 & 0 & 0 \\
BetaTailRisk & 1056 & long & 1 & 0 & 0.2489 & 4.02 & 0 & 0 \\
CompEquIss & 1067 & long & 1 & 0 & 0.2521 & 2.40 & 0 & 0 \\
Coskewness & 1110 & long & 1 & 0 & 0.4143 & 7.00 & 0 & 0 \\
DivOmit & 1080 & long & 1 & 0 & 0.2902 & 4.94 & 0 & 0 \\
DivSeason & 1124 & long & 1 & 0 & 0.2502 & 3.80 & 0 & 0 \\
FirmAge & 1098 & long & 1 & 0 & 0.4262 & 12.31 & 0 & 0 \\
FirmAgeMom & 1104 & long & 1 & 0 & 0.5581 & 14.15 & 0 & 0 \\
High52 & 1122 & long & 1 & 0 & 0.7134 & 5.00 & 0 & 0 \\
IdioVol3F & 1121 & long & 1 & 0 & 0.7678 & 23.54 & 0 & 0 \\
IdioVolAHT & 1110 & long & 1 & 0 & 0.7956 & 19.91 & 0 & 0 \\
Illiquidity & 1110 & long & 1 & 0 & 0.2928 & 4.08 & 0 & 0 \\
LRreversal & 1092 & long & 1 & 0 & 0.2880 & 3.41 & 0 & 0 \\
MaxRet & 1127 & long & 1 & 0 & 0.7658 & 15.21 & 0 & 0 \\
Mom12mOffSeason & 1126 & long & 1 & 0 & 0.7718 & 8.69 & 0 & 0 \\
MomOffSeason06YrPlus & 1067 & long & 1 & 0 & 0.2598 & 4.81 & 0 & 0 \\
MomOffSeason11YrPlus & 1007 & long & 1 & 0 & 0.2962 & 5.46 & 0 & 0 \\
MomSeason & 1104 & long & 1 & 0 & 0.3377 & 4.50 & 0 & 0 \\
MomSeason06YrPlus & 1056 & long & 1 & 0 & 0.2898 & 5.18 & 0 & 0 \\
MomSeason11YrPlus & 996 & long & 1 & 0 & 0.2680 & 8.75 & 0 & 0 \\
MomSeasonShort & 1116 & long & 1 & 0 & 0.2960 & 3.57 & 0 & 0 \\
PriceDelayTstat & 1105 & long & 1 & 0 & 0.3327 & 3.41 & 0 & 0 \\
RIO\_Turnover & 1120 & long & 1 & 0 & 0.2802 & 7.03 & 0 & 0 \\
RIO\_Volatility & 1119 & long & 1 & 0 & 0.2995 & 4.61 & 0 & 0 \\
RealizedVol & 1121 & long & 1 & 0 & 0.7400 & 14.97 & 0 & 0 \\
ResidualMomentum & 1075 & long & 1 & 0 & 0.4038 & 6.77 & 0 & 0 \\
ShareIss5Y & 1062 & long & 1 & 0 & 1.4867 & 27.21 & 0 & 0 \\
ShareVol & 1116 & long & 1 & 0 & 0.2860 & 5.42 & 0 & 0 \\
VolMkt & 1122 & long & 1 & 0 & 0.3154 & 5.59 & 0 & 0 \\
VolSD & 1114 & long & 1 & 0 & 0.3104 & 4.18 & 0 & 0 \\
VolumeTrend & 1098 & long & 1 & 0 & 0.2916 & 3.39 & 0 & 0 \\
zerotrade12M & 1115 & long & 1 & 0 & 0.4275 & 5.75 & 0 & 0 \\
zerotrade1M & 1116 & long & 1 & 0 & 0.3650 & 4.86 & 0 & 0 \\
zerotrade6M & 1121 & long & 1 & 0 & 0.3961 & 6.39 & 0 & 0 \\
Accruals & 810 & medium & 1 & 0 & 0.2449 & 7.96 & 0 & 0 \\
AccrualsBM & 665 & medium & 1 & 0 & 0.3496 & 11.35 & 0 & 0 \\
AdExp & 668 & medium & 1 & 0 & 0.2904 & 4.53 & 0 & 0 \\
BM & 822 & medium & 1 & 0 & 0.8873 & 25.34 & 0 & 0 \\
BMdec & 810 & medium & 1 & 0 & 0.8317 & 27.41 & 0 & 0 \\
BetaLiquidityPS & 673 & medium & 1 & 0 & 0.2392 & 7.50 & 0 & 0 \\
BrandInvest & 654 & medium & 1 & 0 & 1.0525 & 9.38 & 0 & 0 \\
CBOperProf & 690 & medium & 1 & 0 & 0.3474 & 15.76 & 0 & 0 \\
CF & 822 & medium & 1 & 0 & 1.0472 & 26.19 & 0 & 0 \\
ChAssetTurnover & 798 & medium & 1 & 0 & 0.3131 & 11.12 & 0 & 0 \\
ChEQ & 690 & medium & 1 & 0 & 0.3593 & 14.28 & 0 & 0 \\
ChInv & 810 & medium & 1 & 0 & 0.3647 & 12.36 & 0 & 0 \\
ChInvIA & 810 & medium & 1 & 0 & 0.6698 & 27.43 & 0 & 0 \\
ChNNCOA & 810 & medium & 1 & 0 & 0.2650 & 9.20 & 0 & 0 \\
ChNWC & 810 & medium & 1 & 0 & 0.2362 & 5.85 & 0 & 0 \\
ChTax & 687 & medium & 1 & 0 & 0.3142 & 6.64 & 0 & 0 \\
CompositeDebtIssuance & 762 & medium & 1 & 0 & 0.3091 & 15.05 & 0 & 0 \\
ConvDebt & 828 & medium & 1 & 0 & 0.6710 & 30.97 & 0 & 0 \\
DelCOA & 810 & medium & 1 & 0 & 0.4210 & 17.00 & 0 & 0 \\
DelCOL & 810 & medium & 1 & 0 & 0.3118 & 14.11 & 0 & 0 \\
DelEqu & 690 & medium & 1 & 0 & 0.4101 & 12.73 & 0 & 0 \\
DelFINL & 810 & medium & 1 & 0 & 0.3909 & 12.24 & 0 & 0 \\
DelLTI & 798 & medium & 1 & 0 & 1.4881 & 29.39 & 0 & 0 \\
EBM & 690 & medium & 1 & 0 & 0.2881 & 8.88 & 0 & 0 \\
EP & 828 & medium & 1 & 0 & 0.7926 & 12.04 & 0 & 0 \\
EarnSupBig & 679 & medium & 1 & 0 & 1.2746 & 47.12 & 0 & 0 \\
EarningsConsistency & 798 & medium & 1 & 0 & 0.3348 & 8.78 & 0 & 0 \\
EntMult & 822 & medium & 1 & 0 & 0.8737 & 23.38 & 0 & 0 \\
EquityDuration & 678 & medium & 1 & 0 & 0.6275 & 20.61 & 0 & 0 \\
ExchSwitch & 686 & medium & 1 & 0 & 0.4840 & 17.00 & 0 & 0 \\
Frontier & 678 & medium & 1 & 0 & 0.5133 & 17.78 & 0 & 0 \\
GP & 822 & medium & 1 & 0 & 0.5563 & 26.13 & 0 & 0 \\
GrLTNOA & 810 & medium & 1 & 0 & 0.2719 & 8.85 & 0 & 0 \\
GrSaleToGrInv & 810 & medium & 1 & 0 & 0.2764 & 13.18 & 0 & 0 \\
IntanBM & 642 & medium & 1 & 0 & 0.3631 & 9.24 & 0 & 0 \\
IntanEP & 762 & medium & 1 & 0 & 0.6635 & 16.31 & 0 & 0 \\
IntanSP & 762 & medium & 1 & 0 & 1.0547 & 38.14 & 0 & 0 \\
InvGrowth & 810 & medium & 1 & 0 & 0.4546 & 19.50 & 0 & 0 \\
InvestPPEInv & 816 & medium & 1 & 0 & 0.5967 & 23.04 & 0 & 0 \\
Investment & 798 & medium & 1 & 0 & 0.5856 & 12.91 & 0 & 0 \\
MeanRankRevGrowth & 750 & medium & 1 & 0 & 0.2878 & 12.43 & 0 & 0 \\
NetDebtPrice & 678 & medium & 1 & 0 & 1.5318 & 38.87 & 0 & 0 \\
NetPayoutYield & 798 & medium & 1 & 0 & 1.0223 & 27.93 & 0 & 0 \\
NumEarnIncrease & 672 & medium & 1 & 0 & 0.4483 & 19.43 & 0 & 0 \\
OperProfRD & 678 & medium & 1 & 0 & 0.2885 & 13.40 & 0 & 0 \\
OrgCap & 822 & medium & 1 & 0 & 0.4005 & 23.73 & 0 & 0 \\
PayoutYield & 798 & medium & 1 & 0 & 0.4882 & 16.40 & 0 & 0 \\
PctAcc & 666 & medium & 1 & 0 & 0.2315 & 3.43 & 0 & 0 \\
RDAbility & 750 & medium & 1 & 0 & 0.2224 & 5.26 & 0 & 0 \\
RIO\_MB & 681 & medium & 1 & 0 & 0.2595 & 4.13 & 0 & 0 \\
RevenueSurprise & 679 & medium & 1 & 0 & 0.4908 & 12.19 & 0 & 0 \\
SP & 822 & medium & 1 & 0 & 1.0761 & 31.12 & 0 & 0 \\
VarCF & 798 & medium & 1 & 0 & 1.0712 & 25.14 & 0 & 0 \\
cfp & 666 & medium & 1 & 0 & 2.0053 & 47.71 & 0 & 0 \\
dNoa & 690 & medium & 1 & 0 & 0.3156 & 9.40 & 0 & 0 \\
grcapx & 798 & medium & 1 & 0 & 0.4305 & 24.94 & 0 & 0 \\
grcapx3y & 786 & medium & 1 & 0 & 0.5321 & 23.51 & 0 & 0 \\
hire & 654 & medium & 1 & 0 & 0.3527 & 11.12 & 0 & 0 \\
AOP & 522 & short & 1 & 0 & 0.2908 & 12.22 & 0 & 0 \\
AbnormalAccruals & 570 & short & 1 & 0 & 0.4519 & 17.27 & 0 & 0 \\
Activism1 & 137 & short & 1 & 0 & 0.9825 & 44.02 & 0 & 0 \\
Activism2 & 137 & short & 1 & 0 & 0.4013 & 21.11 & 0 & 0 \\
AgeIPO & 468 & short & 1 & 0 & 0.7446 & 29.60 & 0 & 0 \\
CPVolSpread & 264 & short & 1 & 0 & 0.2381 & 5.50 & 0 & 0 \\
ChForecastAccrual & 522 & short & 1 & 0 & 0.4311 & 20.90 & 0 & 0 \\
ChNAnalyst & 498 & short & 1 & 0 & 0.2681 & 2.63 & 0 & 0 \\
ChangeInRecommendation & 313 & short & 1 & 0 & 0.6087 & 9.11 & 0 & 0 \\
CitationsRD & 510 & short & 1 & 0 & 0.3045 & 11.47 & 0 & 0 \\
ConsRecomm & 314 & short & 1 & 0 & 0.4326 & 12.32 & 0 & 0 \\
CredRatDG & 598 & short & 1 & 0 & 0.3489 & 4.77 & 0 & 0 \\
CustomerMomentum & 510 & short & 1 & 0 & 0.9384 & 26.36 & 0 & 0 \\
DebtIssuance & 576 & short & 1 & 0 & 1.3761 & 31.00 & 0 & 0 \\
DelBreadth & 474 & short & 1 & 0 & 1.9968 & 39.88 & 0 & 0 \\
DownRecomm & 313 & short & 1 & 0 & 0.6013 & 9.92 & 0 & 0 \\
EarningsForecastDisparity & 456 & short & 1 & 0 & 0.7396 & 41.19 & 0 & 0 \\
EarningsStreak & 422 & short & 1 & 0 & 0.2753 & 4.91 & 0 & 0 \\
ExclExp & 428 & short & 1 & 0 & 1.1362 & 34.77 & 0 & 0 \\
FEPS & 527 & short & 1 & 0 & 1.3611 & 49.33 & 0 & 0 \\
FR & 462 & short & 1 & 0 & 0.8934 & 29.88 & 0 & 0 \\
ForecastDispersion & 527 & short & 1 & 0 & 1.1421 & 33.19 & 0 & 0 \\
Governance & 137 & short & 1 & 0 & 0.9104 & 27.08 & 0 & 0 \\
IndIPO & 536 & short & 1 & 0 & 0.9523 & 18.35 & 0 & 0 \\
MS & 527 & short & 1 & 0 & 0.2736 & 3.09 & 0 & 0 \\
Mom6mJunk & 598 & short & 1 & 0 & 0.6791 & 17.41 & 0 & 0 \\
NetDebtFinance & 570 & short & 1 & 0 & 0.2961 & 11.59 & 0 & 0 \\
NetEquityFinance & 570 & short & 1 & 0 & 1.3697 & 25.52 & 0 & 0 \\
OptionVolume2 & 270 & short & 1 & 0 & 0.3575 & 2.60 & 0 & 0 \\
OrderBacklog & 582 & short & 1 & 0 & 0.2548 & 8.84 & 0 & 0 \\
PS & 576 & short & 1 & 0 & 1.0468 & 49.93 & 0 & 0 \\
PctTotAcc & 378 & short & 1 & 0 & 0.2992 & 6.10 & 0 & 0 \\
PredictedFE & 438 & short & 1 & 0 & 0.4915 & 14.28 & 0 & 0 \\
ProbInformedTrading & 180 & short & 1 & 0 & 0.9991 & 41.58 & 0 & 0 \\
RDIPO & 512 & short & 1 & 0 & 0.2388 & 3.00 & 0 & 0 \\
RDS & 558 & short & 1 & 0 & 0.3044 & 4.03 & 0 & 0 \\
RIO\_Disp & 527 & short & 1 & 0 & 0.2977 & 4.41 & 0 & 0 \\
RIVolSpread & 264 & short & 1 & 0 & 0.4526 & 5.67 & 0 & 0 \\
Recomm\_ShortInterest & 313 & short & 1 & 0 & 0.3242 & 17.42 & 0 & 0 \\
ShareRepurchase & 570 & short & 1 & 0 & 1.2740 & 48.81 & 0 & 0 \\
ShortInterest & 563 & short & 1 & 0 & 0.4925 & 19.99 & 0 & 0 \\
SmileSlope & 264 & short & 1 & 0 & 0.3696 & 6.94 & 0 & 0 \\
UpRecomm & 313 & short & 1 & 0 & 1.0794 & 23.31 & 0 & 0 \\
betaVIX & 407 & short & 1 & 0 & 0.9134 & 38.68 & 0 & 0 \\
dCPVolSpread & 263 & short & 1 & 0 & 0.3893 & 6.85 & 0 & 0 \\
dVolCall & 263 & short & 1 & 0 & 0.2796 & 2.55 & 0 & 0 \\
fgr5yrLag & 450 & short & 1 & 0 & 0.8056 & 20.04 & 0 & 0 \\
iomom\_cust & 407 & short & 1 & 0 & 1.5715 & 17.68 & 0 & 0 \\
iomom\_supp & 407 & short & 1 & 0 & 0.9206 & 24.11 & 0 & 0 \\
realestate & 594 & short & 1 & 0 & 0.4797 & 14.20 & 0 & 0 \\
retConglomerate & 527 & short & 1 & 0 & 0.4290 & 11.67 & 0 & 0 \\
skew1 & 264 & short & 1 & 0 & 0.2835 & 7.73 & 0 & 0 \\
\end{longtable}
\endgroup

%% file: tabs/prop4_decomp_summary.tex
\begin{table}[htbp]
\centering
\caption{Summary Statistics for Proposition 4 Decomposition}
\label{tab:prop4_decomp_summary}
\begin{tabular}{lr}
\toprule
\textbf{Statistic} & \textbf{Value} \\
\midrule
Number of anomalies ($N$) & 212 \\
\addlinespace
Mean of $\pi_k$ & 0.4420 \\
Standard deviation of $\pi_k$ & 0.0140 \\
\addlinespace
Mean of $\mu_{1,k}$ & 0.1424 \\
Standard deviation of $\mu_{1,k}$ & 0.2099 \\
\addlinespace
Mean of $\mu_{0,k}$ & 0.0953 \\
Standard deviation of $\mu_{0,k}$ & 0.0707 \\
\addlinespace
Mean of $E[\Delta_k]$ & 0.1168 \\
Standard deviation of $E[\Delta_k]$ & 0.1340 \\
\addlinespace
Correlation ($E[\Delta_k]$, $\widehat{E}[\Delta_k]$) & 1.0000 \\
Mean absolute decomposition error & 0.0000 \\
Max absolute decomposition error & 0.0000 \\
Pooled 90th percentile of $\Delta$ ($\tau_{90}$) & 0.3573 \\
\bottomrule
\end{tabular}
\end{table}

%% file: tabs/prop4_decomp_xsec.tex
\begin{table}[htbp]
\centering
\caption{Cross-sectional HC3 regressions for Proposition 4 reduced-form diagnostic}
\label{tab:prop4_decomp_xsec_appendix}
\scriptsize

\resizebox{\textwidth}{!}{%
\begin{tabular}{lccccccccccc}
\toprule
 & (A1) & (A2) & (A3) & (B1) & (B2) & (B3) & (C1) & (C2) & (C3) & (C4) & (C5) \\
\midrule

Dep. var.
& $\mu_1$ & $\mu_1$ & $\mu_1$
& $\pi$ & $\pi$ & $\pi$
& $\mu_1$ & $E\Delta$ & $E\Delta$
& spikefreq & spikefreq \\

\midrule

IVOL
& 0.0623** &  & 0.0624**
& 0.0009 &  & 0.0009
&  &  &  &  &  \\
& (0.0303) &  & (0.0301)
& (0.0021) &  & (0.0021)
&  &  &  &  &  \\

$1-R^2$
&  & -0.0054 & -0.0064
&  & 0.0011 & 0.0011
&  &  &  &  &  \\
&  & (0.0142) & (0.0133)
&  & (0.0008) & (0.0009)
&  &  &  &  &  \\

$\pi$
&  &  & 
&  &  & 
& 6.4378** & 4.5572** & -136.5523
& 2.3980*** & -2.2596 \\
&  &  & 
&  &  & 
& (3.1672) & (1.9453) & (159.6500)
& (0.2762) & (12.8734) \\

$\pi^2$
&  &  & 
&  &  & 
&  &  & 158.8247
&  & 5.2423 \\
&  &  & 
&  &  & 
&  &  & (181.9531)
&  & (14.6386) \\

\midrule

$R^2$
& 0.0885 & 0.0007 & 0.0894
& 0.0046 & 0.0060 & 0.0104
& 0.1834 & 0.2255 & 0.3571
& 0.3688 & 0.3697 \\

N
& 212 & 212 & 212
& 212 & 212 & 212
& 212 & 212 & 212
& 212 & 212 \\

\bottomrule
\end{tabular}%
}

\vspace{0.3em}
\parbox{0.95\textwidth}{\footnotesize
\textit{Notes:} HC3 heteroskedasticity-robust standard errors are reported in parentheses.
*, **, and *** denote statistical significance at the 10\%, 5\%, and 1\% levels, respectively.}
\end{table}

%% file: tabs/prop4_decomp_signsummary.tex
\begin{table}[htbp]
\centering
\caption{Compact Sign Summary for Proposition 4 Reduced-Form Diagnostics.}
\label{tab:prop4_decomp_signsummary}
\begin{tabular}{ll l c c}
\toprule
\textbf{Regression} & \textbf{Model} & \textbf{Sign} & \textbf{Coefficient} & \textbf{$p$-value} \\
\midrule
$\mu_1$ on $\mathit{IVOL}_z$ & A1 & Positive & 0.0623 & 0.0401 \\
$\mu_1$ on $(1-R^2)_z$ & A2 & Negative & -0.0054 & 0.7044 \\
$\pi$ on $\mathit{IVOL}_z$ & B1 & Positive & 0.0009 & 0.6457 \\
$\pi$ on $(1-R^2)_z$ & B2 & Positive & 0.0011 & 0.2047 \\
$\mu_1$ on $\pi$ & C1 & Positive & 6.4378 & 0.0421 \\
$E[\Delta]$ on $\pi$ & C3 & Negative & -136.5523 & 0.3924 \\
$E[\Delta]$ on $\pi^2$ & C3 & Positive & 158.8247 & 0.3827 \\
\bottomrule
\end{tabular}
\end{table}

%% file: tabs/prop4_tertile_regressions.tex
\begin{table}[htbp]
\centering
\small
\caption{IVOL-tertile regressions of average mislearning and spike frequency on break-proneness.}
\label{tab:app-prop4-tertile-regressions}
\begin{tabular}{llcccccc}
\toprule
Tertile & Dep. var. & Coef. & SE & $t$ & $p$ & $R^2$ & $N$ \\
\midrule

\multicolumn{8}{l}{\textit{Low IVOL}} \\
 & Mislearning ($\Delta_{k}$) & 2.0278 & 0.4552 & 4.455 & 0.0000 & 0.2733 & 71 \\
 & Spike freq. & 1.6737 & 0.2084 & 8.033 & 0.0000 & 0.5443 & 71 \\

\addlinespace
\multicolumn{8}{l}{\textit{Medium IVOL}} \\
 & Mislearning ($\Delta_{k}$) & 4.4332 & 0.5198 & 8.529 & 0.0000 & 0.5032 & 70 \\
 & Spike freq. & 2.8410 & 0.4422 & 6.425 & 0.0000 & 0.4693 & 70 \\

\addlinespace
\multicolumn{8}{l}{\textit{High IVOL}} \\
 & Mislearning ($\Delta_{k}$) & 8.1121 & 6.4139 & 1.265 & 0.2060 & 0.2729 & 71 \\
 & Spike freq. & 2.9597 & 0.8047 & 3.678 & 0.0002 & 0.3237 & 71 \\

\bottomrule
\end{tabular}

\vspace{0.4em}
\begin{minipage}{0.88\textwidth}
\footnotesize
\textit{Notes:} Dependent variables are average mislearning ($\Delta_{k}$) and spike frequency. HC3 robust standard errors reported.
\end{minipage}
\end{table}

%% file: tabs/prop4_tertile_descriptives.tex
\begin{table}[htbp]
\centering
\small
\caption{Descriptive statistics for the IVOL-tertile validation sample.}
\label{tab:app-prop4-tertile-descriptives}
\begin{tabular}{lcccccccc}
\toprule
Tertile & $N$ & IVOL & $\pi_k$ & $\mathbb{E}[\Delta_{k,t}]$ & $\Pr(\Delta_{k,t} > c)$ & $\text{SD}(\mu_{1,k})$ & $\text{SD}(\mu_{0,k})$ & $\text{SD}(\Delta_{k,t})$ \\
\midrule
Low IVOL    & 71 & 1.7488 & 0.4419 & 0.0794 & 0.0800 & 0.0968 & 0.0407 & 0.0623 \\
Medium IVOL & 70 & 2.8683 & 0.4397 & 0.1142 & 0.0956 & 0.1121 & 0.0543 & 0.0761 \\
High IVOL   & 71 & 4.8653 & 0.4445 & 0.1569 & 0.1161 & 0.3251 & 0.0943 & 0.2036 \\
\bottomrule
\end{tabular}

\vspace{0.4em}
\begin{minipage}{1.0\textwidth}
\footnotesize
\textit{Notes:} IVOL is average idiosyncratic volatility. $\pi_k$ denotes break-proneness. $\mathbb{E}[\Delta_{k,t}]$ is average mislearning, and $\Pr(\Delta_{k,t} > c)$ is spike frequency. $\text{SD}(\mu_{1,k})$ and $\text{SD}(\mu_{0,k})$ denote the standard deviations of the state parameters $\mu_{1,k}$ and $\mu_{0,k}$, respectively. $\text{SD}(\Delta_{k,t})$ is the standard deviation of mislearning.
\end{minipage}
\end{table}

%% file: tabs/anomaly_stage5diag_alt_inference.tex
\begin{table}[htbp]
\centering
\caption{Alternative Inference Checks for 12-Month Anomaly Predictive Regressions}
\label{tab:anomaly_stage5diag_alt_inference}

\small

\begin{tabular}{lccccc}
\toprule
 & Stage5 & Time Clust. & Anom. Clust. & Double Clust. & NW (HAC) \\
\midrule

\multicolumn{6}{l}{\textbf{Panel A: Baseline}} \\
\midrule

\multicolumn{6}{l}{Future Sharpe (12m)} \\
Coefficient & -0.0025 & -0.0025 & -0.0025 & -0.0025 & -0.0025 \\
$p$   & 0.2338  & 0.2338  & 0.2619  & 0.3313  & 0.3372 \\
$N$         & \multicolumn{5}{c}{158{,}038} \\

\addlinespace
\multicolumn{6}{l}{Future Cumulative Return (12m)} \\
Coefficient & 0.0022 & 0.0022 & 0.0022 & 0.0022 & 0.0022 \\
$p$   & 0.1665 & 0.1665 & 0.2385 & 0.2482 & 0.3164 \\
$N$         & \multicolumn{5}{c}{158{,}038} \\

\addlinespace
\multicolumn{6}{l}{Future Volatility (12m)} \\
Coefficient & 0.0048 & 0.0048 & 0.0048 & 0.0048 & 0.0048 \\
$p$   & 0.0011 & 0.0011 & 0.0003 & 0.0008 & 0.0117 \\
$N$         & \multicolumn{5}{c}{158{,}038} \\

\midrule
\multicolumn{6}{l}{\textbf{Panel B: Controlled (Lagged Variables)}} \\
\midrule

\multicolumn{6}{l}{Future Sharpe (12m)} \\
Coefficient & -0.0012 & -0.0012 & -0.0012 & -0.0012 & -0.0012 \\
$p$   & 0.5873  & 0.5873  & 0.5868  & 0.6339  & 0.6551 \\
$N$         & \multicolumn{5}{c}{82{,}887} \\

\addlinespace
\multicolumn{6}{l}{Future Cumulative Return (12m)} \\
Coefficient & 0.0010 & 0.0010 & 0.0010 & 0.0010 & 0.0010 \\
$p$   & 0.4957 & 0.4957 & 0.5157 & 0.5256 & 0.5870 \\
$N$         & \multicolumn{5}{c}{82{,}887} \\

\addlinespace
\multicolumn{6}{l}{Future Volatility (12m)} \\
Coefficient & 0.0027 & 0.0027 & 0.0027 & 0.0027 & 0.0027 \\
$p$   & 0.0036 & 0.0036 & 0.0010 & 0.0037 & 0.0262 \\
$N$         & \multicolumn{5}{c}{82{,}887} \\

\bottomrule
\end{tabular}

\vspace{0.5em}
\begin{minipage}{0.95\textwidth}
\footnotesize
\textit{Notes:} Stage5 = baseline Stage 5 inference; Time Clust. = time-clustered standard errors; 
Anom. Clust. = anomaly-clustered standard errors; Double Clust. = two-way clustering (time and anomaly); 
NW (HAC) = Newey--West heteroskedasticity- and autocorrelation-consistent standard errors.
\end{minipage}

\end{table}

%% file: tabs/anomaly_stage5diag_alt_family_summary.tex
\begingroup
\footnotesize
\setlength{\tabcolsep}{4.5pt}

\begin{center}
\begin{longtable}{p{3.6cm} p{5.2cm} r r r r}
\caption{Appendix summary of family-level anomaly predictive patterns under alternative 12-month outcomes}
\label{tab:anomaly-alt-family} \\
\toprule
Anomaly family & Outcome & Coef. & $p$ & Anomalies & Obs. \\
\midrule
\endfirsthead

\multicolumn{6}{l}{\textit{Table \thetable\ (continued)}} \\
\toprule
Anomaly family & Outcome & Coef. & $p$ & Anomalies & Obs. \\
\midrule
\endhead

\midrule
\multicolumn{6}{r}{\textit{Continued on next page}} \\
\endfoot

\bottomrule
\multicolumn{6}{p{0.96\textwidth}}{\footnotesize \textit{Notes:} This table presents an appendix summary of anomaly predictive patterns at the family level under alternative 12-month future outcomes. “Coef.” denotes the estimated coefficient, “$p$” denotes the $p$-value, “Anomalies” denotes the number of distinct anomalies within the respective group, and “Obs.” denotes the number of observations.} \\
\endlastfoot

\multicolumn{6}{l}{\textit{Panel A. Pooled}} \\
\midrule
Pooled & 12-month future Sharpe ratio & -0.0012 & 0.5873 & 212 & 82,887 \\
Pooled & 12-month future cumulative return & 0.0010 & 0.4957 & 212 & 82,887 \\
Pooled & 12-month future break-state cum. return & 0.0007 & 0.6360 & 212 & 17,427 \\
Pooled & 12-month future downside semivolatility & 0.0019 & 0.0025 & 212 & 82,887 \\
Pooled & 12-month future maximum drawdown & 0.0018 & 0.0024 & 212 & 82,887 \\
Pooled & 12-month future volatility ratio & 0.0118 & 0.0131 & 212 & 82,887 \\
Pooled & 12-month future volatility & 0.0027 & 0.0036 & 212 & 82,887 \\

\midrule
\multicolumn{6}{l}{\textit{Panel B. Profitability and quality}} \\
\midrule
Profitability and quality & 12-month future Sharpe ratio & 0.0588 & 0.0000 & 10 & 3,788 \\
Profitability and quality & 12-month future cumulative return & 0.0235 & 0.0000 & 10 & 3,788 \\
Profitability and quality & 12-month future break-state cum. return & 0.0205 & 0.0000 & 10 & 688 \\
Profitability and quality & 12-month future downside semivolatility & 0.0028 & 0.0002 & 10 & 3,788 \\
Profitability and quality & 12-month future maximum drawdown & 0.0013 & 0.1020 & 10 & 3,788 \\
Profitability and quality & 12-month future volatility ratio & 0.0447 & 0.0000 & 10 & 3,788 \\
Profitability and quality & 12-month future volatility & 0.0062 & 0.0000 & 10 & 3,788 \\

\midrule
\multicolumn{6}{l}{\textit{Panel C. Momentum}} \\
\midrule
Momentum & 12-month future Sharpe ratio & 0.0176 & 0.0725 & 33 & 12,878 \\
Momentum & 12-month future cumulative return & 0.0091 & 0.0014 & 33 & 12,878 \\
Momentum & 12-month future break-state cum. return & 0.0081 & 0.0025 & 33 & 2,828 \\
Momentum & 12-month future downside semivolatility & 0.0034 & 0.1281 & 33 & 12,878 \\
Momentum & 12-month future maximum drawdown & 0.0025 & 0.1682 & 33 & 12,878 \\
Momentum & 12-month future volatility ratio & 0.0287 & 0.0219 & 33 & 12,878 \\
Momentum & 12-month future volatility & 0.0058 & 0.0312 & 33 & 12,878 \\

\midrule
\multicolumn{6}{l}{\textit{Panel D. Investment}} \\
\midrule
Investment & 12-month future Sharpe ratio & -0.0256 & 0.0034 & 16 & 6,512 \\
Investment & 12-month future cumulative return & -0.0036 & 0.0014 & 16 & 6,512 \\
Investment & 12-month future break-state cum. return & -0.0027 & 0.0014 & 16 & 1,424 \\
Investment & 12-month future downside semivolatility & 0.0019 & 0.0000 & 16 & 6,512 \\
Investment & 12-month future maximum drawdown & 0.0024 & 0.0018 & 16 & 6,512 \\
Investment & 12-month future volatility ratio & 0.0188 & 0.0506 & 16 & 6,512 \\
Investment & 12-month future volatility & 0.0027 & 0.0186 & 16 & 6,512 \\

\midrule
\multicolumn{6}{l}{\textit{Panel E. Reversal, seasonality, and microstructure}} \\
\midrule
Reversal, seasonality, and microstructure & 12-month future Sharpe ratio & -0.0374 & 0.0954 & 34 & 13,587 \\
Reversal, seasonality, and microstructure & 12-month future cumulative return & 0.0014 & 0.8222 & 34 & 13,587 \\
Reversal, seasonality, and microstructure & 12-month future break-state cum. return & 0.0055 & 0.5230 & 34 & 2,520 \\
Reversal, seasonality, and microstructure & 12-month future downside semivolatility & 0.0018 & 0.0352 & 34 & 13,587 \\
Reversal, seasonality, and microstructure & 12-month future maximum drawdown & 0.0024 & 0.0990 & 34 & 13,587 \\
Reversal, seasonality, and microstructure & 12-month future volatility ratio & 0.0097 & 0.2162 & 34 & 13,587 \\
Reversal, seasonality, and microstructure & 12-month future volatility & 0.0023 & 0.1731 & 34 & 13,587 \\

\end{longtable}
\end{center}
\endgroup

%% file: tabs/anomaly_stage5diag_fit_extremes.tex
\begingroup
\footnotesize
\setlength{\tabcolsep}{4.5pt}

\begin{center}
\begin{longtable}{p{5.5cm} p{4.0cm} r r r}
\caption{Appendix robustness checks for anomaly predictive regressions under extreme-value cleaning and sample restrictions}
\label{tab:anomaly-fit-extremes} \\
\toprule
Outcome & Cleaning method & Coef. & $p$ & Obs. \\
\midrule
\endfirsthead

\multicolumn{5}{l}{\textit{Table \thetable\ (continued)}} \\
\toprule
Outcome & Cleaning method & Coef. & $p$ & Obs. \\
\midrule
\endhead

\midrule
\multicolumn{5}{r}{\textit{Continued on next page}} \\
\endfoot

\bottomrule
\multicolumn{5}{p{0.96\textwidth}}{\footnotesize \textit{Notes:} This table reports robustness checks for anomaly predictive regressions, exploring the sensitivity of the results to different extreme-value cleaning methods and sample horizons (all, long, medium, and short legs). “Original” denotes the unadjusted data. “Winsorized (1\%, 99\%)” denotes data winsorized at the 1st and 99th percentiles. “Trimmed (1\%, 99\%)” denotes data trimmed at the 1st and 99th percentiles. “Leave top 1\% abs.” denotes removing observations with absolute values in the top 1\%. “Coef.” denotes the estimated coefficient, “$p$” denotes the $p$-value, and “Obs.” denotes the number of observations.} \\
\endlastfoot

\multicolumn{5}{l}{\textit{Panel A. All samples}} \\
\midrule
12-month future Sharpe ratio & Original & -0.0012 & 0.5873 & 82,887 \\
12-month future Sharpe ratio & Winsorized (1\%, 99\%) & 0.0017 & 0.9260 & 82,887 \\
12-month future Sharpe ratio & Trimmed (1\%, 99\%) & -0.0113 & 0.6414 & 81,229 \\
12-month future Sharpe ratio & Leave top 1\% abs. & -0.0181 & 0.4251 & 82,058 \\
12-month future maximum drawdown & Original & 0.0018 & 0.0024 & 82,887 \\
12-month future maximum drawdown & Winsorized (1\%, 99\%) & 0.0098 & 0.0000 & 82,887 \\
12-month future maximum drawdown & Trimmed (1\%, 99\%) & 0.0071 & 0.0003 & 81,229 \\
12-month future maximum drawdown & Leave top 1\% abs. & 0.0056 & 0.0035 & 82,058 \\
12-month future downside semivolatility & Original & 0.0019 & 0.0025 & 82,887 \\
12-month future downside semivolatility & Winsorized (1\%, 99\%) & 0.0108 & 0.0000 & 82,887 \\
12-month future downside semivolatility & Trimmed (1\%, 99\%) & 0.0076 & 0.0000 & 81,229 \\
12-month future downside semivolatility & Leave top 1\% abs. & 0.0058 & 0.0008 & 82,058 \\
12-month future break-state cum. return & Original & 0.0007 & 0.6360 & 17,427 \\
12-month future break-state cum. return & Winsorized (1\%, 99\%) & 0.0138 & 0.0001 & 17,427 \\
12-month future break-state cum. return & Trimmed (1\%, 99\%) & 0.0123 & 0.0038 & 17,077 \\
12-month future break-state cum. return & Leave top 1\% abs. & 0.0118 & 0.0034 & 17,252 \\

\midrule
\multicolumn{5}{l}{\textit{Panel B. Long leg}} \\
\midrule
12-month future Sharpe ratio & Original & 0.0091 & 0.2884 & 25,221 \\
12-month future Sharpe ratio & Winsorized (1\%, 99\%) & -0.0223 & 0.4700 & 25,221 \\
12-month future Sharpe ratio & Trimmed (1\%, 99\%) & -0.0338 & 0.3742 & 24,715 \\
12-month future Sharpe ratio & Leave top 1\% abs. & -0.0468 & 0.2042 & 24,968 \\
12-month future maximum drawdown & Original & 0.0018 & 0.1775 & 25,221 \\
12-month future maximum drawdown & Winsorized (1\%, 99\%) & 0.0084 & 0.0224 & 25,221 \\
12-month future maximum drawdown & Trimmed (1\%, 99\%) & 0.0054 & 0.1883 & 24,715 \\
12-month future maximum drawdown & Leave top 1\% abs. & 0.0048 & 0.1871 & 24,968 \\
12-month future downside semivolatility & Original & 0.0029 & 0.0866 & 25,221 \\
12-month future downside semivolatility & Winsorized (1\%, 99\%) & 0.0107 & 0.0042 & 25,221 \\
12-month future downside semivolatility & Trimmed (1\%, 99\%) & 0.0064 & 0.0780 & 24,715 \\
12-month future downside semivolatility & Leave top 1\% abs. & 0.0057 & 0.0774 & 24,968 \\
12-month future break-state cum. return & Original & 0.0114 & 0.0085 & 4,874 \\
12-month future break-state cum. return & Winsorized (1\%, 99\%) & 0.0238 & 0.0059 & 4,874 \\
12-month future break-state cum. return & Trimmed (1\%, 99\%) & 0.0136 & 0.1050 & 4,776 \\
12-month future break-state cum. return & Leave top 1\% abs. & 0.0122 & 0.1298 & 4,825 \\

\midrule
\multicolumn{5}{l}{\textit{Panel C. Medium leg}} \\
\midrule
12-month future Sharpe ratio & Original & -0.0085 & 0.0094 & 33,781 \\
12-month future Sharpe ratio & Winsorized (1\%, 99\%) & -0.0274 & 0.2752 & 33,781 \\
12-month future Sharpe ratio & Trimmed (1\%, 99\%) & -0.0311 & 0.3782 & 33,105 \\
12-month future Sharpe ratio & Leave top 1\% abs. & -0.0301 & 0.3715 & 33,443 \\
12-month future maximum drawdown & Original & 0.0020 & 0.0001 & 33,781 \\
12-month future maximum drawdown & Winsorized (1\%, 99\%) & 0.0077 & 0.0006 & 33,781 \\
12-month future maximum drawdown & Trimmed (1\%, 99\%) & 0.0056 & 0.0022 & 33,105 \\
12-month future maximum drawdown & Leave top 1\% abs. & 0.0048 & 0.0044 & 33,443 \\
12-month future downside semivolatility & Original & 0.0020 & 0.0000 & 33,781 \\
12-month future downside semivolatility & Winsorized (1\%, 99\%) & 0.0078 & 0.0001 & 33,781 \\
12-month future downside semivolatility & Trimmed (1\%, 99\%) & 0.0052 & 0.0000 & 33,105 \\
12-month future downside semivolatility & Leave top 1\% abs. & 0.0043 & 0.0002 & 33,443 \\
12-month future break-state cum. return & Original & 0.0002 & 0.7075 & 7,694 \\
12-month future break-state cum. return & Winsorized (1\%, 99\%) & 0.0045 & 0.0503 & 7,694 \\
12-month future break-state cum. return & Trimmed (1\%, 99\%) & 0.0056 & 0.0601 & 7,540 \\
12-month future break-state cum. return & Leave top 1\% abs. & 0.0049 & 0.0872 & 7,617 \\

\midrule
\multicolumn{5}{l}{\textit{Panel D. Short leg}} \\
\midrule
12-month future Sharpe ratio & Original & -0.0007 & 0.7606 & 23,885 \\
12-month future Sharpe ratio & Winsorized (1\%, 99\%) & 0.0350 & 0.1328 & 23,885 \\
12-month future Sharpe ratio & Trimmed (1\%, 99\%) & 0.0291 & 0.3735 & 23,407 \\
12-month future Sharpe ratio & Leave top 1\% abs. & 0.0353 & 0.2452 & 23,646 \\
12-month future maximum drawdown & Original & 0.0017 & 0.0376 & 23,885 \\
12-month future maximum drawdown & Winsorized (1\%, 99\%) & 0.0159 & 0.0000 & 23,885 \\
12-month future maximum drawdown & Trimmed (1\%, 99\%) & 0.0126 & 0.0000 & 23,407 \\
12-month future maximum drawdown & Leave top 1\% abs. & 0.0099 & 0.0004 & 23,646 \\
12-month future downside semivolatility & Original & 0.0018 & 0.0439 & 23,885 \\
12-month future downside semivolatility & Winsorized (1\%, 99\%) & 0.0166 & 0.0000 & 23,885 \\
12-month future downside semivolatility & Trimmed (1\%, 99\%) & 0.0139 & 0.0000 & 23,407 \\
12-month future downside semivolatility & Leave top 1\% abs. & 0.0111 & 0.0000 & 23,646 \\
12-month future break-state cum. return & Original & -0.0009 & 0.6650 & 4,859 \\
12-month future break-state cum. return & Winsorized (1\%, 99\%) & 0.0138 & 0.0053 & 4,859 \\
12-month future break-state cum. return & Trimmed (1\%, 99\%) & 0.0145 & 0.0327 & 4,761 \\
12-month future break-state cum. return & Leave top 1\% abs. & 0.0145 & 0.0244 & 4,810 \\

\end{longtable}
\end{center}
\endgroup